\newif\ifwordcount
\wordcountfalse
\documentclass[10pt,onecolumn,amsmath,nofootinbib,longbibliography,notitlepage,superscriptaddress,prd]{revtex4-1}

\usepackage{graphicx}
\usepackage{float}
\usepackage{natbib,ifthen}
\usepackage{hyperref}
\usepackage[usenames,dvipsnames]{pstricks}
\usepackage{epsfig}
\usepackage{pst-grad}
\usepackage{pst-plot}
\usepackage{listings}
\usepackage{color}
\usepackage{amsmath,amsfonts,amssymb}
\usepackage{simplewick}
\usepackage{dsfont}

\definecolor{dkgreen}{rgb}{0,0.6,0}
\definecolor{gray}{rgb}{0.5,0.5,0.5}
\definecolor{mauve}{rgb}{0.58,0,0.82}
\definecolor{orange}{rgb}{255,119,0}

\newcommand{\planck}{\textsc{Planck}}

\newcommand{\cosmomc}{\textsc{CosmoMC}}

\newcommand{\nhat}{\hat{\textbf{n}}}
\newcommand{\balpha}{\boldsymbol \alpha}
\newcommand{\bnabla}{\boldsymbol \nabla}

\newcommand{\comment}[1]{{}}
\newcommand{\wigner}[6]{\begin{pmatrix} #1 & #2 & #3 \\ #4 & #5 & #6 \end{pmatrix}}

\newcommand{\wignerd}[3]{d_{#1 #2}^{#3}}
\newcommand{\covlensCMB}[4]{\text{cov}(\hat{C}^{\tilde{#1} \tilde{#2}}_{\ell_1,\text{expt}}, \hat{C}^{\tilde{#3} \tilde{#4}}_{\ell_2,\text{expt}})}
\newcommand{\covlensCMBells}[6]{\text{cov}(\hat{C}^{\tilde{#3} \tilde{#4}}_{#1,\text{expt}}, \hat{C}^{\tilde{#5} \tilde{#6}}_{#2,\text{expt}})}
\newcommand{\covCMBNoLens}[6]{\text{cov}(\hat{C}^{#3 #4}_{#1}, \hat{C}^{#5 #6}_{#2})}

\newcommand{\corrCMBNoLens}[6]{\text{corr}(\hat{C}^{#3 #4}_{#1}, \hat{C}^{#5 #6}_{#2})}
\newcommand{\covPhi}[6]{\text{cov}(\hat{C}^{\phi^{#3} \phi^{#4}}_{#1}, \hat{C}^{\phi^{#5} \phi^{#6}}_{#2})}
\newcommand{\covPhiCMB}[6]{\text{cov}(\hat{C}^{\phi^{#3} \phi^{#4}}_{#1}, \hat{C}_{#2,\text{expt}}^{\tilde{#5} \tilde{#6}})}
\newcommand{\covPhiCMBtransp}[6]{\text{cov}(\hat{C}_{#2,\text{expt}}^{\tilde{#5} \tilde{#6}}, \hat{C}^{\phi^{#3} \phi^{#4}}_{#1})}
\newcommand{\Cell}[3]{C_{#1}^{#2 #3}}
\newcommand{\CellHat}[3]{\hat{C}_{#1}^{#2 #3}}
\newcommand{\CellHatRDN}[3]{\hat{C}_{#1,\text{RDN0}}^{#2 #3}}
\newcommand{\Nzero}[3]{N_{#1}^{(0), #2 #3}}
\newcommand{\NzeroHat}[3]{\hat{N}_{#1}^{(0), #2 #3}}
\newcommand{\NzeroHatCAL}[3]{\hat{\mathcal{N}}_{#1}^{(0), #2 #3}}
\newcommand{\Nouane}[3]{N_{#1}^{(1), #2 #3}}
\newcommand{\CellNoHatLensed}[3]{C_{#1}^{\tilde{#2} \tilde{#3}}}
\newcommand{\CellLensednoisy}[3]{\hat{C}_{#1,\text{expt}}^{\tilde{#2} \tilde{#3}}}
\newcommand{\CellNoHatLensednoisy}[3]{C_{#1,\text{expt}}^{\tilde{#2} \tilde{#3}}}
\newcommand{\CellNoHatLensednoisyFid}[3]{C_{#1,\text{expt}}^{\tilde{#2} \tilde{#3} (\rm{fid})}}
\newcommand{\Amplitude}[2]{{\cal{A}}_{#1}^{#2}}
\newcommand{\gweight}[4]{\tilde{g}_{#1 #2}^{#4}(#3)}
\newcommand{\fweightLensed}[4]{\tilde{f}_{#1 #2 #3}^{#4}}
\newcommand{\Fweight}[4]{{}_{#1}F_{#2 #3 #4}}
\newcommand{\OrderStop}[3]{{\cal{O}}([C^{#1 #2}_\ell]^{#3})}
\newcommand{\tT}{\tilde{T}}
\newcommand{\tE}{\tilde{E}}
\newcommand{\tB}{\tilde{B}}
\newcommand{\tX}{\tilde{X}}
\newcommand{\tY}{\tilde{Y}}
\newcommand{\tZ}{\tilde{Z}}
\newcommand{\tW}{\tilde{W}}
\newcommand{\tU}{\tilde{U}}
\newcommand{\tV}{\tilde{V}}
\newcommand{\mC}{{\bold C}}
\newcommand{\barb}{{\bar b}}
\newcommand{\Cov}{\textrm{Cov}}

\newcommand{\la}{\langle}
\newcommand{\ra}{\rangle}

\newcommand{\Sussex}{Department of Physics \& Astronomy, University of Sussex, Brighton BN1 9QH, UK}

\begin{document}

\title{Full covariance of CMB and lensing reconstruction power spectra}
\author{Julien Peloton}
\affiliation{\Sussex}
\author{Marcel Schmittfull}
\affiliation{Institute for Advanced Study, Einstein Drive, Princeton, NJ 08540, USA}
\affiliation{Berkeley Center for Cosmological Physics, University of California,
Berkeley, CA 94720, USA}
\author{Antony Lewis}
\affiliation{\Sussex}
\author{Julien Carron}
\affiliation{\Sussex}
\author{Oliver Zahn}
\affiliation{Berkeley Center for Cosmological Physics, University of California,
Berkeley, CA 94720, USA}

\date{\today}

\begin{abstract}
CMB and lensing reconstruction power spectra are powerful probes of cosmology.
However they are correlated, since the CMB power spectra are lensed
and the lensing reconstruction is constructed using CMB multipoles.
We perform a full analysis of the auto- and cross-covariances, including polarization power spectra and minimum variance lensing estimators, and compare with simulations of idealized future CMB-S4 observations.
Covariances sourced by fluctuations in the unlensed CMB and instrumental noise can largely be removed by using a realization-dependent subtraction of lensing reconstruction noise, leaving a
relatively simple covariance model that is dominated by lensing-induced terms and well described by a small number of principal components.
The correlations between the CMB and lensing power spectra will be detectable at the level of $\sim 5\sigma$ for a CMB-S4 mission, and neglecting those could underestimate some parameter error bars by several tens of percent.
However we found that the inclusion of external priors or data sets to estimate parameter error bars can make the impact of the correlations almost negligible.
\end{abstract}


\ifwordcount
\else
\maketitle
\fi

\section{Introduction} \label{sec:introduction}

Gravitational lensing is the leading non-linear effect on the observed CMB anisotropies on intermediate and large scales.
The lensing smooths out acoustic peaks in the temperature and E-polarization power spectra, generates B-mode polarization by lensing of E modes, and transfers power into the damping tail at very high $\ell$. The non-Gaussianity of the signal can also be used to reconstruct the lensing potential, which then has its own power spectrum which can be a powerful cosmological probe.

Several works \cite{Smith:2004up,Smith:2005ue,Smith:2006nk,Li:2006pu, BenoitLevy:2012va, Schmittfull:2013uea} have studied the auto-covariance of the lensed CMB power spectra. The lensing-induced peak smoothing correlates different multipoles, since when the lensing power fluctuates high the smoothing increases everywhere: all the CMB power spectrum peaks go down, and all the troughs go up. However, lensing is only a small part of the signal in the T and E-polarization, and quite a large number of lensing modes contribute (at $\ell \alt 100$), so the off-diagonal lensing covariance is small compared to the total. For the B-mode power spectrum the effect is much more important, since (except possibly on very large scale) all of the signal is expected to be generated by lensing.

Since lensing reconstruction also probes the lensing modes more directly, it is not independent of the lensing effect on the CMB power spectra. Furthermore, the four-point estimator for the lensing reconstruction power spectrum uses the lensed CMB modes, so the reconstruction is also not independent of fluctuations in the unlensed CMB. Both effects lead to covariance between the lensing reconstruction power and the observed CMB power spectra. The correlations have been studied in detail by Ref.~\cite{Schmittfull:2013uea} for the CMB temperature, where the effect is shown to be small at \planck\ sensitivity. However, future observations will have much higher signal-to-noise reconstructions, and will also observe the CMB polarization at much higher sensitivity and resolution, so a full polarization analysis is timely to avoid potential double counting of information. With high signal-to-noise lensing reconstructions, ultimately it may be possible to delens most of the lensing contributions to the CMB power spectra, rendering the spectra more independent~\cite{Green:2016cjr}. However even with delensing there will be some residual correlation that needs to be modelled, and for the foreseeable future it will remain an important consistency check that compatible results can be obtained using the lensed spectra.

Considering just the lensing reconstruction alone, the complicated four-point nature of the estimator means that there may be non-trivial correlations between multipoles that need to be modelled consistently to construct a reliable likelihood. Refs.~\cite{Hanson:2010rp,Namikawa:2012pe,Schmittfull:2013uea} have shown that by using knowledge of the observed CMB power spectra it is possible to use a realization-dependent reconstruction noise subtraction that removes the dominant off-diagonal correlation due to fluctuations in the CMB power (and hence also much of the correlation with the CMB power). A generalization to realistic cut sky filters, cross-spectra and polarization was used by the \planck\ analysis~\cite{Ade:2015zua}, and can be motivated by the form of the optimal four-point estimator~\cite{Regan:2010cn,Schmittfull:2013uea,Ade:2013mta} in temperature.

As data accuracy improves, polarization will become relatively more important, as EB lensing reconstruction becomes better than TT reconstruction once the noise levels are low enough due to the absence of intrinsic small scale fluctuations in B. Ongoing CMB experiments already started to use minimum variance reconstructions of the lensing potential power spectrum in intensity and in polarization, although the intensity measurement dominates the total reconstruction so far~\cite{Story:2014hni,Ade:2015zua,Array:2016afx}.
Already ground-based high-sensitivity CMB experiments such as the \textsc{Simons Array}\footnote{\url{http://cosmology.ucsd.edu/simonsarray.html}}, the South Pole Telescope (\textsc{SPT-3G})\footnote{\url{ https://pole.uchicago.edu/spt/}}, the Advanced Atacama Cosmology Telescope (\textsc{AdvACT})\footnote{\url{ https://act.princeton.edu}}, and the \textsc{Simons Observatory}\footnote{\url{ https://simonsobservatory.org}} are under deployment and future ground-based facilities such as \textsc{CMB-S4}\footnote{\url{http://CMB-S4.org}}, as well a possible space satellite,
are being proposed to further increase the sensitivity at high resolution.
See Fig.~\ref{fig:n0_n1_bias} for the different levels of reconstruction noise in the case of a CMB-S4 like experiment, which has sufficiently low noise that the EB reconstruction contains most, but not all, of the information.
In this paper, we focus on the minimum variance reconstruction of the lensing potential power-spectrum, including semi-optimally weighted combinations of all estimators between T, E, and B.

Our paper is organized as follows. We start with a review of CMB lensing and its reconstruction in Sec.~\ref{sec:background} to lay out our notation.
We then describe our analytical model for the auto- and cross-covariances of CMB and lensing power spectra in Sec.~\ref{sec:covariances}, and compare it against simulations described in Sec.~\ref{sec:simulations}.
We discuss the impact of the correlations on parameter estimation in Sec.~\ref{sec:impact}, and conclude in Sec.~\ref{sec:conclusions}.
Details of analytical calculations and numerical evaluations are presented in a series of appendices.

\section{CMB lensing and reconstruction} \label{sec:background}

\subsection{Weak lensing of the CMB}

At the epoch of recombination the Universe becomes mostly transparent to photons, and is well approximated as a single source plane for CMB photons.
Weak gravitational lensing by large-scale structure along the line of sight gives small but important distortions to the primordial anisotropies of the CMB. We can relate the lensed CMB field $\tX(\nhat)$ along direction $\nhat$ to the unlensed field $X(\nhat)$ by the deflection angle $\balpha(\nhat)$:

\begin{equation}\label{eq:lensed-CMB-multipoles}
\tX(\nhat) = X(\nhat+\balpha(\nhat)).
\end{equation}
where $X \in \{ T, E, B \}$.
In the Born approximation\footnote{Throughout this paper we neglect the corrections introduced by post-Born lensing and large-scale structure non-Gaussianity; these have a negligible impact on the CMB power spectra~\cite{Pratten:2016dsm,Lewis:2016tuj}, though small biases in the quadratic estimators may ultimately need to be separately corrected to avoid biased estimates~\cite{Bohm:2016gzt}.}, we define the (projected) lensing potential $\phi$ as

\begin{equation}\label{eq:lensing-potential-phi}
\phi(\nhat) = - 2 \int_0^{\chi_*} d\chi \dfrac{f_K(\chi_* - \chi)}{f_K(\chi_*)f_K(\chi)} \Psi (\chi \nhat, \eta_0 - \chi),
\end{equation}
where $\Psi$ is the (Weyl) gravitational potential.
The deflection angle is given by the angular derivative of the lensing potential, $\balpha(\nhat) = \bnabla \phi(\nhat)$.
The lensing potential is an integrated measure of the mass distribution back to the moment of recombination, including geometrical effects of the background through $f_K$, which is the angular diameter distance and encodes the relationship between comoving distance and angle.

The lensing potential can be decomposed into multipole moments
\begin{equation}
\phi(\nhat) = \sum_{\ell m} \phi_{\ell m} Y_{\ell}^m(\nhat),
\end{equation}
and the effect of lensing on the unlensed CMB field can be expressed perturbatively by Taylor expanding Eq.~\ref{eq:lensed-CMB-multipoles} in the harmonic domain \cite{Hu:2000ee}
\begin{equation} \label{eq:taylor-CMB-multipoles}
\tX_{\ell m} = X_{\ell m} + \delta X_{\ell m} + \delta^2 X_{\ell m} + ...,
\end{equation}
where $\tX_{\ell m}$ are the multipoles of the lensed CMB.
High order terms ($\delta^nX_{\ell m}$) are due to the effect of the lensing, and are of the order ${\cal O} (\phi^n)$ and linear in the unlensed CMB field.
For example, we have at first order \cite{Okamoto03}

\begin{align}\label{eq:taylor-CMB-multipoles-order1}
\delta X_{\ell m} &= \sum_{\ell_1 m_1}\sum_{\ell_2 m_2} (-1)^m \phi_{\ell_1 m_1} \wigner{\ell}{\ell_2}{\ell_1}{m}{-m_2}{-m_1} \Fweight{s_{X}}{\ell}{\ell_1}{\ell_2} [ \epsilon_{\ell \ell_1 \ell_2} X_{\ell_2 m_2} + \beta_{\ell \ell_1 \ell_2} \bar{X}_{\ell_2 m_2} ],
\end{align}
where we used parity complements $\bar{T} = 0$, $\bar{E}=-B$, and $\bar{B} = E$.
The expression for the second order term can be found for example in Ref.~\cite{Hanson:2010rp}.
The $\epsilon$ and $\beta$ parity terms are defined as

\begin{equation}
\epsilon_{\ell \ell_1 \ell_2} = \dfrac{1+(-1)^{\ell+\ell_1+\ell_2}}{2}, \qquad \beta_{\ell \ell_1 \ell_2} = \dfrac{1-(-1)^{\ell+\ell_1+\ell_2}}{2i},
\end{equation}
and the function $F$ is defined as

\begin{equation}
\Fweight{s_{X}}{\ell}{\ell_1}{\ell_2} = [- \ell(\ell+1) + \ell_1(\ell_1+1) + \ell_2(\ell_2+1)] \sqrt{\dfrac{(2\ell+1)(2\ell_1+1)(2\ell_2+1)}{16\pi}}\wigner{\ell}{\ell_1}{\ell_2}{s_{X}}{0}{-s_{X}},
\end{equation}
where $s_{X}$ is the spin number of CMB field $X$ (zero for temperature, two for polarization)..

\subsection{Lensing reconstruction}\label{sec:lensing_reconstruction}

The lensing potential is approximately Gaussian and its power spectrum on the full sky is given by:

\begin{equation}\label{eq:phiphi2cl}
\langle \phi_{\ell m} \phi_{\ell' m'}^{*} \rangle = \delta_{\ell \ell'} \delta_{m m'} C_{\ell}^{\phi \phi}.
\end{equation}
We can define an estimator $\CellHat{\ell}{\phi}{\phi}$ for this power spectrum as

\begin{equation}\label{eq:estimator-phiphi_theo}
\CellHat{\ell}{\phi}{\phi} = \dfrac{1}{2\ell +1} \displaystyle \sum_{m=-\ell}^{\ell} | \hat{\phi}_{\ell m}|^2.
\end{equation}
We can reconstruct\footnote{neglecting $X-\phi$ correlations.} $\hat{\phi}_{\ell m}$ from lensed CMB modes by noting that fixed lenses introduce correlations between CMB modes $X,Y \in \{ T, E, B \}$.
In the context of quadratic estimators \cite{Seljak:1998aq,Hu:2001tn,Hu:2001kj,Okamoto03}, we have

\begin{equation}\label{eq:estimator-phi}
\hat{\phi}_{\ell m}^{XY} = \Amplitude{\ell}{XY} \displaystyle \sum_{\ell_1 m_1}\sum_{\ell_2 m_2} (-1)^m \wigner{\ell_1}{\ell_2}{\ell}{m_1}{m_2}{-m} \gweight{\ell_1}{\ell_2}{\ell}{XY} \tX_{\ell_1 m_1}\tY_{\ell_2 m_2},
\end{equation}
where $\Amplitude{\ell}{XY}$ is a normalization factor which ensures that $\hat{\phi}_{\ell m}^{XY}$ is unbiased

\begin{equation} \label{eq:amplitude-estimator}
\Amplitude{\ell}{XY} = (2\ell+1) \Big( \sum_{\ell_1 \ell_2} \fweightLensed{\ell_1}{\ell}{\ell_2}{XY}\gweight{\ell_1}{\ell_2}{\ell}{XY}  \Big)^{-1},
\end{equation}
with the response functions $f$ given by
\begin{equation}\label{eq:fweights}
\fweightLensed{\ell_1}{\ell}{\ell_2}{XY}  \approx \Fweight{s_{X}}{\ell_1}{\ell}{\ell_2} \Big( \epsilon_{\ell \ell_1 \ell_2}  \Cell{\ell_2}{\tX}{\tY} + \beta_{\ell \ell_1 \ell_2} \Cell{\ell_2}{\bar{\tX}}{\tY} \Big) + \Fweight{s_Y}{\ell_2}{\ell}{\ell_1} \Big( \epsilon_{\ell \ell_1 \ell_2}  \Cell{\ell_1}{\tX}{\tY} - \beta_{\ell \ell_1 \ell_2} \Cell{\ell_1}{\tX}{\bar{\tY}} \Big).
\end{equation}
For estimators involving polarization we use an approximation for the non-perturbative response function that follows the form of the lowest-order perturbative result but uses lensed spectra in the expression for $\tilde{f}$ (as written here, denoted by a tilde, following Refs.~\cite{Hanson:2010rp,Lewis:2011fk}).
For unbiased results from very small-scale temperature we found it was necessary to replace the lensed power spectra by $C_{\ell}^{\tT \nabla \tT}$,
the lensed temperature-gradient power spectrum that appears in the non-perturbative response function~\cite{Lewis:2011fk}\footnote{To compute $C_{\ell}^{\tT \nabla \tT}$, we follow the full-sky derivation using correlation functions in the Appendix C in Ref.~\cite{Lewis:2011fk}. We discuss the magnitude of this effect in Sec.~\ref{sec:simulations}.}.

\begin{figure}[t]
\begin{center}
\includegraphics[width=0.8\textwidth]{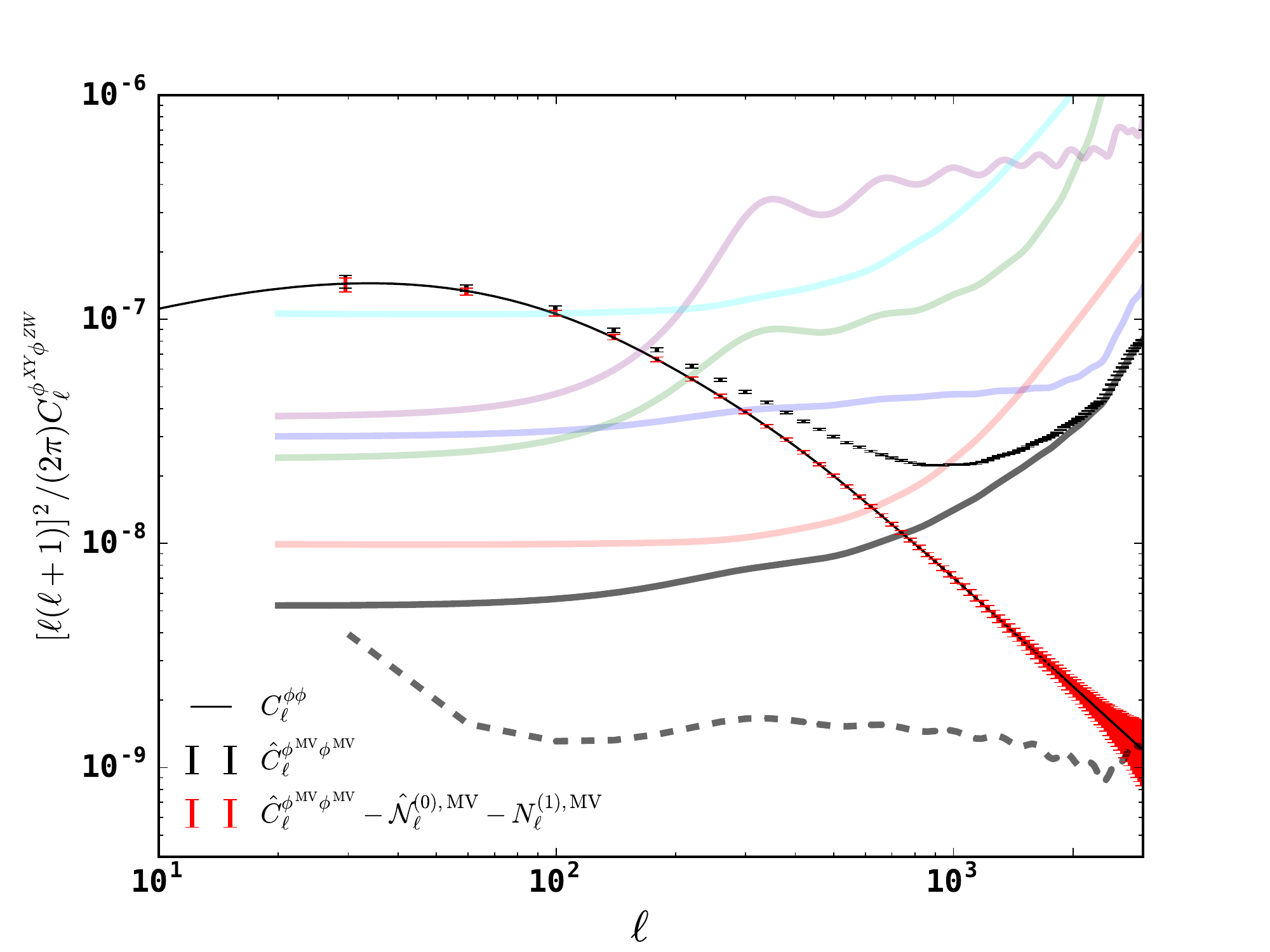}
\caption{Minimum-variance lensing power spectrum measured from our set of simulations (red points with error bars) and theoretical expectation (solid thin black line),
assuming a possible CMB-S4 configuration with 1.5 $\mu$K.arcmin white noise for temperature, 3 arcmin beam, multipole range $20 \leq \ell \leq 3000$ and sky coverage $f_\mathrm{sky}=0.4$.
The measurement is obtained by subtracting the realization-dependent noise bias $\NzeroHatCAL{ }{\rm MV}{ }$ and analytical $N^{(1),\rm{MV}}$ bias from the uncorrected measured lensing power spectrum (black points with error bars).
Coloured lines show the analytically-calculated $N^{(0)}$ biases for various combinations (see Eq.~\ref{eq:gaussian_bias}): TTTT (blue), EEEE (green), TETE (purple), TBTB (cyan), EBEB (red), and the minimum variance noise biases (solid thick black line for $N^{(0),\rm{MV}}$ and dashed thick black line for $N^{(1),\rm{MV}}$).
}
\label{fig:n0_n1_bias}
\end{center}
\end{figure}

The optimal weights $\tilde{g}$ can be found by minimising the variance of the estimator for a fiducial model\,\footnote{Sometimes it is useful to maintain separability of the individual polarization estimators, in which case the cross-correlation term in the denominator can be dropped (as in the \planck\ analysis), with a small loss of optimality~\cite{Hu:2001kj}. Here we use the full joint-analysis result.
}

\begin{equation} \label{eq:gweights}
\gweight{\ell_1}{\ell_2}{\ell}{XY} = \dfrac{\CellNoHatLensednoisyFid{\ell_2}{X}{X} \CellNoHatLensednoisyFid{\ell_1}{Y}{Y} \fweightLensed{\ell_1}{\ell}{\ell_2}{XY(\rm{fid})*} - (-1)^{\ell+\ell_1+\ell_2} \CellNoHatLensednoisyFid{\ell_1}{X}{Y} \CellNoHatLensednoisyFid{\ell_2}{X}{Y} \fweightLensed{\ell_2}{\ell}{\ell_1}{XY(\rm{fid})*}}{\CellNoHatLensednoisyFid{\ell_1}{X}{X} \CellNoHatLensednoisyFid{\ell_2}{X}{X}\CellNoHatLensednoisyFid{\ell_1}{Y}{Y} \CellNoHatLensednoisyFid{\ell_2}{Y}{Y} - (\CellNoHatLensednoisyFid{\ell_1}{X}{Y} \CellNoHatLensednoisyFid{\ell_2}{X}{Y})^2}.
\end{equation}
Here $\CellNoHatLensednoisyFid{\ell}{X}{Y}$ is the fiducial expectation of the total lensed CMB power spectrum, including signal and the noise
\begin{equation}\label{eq:lensed_theo_spectra}
\CellNoHatLensednoisyFid{\ell}{X}{Y} = \Cell{\ell}{\tilde X}{\tilde Y (\rm{fid})} + \sigma_{XY}^{2}\exp[\ell(\ell +1)\sigma_{\rm{FWHM}}^2/(8\ln2)],
\end{equation}
and $\sigma^2_{XY}$ is the level of instrumental noise (assumed homogeneous and isotropic), and $\sigma_{\rm{FWHM}}$ is the full-width half maximum of the optical beam (assumed perfectly gaussian). The response functions are also evaluated with the fiducial theoretical lensed power spectra $\Cell{\ell}{\tilde X}{\tilde Y (\rm{fid})}$.
Throughout this manuscript, we focus on a possible \textsc{CMB-S4} configuration, and we set the level of noise in temperature to $\sigma_{TT} = 1.5$ $\mu$K.arcmin ($\sqrt{2}$ bigger in polarization), and we assume a beam width $\sigma_{\rm{FWHM}} = 3$ arcmin.
We restrict our analysis to 40$\%$ of the sky to take into account the fact that CMB-S4 will not be able to cover the full sky from the ground. Furthermore, we restrict the multipole range to the interval $20 \leq \ell \leq 3000$, to mimic the difficulty for a ground-based experiment to deal with foreground contamination, atmosphere contamination, or masking effects. We later discuss the effect of extending the multipole range.
Using Eq.~\ref{eq:estimator-phiphi_theo} and \ref{eq:estimator-phi}, we obtain

\begin{equation}\label{eq:estimator-phiphi}
\CellHat{\ell}{\phi^{XY}}{\phi^{ZW}} = \dfrac{1}{2\ell +1}  \sum_{m=-\ell}^{\ell} (-1)^m \hat{\phi}_{\ell m}^{XY} \hat{\phi}_{\ell, -m}^{ZW},
\end{equation}
for which the expectation value is given by

\begin{equation} \label{eq:phiphi_expectation_value}
\langle \CellHat{\ell}{\phi^{XY}}{\phi^{ZW}} \rangle = \Nzero{\ell}{XY}{ZW} + \Cell{\ell}{\phi}{\phi} + \Nouane{\ell}{XY}{ZW} + \OrderStop{\phi}{\phi}{3}.
\end{equation}
where $N^{(n)}$ is the reconstruction ``noise'' of order $\OrderStop{\phi}{\phi}{n}$.
The first term is called the Gaussian reconstruction noise (or disconnected part of the lensed CMB 4-point function), and its general expression is

\begin{equation}\label{eq:gaussian_bias}
\Nzero{\ell}{XY}{ZW} = \dfrac{\Amplitude{\ell}{XY} \Amplitude{\ell}{ZW}}{2\ell +1} \sum_{\ell_1 \ell_2} \gweight{\ell_1}{\ell_2}{\ell}{XY} \Big[ (-1)^{\ell+\ell_1+\ell_2} \gweight{\ell_1}{\ell_2}{\ell}{ZW}\CellNoHatLensednoisy{\ell_1}{X}{Z}\CellNoHatLensednoisy{\ell_2}{Y}{W} + \gweight{\ell_2}{\ell_1}{\ell}{ZW}\CellNoHatLensednoisy{\ell_1}{X}{W}\CellNoHatLensednoisy{\ell_2}{Y}{Z}  \Big].
\end{equation}
The $N^{(1)}$ bias (linear in the lensing potential spectrum) is taken into account following Refs.~\cite{Kesden:2003cc,Ade:2015zua}.
Notice that due to the complex form of this term, we compute it in the flat-sky approximation, which is valid on large angular scales $\ell < 20$.
Given that in this work we are interested in small scales ($\ell > 100$) and the influence of the $N^{(1)}$ bias is mostly at small scales (see Fig.~\ref{fig:n0_n1_bias}), we do not expect any difference with respect to the curved-sky result.

As for the temperature case \cite{Hanson:2010rp,Schmittfull:2013uea} (also see \cite{Dvorkin:2008tf}), we can define realization-dependent noise-subtracted power spectra by forming
\begin{align}\label{eq:clphiphi-RDN0}
\CellHatRDN{\ell}{\phi^{XY}}{\phi^{ZW}} &=
\CellHat{\ell}{\phi^{XY}}{\phi^{ZW}} - \Nzero{\ell}{XY}{ZW}  -\sum_{ab, \ell'} \frac{\partial \Nzero{\ell}{XY}{ZW}}{\partial \CellNoHatLensednoisy{\ell'}{a}{b}} (\CellLensednoisy{\ell'}{a}{b} - \CellNoHatLensednoisy{\ell'}{a}{b})  \nonumber\\ &=
\CellHat{\ell}{\phi^{XY}}{\phi^{ZW}} - 2\NzeroHat{\ell}{XY}{ZW}  + \Nzero{\ell}{XY}{ZW} ,
\end{align}
where the realization-dependent $\NzeroHat{\phantom{\ell}}{XY}{ZW}$ (RDN0 hereafter) is defined by replacing some of the lensed CMB spectra in Eq.~\ref{eq:gaussian_bias} by their observed realization:
\begin{align} \label{eq:RDN0_gaussian_bias}
2\NzeroHat{\ell}{XY}{ZW} &=
\sum_{ab, \ell'} \frac{\partial \Nzero{\ell}{XY}{ZW}}{\partial \CellNoHatLensednoisy{\ell'}{a}{b}} \CellLensednoisy{\ell'}{a}{b} \nonumber \\
&=
\dfrac{\Amplitude{\ell}{XY} \Amplitude{\ell}{ZW}}{2\ell +1} \sum_{\ell_1 \ell_2} \gweight{\ell_1}{\ell_2}{\ell}{XY} \Big[ (-1)^{\ell+\ell_1+\ell_2} \gweight{\ell_1}{\ell_2}{\ell}{ZW} \Big( \CellLensednoisy{\ell_1}{X}{Z}\CellNoHatLensednoisy{\ell_2}{Y}{W}  + \CellNoHatLensednoisy{\ell_1}{X}{Z}\CellLensednoisy{\ell_2}{Y}{W} \Big)
\nonumber \\&
\phantom{= \dfrac{\Amplitude{\ell}{XY} \Amplitude{\ell}{ZW}}{2\ell +1} \sum_{\ell_1 \ell_2} \gweight{\ell_1}{\ell_2}{\ell}{XY} \Big[ }
+ \gweight{\ell_2}{\ell_1}{\ell}{ZW} \Big( \CellLensednoisy{\ell_1}{X}{W}\CellNoHatLensednoisy{\ell_2}{Y}{Z} + \CellNoHatLensednoisy{\ell_1}{X}{W}\CellLensednoisy{\ell_2}{Y}{Z} \Big)  \Big].
\end{align}
Notice that in the fiducial model $\langle \NzeroHat{\ell}{XY}{ZW} \rangle = \Nzero{\ell}{XY}{ZW}$, but with the advantage
that the realization-dependent subtraction takes out fluctuations in $\CellHat{\ell}{\phi^{XY}}{\phi^{ZW}}$ due
to reconstruction noise fluctuations from the realization of the CMB and noise power (and also any leading error from inaccuracy in the fiducial  $\CellNoHatLensednoisyFid{\ell_1}{X}{Z}$ assumed), as can be seen
from the first line of Eq.~\eqref{eq:clphiphi-RDN0}.
In order to simplify notation, we define
\begin{equation}
\CellHatRDN{\ell}{\phi^{XY}}{\phi^{ZW}} = \CellHat{\ell}{\phi^{XY}}{\phi^{ZW}} - \NzeroHatCAL{\ell}{XY}{ZW}
\end{equation}
where $\NzeroHatCAL{\ell}{XY}{ZW} \equiv 2\NzeroHat{\ell}{XY}{ZW}  - \Nzero{\ell}{XY}{ZW}$.
As we shall see later, the data-dependent noise mitigation of Eq.~\ref{eq:clphiphi-RDN0} also simplifies covariances, removing almost all of the noise correlations (see Sec.~\ref{sec:simulations}). At low noise levels iterative estimators may be able to do significantly better than the simple quadratic estimators, but for simplicity we restrict to quadratic estimator reconstruction here.

\subsection{Minimum variance reconstruction}

The reconstructed minimum variance (MV) lensing potential $\hat{\phi}^{\rm MV}$ can be expressed in term of the individual reconstructed lensing potentials $\hat{\phi}^{XY}$ as

\begin{equation}\label{eq:MV_phi}
\hat{\phi}^{\rm MV}_{\ell m} = \sum_{XY} w_\ell^{XY} \hat{\phi}^{XY}_{\ell m}.
\end{equation}
The summation over XY is done over the 6 pairs TT, EE, BB, TE, TB, and EB.
The weights $w^{XY}$ depend on the reconstruction noise \cite{Okamoto03}, and are given by

\begin{equation}\label{eq:weights_for_MV}
w_\ell^{XY} = N_\ell^{(0),\rm MV} \sum_{ZW} (\textbf{N}_\ell^{(0)}{}^{-1})^{XYZW},
\end{equation}
where $\textbf{N}^{(0)}$ is a matrix containing all the individual reconstruction noises and the minimum variance reconstruction noise given by

\begin{equation}\label{eq:MV_N0}
N_\ell^{(0),\rm MV} = \dfrac{1}{\displaystyle \sum_{XYZW} (\textbf{N}_\ell^{(0)}{}^{-1})^{XYZW}}.
\end{equation}
Using Eq.~\ref{eq:MV_phi}, we obtain the minimum variance lensing potential power-spectrum

\begin{equation}
\CellHat{\ell}{\phi^{\rm MV}}{\phi^{\rm MV}} = \sum_{XYZW} w_\ell^{XY} w_\ell^{ZW} \CellHat{\ell}{\phi^{XY}}{\phi^{ZW}}.
\end{equation}
Building on this, the auto-covariance matrix for the reconstructed minimum variance lensing potential power spectrum is given in terms of all individual covariances by

\begin{equation}
\covPhi{\ell_1}{\ell_2}{\rm MV}{\rm MV}{\rm MV}{\rm MV} =\sum_{XY,ZW} \sum_{X^\prime Y^\prime, Z^\prime W^\prime} w_{\ell_1}^{XY} w_{\ell_1}^{ZW} \covPhi{\ell_1}{\ell_2}{XY}{ZW}{X^\prime Y^\prime}{Z^\prime W^\prime} w_{\ell_2}^{X^\prime Y^{\prime}} w_{\ell_2}^{Z^\prime W^{\prime}},
\end{equation}
for XY, ZW, X$^\prime$Y$^\prime$, Z$^\prime$W$^\prime$ running over $\{ TT, EE, TE, BB, TB, EB \}$.
We will include all covariance terms, including off-diagonal contributions (see Sec.~\ref{sec:four-point-covariances}).
Similarly, the cross-covariance between the reconstructed minimum variance lensing potential power spectrum and the estimated lensed CMB spectra can be written as

\begin{equation}
\covPhiCMB{\ell_1}{\ell_2}{\rm MV}{\rm MV}{U}{V} =\sum_{XY,ZW} w_{\ell_1}^{XY} w_{\ell_1}^{ZW} \covPhiCMB{\ell_1}{\ell_2}{XY}{ZW}{U}{V}.
\end{equation}

\section{Covariances} \label{sec:covariances}

In this section we describe our approximate analytic model for the covariances of the measured CMB and lensing power spectra and compare it against simulations.
We will start with covariances between CMB power spectra, proceed with cross-covariances between CMB and lensing power spectra, and finally discuss covariances between lensing power spectra.
These three power covariances involve the lensed CMB 4-point, 6-point and 8-point functions, respectively, because CMB power spectra involve products of two CMB fields, while lensing power spectra involve products of four CMB fields (assuming quadratic lensing reconstruction).

To model these CMB N-point functions, we make a number of assumptions.
We first assume the underlying unlensed CMB, lensing potential, and noise fields are Gaussian.
In principle, their covariance can then be evaluated exactly in terms of correlation functions following a similar method as for the calculation of the lensed power spectrum~\cite{Seljak:1996ve, Lewis:2006fu}. In practice, full evaluation becomes numerically prohibitive, so in the following we will instead adopt a perturbative approach and identify the leading contributions.
We still aim to keep the disconnected Gaussian covariance and other relevant connected subterms fully non-perturbative wherever possible, but only keep numerically important contributions that are connected by up to four underlying Gaussian fields (e.g.~up to second order in the lensing potential power spectrum), dropping various additional complex terms (typically involving more nested sums) that are not required to match simulations to good accuracy.

We neglect correlation between the lensing potential and the CMB, which should be true to very high accuracy for high-$\ell$ CMB modes where the ISW temperature and reionization-sourced polarization is negligible.
Since we also assume the unlensed CMB, lensing potential and noise to be Gaussian we can neglect all odd connected correlations.
Finally we assume no primordial B-mode contribution.
In some calculations we use the fact that the lensed CMB has zero mean, that is $\langle \tilde{X}_{\ell m} \rangle = 0$, and when averaged over realizations of large-scale structure, the lensing potential has also zero mean, $\langle \phi_{\ell m} \rangle = 0$.

\subsection{Lensed CMB power spectrum correlations}\label{sec:two-point-covariances}

Several works \cite{Smith:2006nk,Li:2006pu, Rocher:2006fh, BenoitLevy:2012va, Schmittfull:2013uea} already probed the correlation of lensed CMB power spectra for various combinations of terms, either using a series expansion or generalizing to non-perturbative forms in an ad hoc way. In the following we mainly follow their work, summarising the important steps. In the Appendix~\ref{app:two-point-covariances} we
give some notes on how the various terms can be derived.

\subsubsection{Covariance model}

Assuming no primordial B modes, the correlation of the lensed CMB power spectrum for temperature or E modes contains three main contributions up to second order in the lensing potential power spectrum

\begin{align} \label{eq:full-CMB-covariance-UUUU}
\covlensCMB{U}{V}{U^\prime}{V^\prime} \simeq\, & \text{cov}_G(\CellLensednoisy{\ell_1}{U}{V}, \CellLensednoisy{\ell_2}{U^\prime}{V^\prime}) + \dfrac{1}{(2\ell_1+1)(2\ell_2+1)} \sum_{\ell_3} \Cell{\ell_3}{\phi}{\phi} (\tilde{f}_{\ell_1 \ell_3 \ell_2}^{UU^\prime}\tilde{f}_{\ell_1 \ell_3 \ell_2}^{VV^\prime} + \tilde{f}_{\ell_1 \ell_3 \ell_2}^{UV^\prime}\tilde{f}_{\ell_1 \ell_3 \ell_2}^{VU^\prime}) \nonumber \\
&+ \sum_{\ell_3} \dfrac{\partial \Cell{\ell_1}{\tU}{\tV}}{\partial \Cell{\ell_3}{\phi}{\phi}} \dfrac{2}{2\ell_3 +1} (\Cell{\ell_3}{\phi}{\phi})^2 \dfrac{\partial \Cell{\ell_2}{\tU^\prime}{\tV^\prime}}{\partial \Cell{\ell_3}{\phi}{\phi}},
\end{align}
where $U,V,U^\prime,V^\prime$ can be $T$ or $E$.
The first term of the right-hand side is the disconnected piece of the covariance, or Gaussian variance, given by
\begin{equation}\label{eq:variance-CMB}
\text{cov}_G(\CellLensednoisy{\ell_1}{U}{V}, \CellLensednoisy{\ell_2}{U^\prime}{V^\prime}) = \delta_{\ell_1 \ell_2} \dfrac{1}{2\ell_1 + 1} \Big ( \CellNoHatLensednoisy{\ell_1}{U}{U^\prime} \CellNoHatLensednoisy{\ell_1}{V}{V^\prime} + \CellNoHatLensednoisy{\ell_1}{U}{V^\prime} \CellNoHatLensednoisy{\ell_1}{V}{U^\prime} \Big).
\end{equation}
The second term in Eq.~\ref{eq:full-CMB-covariance-UUUU} is first order in $C^{\phi \phi}_\ell$ and is numerically small; it is related to the trispectrum contribution to the covariance (see e.g.~\cite{Hu:2001fa}).
The third term is of order $(C^{\phi \phi}_\ell)^2$, and arises from the fact that two lensed band powers are connected by the covariance of the $\phi$ field they share.
More specifically, the derivatives of lensed CMB spectra with respect to the lensing potential power spectrum tell us how the fluctuations in the lensing power propagate to the lensed CMB power spectra.

\begin{figure}[!htbp]
\begin{center}
\hspace*{-0.8cm}
\includegraphics[width=0.64\textwidth]{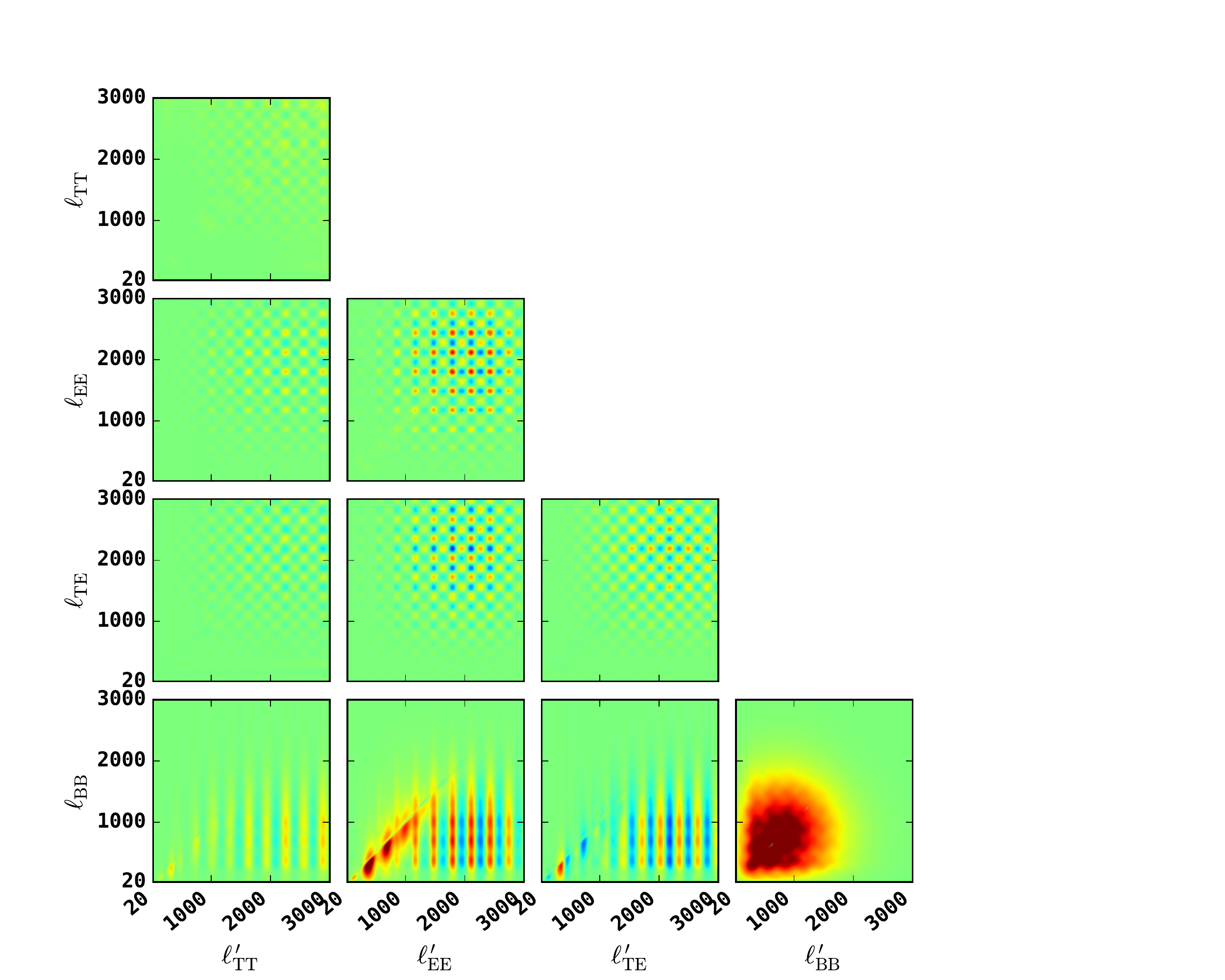}\hspace*{-2.9cm}
\includegraphics[width=0.64\textwidth]{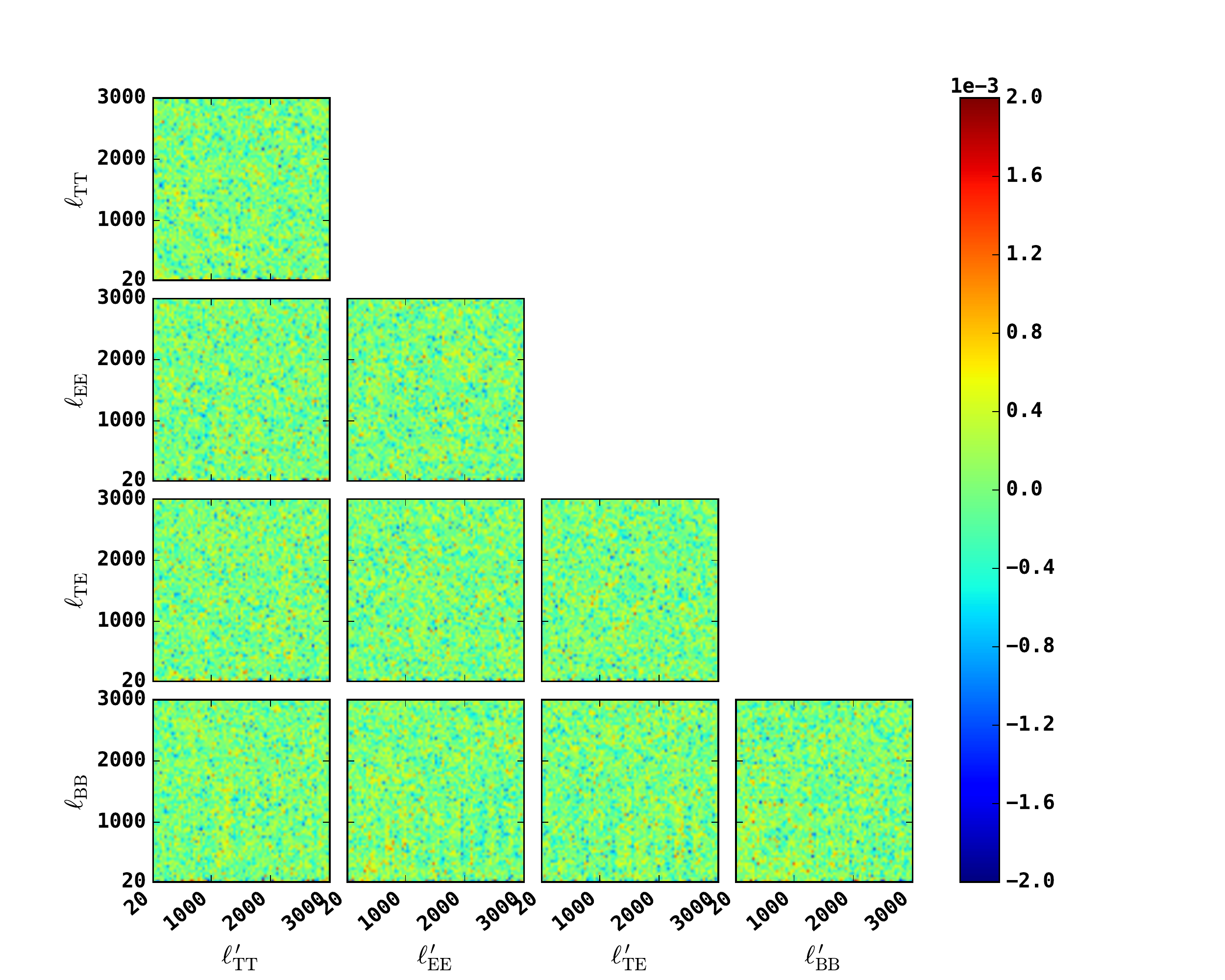}
\caption{Correlation matrices (as defined in Eq.~\ref{eq:corrmat_def}) of lensed CMB power spectra used in this paper (TT, EE, TE, BB) up to second order in $C^{\phi \phi}$, in the case of a possible CMB-S4 experiment. The left panel shows the results obtained using the analytical expressions derived in this section, while the right panel shows the difference between the analytical estimates and the results obtained on simulations. There is a good agreement between both. For visual purposes, the diagonal elements have been set to zero. See text for more discussions.}
\label{fig:corrmat_CMBxCMB_CMBS4}
\end{center}
\end{figure}

The auto-covariance of the B-mode power spectrum is slightly different from the temperature and E-mode. We assume no primordial B-mode contribution in this paper, so they are entirely generated by lensing of E modes. The dominant terms in the covariance are
\begin{align}
\covlensCMB{B}{B}{B}{B} \simeq\, &\text{cov}_G(\CellLensednoisy{\ell_1}{B}{B}, \CellLensednoisy{\ell_2}{B}{B}) + \sum_{\ell_3} \dfrac{\partial \Cell{\ell_1}{\tB}{\tB}}{\partial \Cell{\ell_3}{E}{E}} \dfrac{2}{2\ell_3 +1} (\Cell{\ell_3}{E}{E})^2 \dfrac{\partial \Cell{\ell_2}{\tB}{\tB}}{\partial \Cell{\ell_3}{E}{E}} \nonumber \\
&+ \sum_{\ell_3} \dfrac{\partial \Cell{\ell_1}{\tB}{\tB}}{\partial \Cell{\ell_3}{\phi}{\phi}} \dfrac{2}{2\ell_3 +1} (\Cell{\ell_3}{\phi}{\phi})^2 \dfrac{\partial \Cell{\ell_2}{\tB}{\tB}}{\partial \Cell{\ell_3}{\phi}{\phi}}.
\end{align}
All terms in the auto-covariance for the B modes are at least second order in the lensing potential power spectrum.
The second and third term in the RHS reflect the fact that two lensed B-mode band powers are connected by the covariance of the unlensed E-mode field and the covariance of the $\phi$ field they share.
We neglect one term second order in $C^{\phi \phi}_\ell$ that involves a Wigner-6j symbol, which has a complex form but has been found to be small compared to the other terms~\cite{Li:2006pu} (and we
find that it is not needed at the level of precision required in this paper; see e.g.~Fig.~\ref{fig:corrmat_CMBxCMB_CMBS4}).

We can also write down the cross-covariance between lensed temperature or E-mode and lensed B-mode power spectra following Ref.~\cite{BenoitLevy:2012va}

\begin{equation}\label{eq:covariance_UV-BB}
\covlensCMB{U}{V}{B}{B} \approx \sum_{\ell_3} \dfrac{\partial \Cell{\ell_1}{\tU}{\tV}}{\partial \Cell{\ell_3}{U}{V}} \covCMBNoLens{\ell_3}{\ell_3}{U}{V}{E}{E} \dfrac{\partial \Cell{\ell_2}{\tB}{\tB}}{\partial \Cell{\ell_3}{E}{E}}
+ \sum_{\ell_3} \dfrac{\partial \Cell{\ell_1}{\tU}{\tV}}{\partial \Cell{\ell_3}{\phi}{\phi}} \dfrac{2}{2\ell_3 +1} (\Cell{\ell_3}{\phi}{\phi})^2 \dfrac{\partial \Cell{\ell_2}{\tB}{\tB}}{\partial \Cell{\ell_3}{\phi}{\phi}},
\end{equation}
where $UV \in \{ TT,EE,TE \}$.
Notice that this does not have a disconnected component.

\subsubsection{Correlation matrix}

For visualization purpose, we show in Fig.~\ref{fig:corrmat_CMBxCMB_CMBS4} the correlation matrices between all lensed CMB spectra used in this paper.
The elements of the correlation matrix $\corrCMBNoLens{\ell_1}{\ell_2}{U}{V}{U^\prime}{V^\prime}$ corresponding to the covariance matrix $\covCMBNoLens{\ell_1}{\ell_2}{U}{V}{U^\prime}{V^\prime}$ are defined by

\begin{equation}\label{eq:corrmat_def}
\corrCMBNoLens{\ell_1}{\ell_2}{U}{V}{U^\prime}{V^\prime} = \dfrac{\covCMBNoLens{\ell_1}{\ell_2}{U}{V}{U^\prime}{V^\prime}}{\sqrt{\covCMBNoLens{\ell_1}{\ell_1}{U}{V}{U}{V}\covCMBNoLens{\ell_2}{\ell_2}{U^\prime}{V^\prime}{U^\prime}{V^\prime}}},
\end{equation}
for $UV, U^\prime V^\prime \in \{ TT, EE, TE, BB, \phi \phi \}$.
The temperature and E-mode auto-correlations are dominated by checkerboard structures, as already seen in Refs.~\cite{BenoitLevy:2012va, Schmittfull:2013uea}, which correspond to the position of the acoustic peaks and troughs of the CMB spectra that are most affected by lensing.

For very high multipoles ($\ell > 2500$), the correlation is fainter due to the rise of the noise.
We found that the temperature power spectrum exhibits large correlations between large and very small scales if we go beyond $\ell = 3000$.
These correlations reflect the fact that the lensing transfers power from large scales to small scales, and if we include scales beyond $\ell=3000$ the constraints from temperature become more important than all other spectra (at these noise and beam levels).
However, such small scales would in practice be dominated by other secondary anisotropies, and therefore we do not include them in the analysis (note that this cut also significantly reduces the information from the temperature lensing reconstruction).

The B power auto-correlation is much broader and much stronger than the others, reflecting the fact that the B modes are entirely generated by lensing, and are produced by a very non-local coupling in $\ell$ between $E$ and $\phi$, see e.g.~Ref.~\cite{Fabbian:2013owa} and references therein.
The BB spectrum only shows significant correlation with the other spectra for $\ell_{BB} \leq \ell_{UV}$, where there is a smoothing effect on the acoustic peaks generated by relatively large-scale lensing modes.
In order to understand this pattern, let us take the example of the correlation between lensed EE and lensed BB power-spectra. First, we notice the fact that the correlation is systematically weak for $\ell_{EE} \leq \ell_{BB}$ is mainly driven by the fact that the $\ell^2 C_\ell^{EE}$ power is quite blue. Then there are two regimes: $\ell_{EE} \leq 1000$, and $\ell_{EE} \geq 1000$.
For $\ell_{EE} \leq 1000$, we start from the fact that the B-modes power-spectrum is generated by the lensing of the unlensed E-modes (first term in the RHS of Eq.~\ref{eq:covariance_UV-BB}), with $C_\ell^{\tB \tB} \sim \int {\rm{d}(\log \ell)} \ell^4 C^{\phi\phi}_\ell \ell^2 C^{EE}_\ell$, which peaks at $500 \leq \ell_{EE} \leq 1000$.
For $\ell_{EE} \geq 1000$, we are now interested in the term associated with the cosmic variance of the lens power spectrum (second term in the RHS of Eq.~\ref{eq:covariance_UV-BB}). This term causes a band structure from the EE spectrum derivatives (fluctuations in smoothing), and the correlation is fainter for $\ell_{BB} \geq 1000$, when the BB spectrum drops and becomes noise dominated.

The right panel of Fig.~\ref{fig:corrmat_CMBxCMB_CMBS4} shows the difference between the model and the simulations.
The agreement is on overall good, with some differences for covariances involving one B-mode spectrum. These differences appear to be unimportant for this paper. We note that the diagonal elements between the model and the simulations are in sub-percent agreement.

\subsubsection{Evaluating the derivatives}

To compute the covariances listed above, we need to evaluate the derivatives of lensed CMB power spectra with respect to the lensing potential power spectrum and with respect to the unlensed CMB power spectra.
Some previous works such as Ref.~\cite{Schmittfull:2013uea} made use of the series-expansion of the lensed CMB spectra in terms of $C^{\phi \phi}_\ell$.
This method has the advantage of being fast and giving reasonable results but it may not be sufficiently accurate for the level of precision that future experiments will reach.
Therefore throughout this paper, we evaluate the derivatives of spectra using the more accurate correlation function methods \cite{Lewis:2006fu,Seljak:1996ve}.
We typically found that the correlation between the lensing amplitude estimates $\hat{A}_{\phi^{\rm MV} \phi^{\rm MV}}$ and $\hat{A}_{\tU \tV}$ discussed in Sec.~\ref{sec:lensing_amplitude} is artificially enhanced by up to 30-40$\%$ if we use the series-expansion to compute the derivatives rather than the correlation function method.
We detail the computation of this in Appendix~\ref{app:corr_func}.
A similar technique using the flat-sky approximation has been used recently in the work of Ref.~\cite{Green:2016cjr}, and an alternative scheme to estimate these derivatives is described in the Appendix of Ref.~\cite{BenoitLevy:2012va}.

\subsection{Cross-correlation between observed lensed CMB and reconstructed lensing potential power spectra} \label{sec:cross-covariances}

Using the quadratic estimator for the lensing potential defined in Eq.~\ref{eq:estimator-phi}, cross-covariances between observed lensed CMB power spectra and reconstructed lensing potential power spectra involve covariances between CMB 2-point and 4-point functions. In their most general form, they can be expressed as

\begin{align} \label{eq:full-phiphiCMB}
\covPhiCMB{\ell_1}{\ell_2}{XY}{ZW}{U}{V} &= \dfrac{\Amplitude{\ell_1}{XY} \Amplitude{\ell_1}{ZW}}{(2\ell_1 +1)(2\ell_2 + 1)} \sum_{\underline{\ell}_3,...,\underline{\ell}_{6}, m_1, m_2} (-1)^{m_1+m_2}  \wigner{\ell_3}{\ell_4}{\ell_1}{m_3}{m_4}{-m_1} \wigner{\ell_5}{\ell_6}{\ell_1}{m_5}{m_6}{m_1} \nonumber \\
&\times \gweight{\ell_3}{\ell_4}{\ell_1}{XY} \gweight{\ell_5}{\ell_6}{\ell_1}{ZW}  \Big [ \langle \tX_{\underline{\ell}_3} \tY_{\underline{\ell}_4} \tZ_{\underline{\ell}_5} \tW_{\underline{\ell}_6} \tU_{\ell_2 m_2} \tV_{\ell_2, -m_2}  \rangle - \langle \tX_{\underline{\ell}_3} \tY_{\underline{\ell}_4} \tZ_{\underline{\ell}_5} \tW_{\underline{\ell}_6} \rangle \langle \tU_{\ell_2 m_2} \tV_{\ell_2, -m_2} \rangle \Big],
\end{align}
where $\underline{\ell}_i = (\ell_i, m_i)$.
We identify four main contributions to the cross-covariance which are detailed in the following subsections:
\begin{align} \label{eq:contrib-xcorr}
\covPhiCMB{\ell_1}{\ell_2}{XY}{ZW}{U}{V} &\approx \covPhiCMB{\ell_1}{\ell_2}{XY}{ZW}{U}{V}_{\text{noise}} + \covPhiCMB{\ell_1}{\ell_2}{XY}{ZW}{U}{V}_{\text{trispectrum}}^{\text{Type A}} \nonumber\\
&+ \covPhiCMB{\ell_1}{\ell_2}{XY}{ZW}{U}{V}_{\text{signal}} + \covPhiCMB{\ell_1}{\ell_2}{XY}{ZW}{U}{V}_{\text{trispectrum}}^{\text{Type B - primary}}.
\end{align}
The different terms follow different ways of expressing the 6-point function: 2+2+2-point (disconnected Gaussian piece), 2+4-point (Trispectrum A and B) and connected 6-point function (signal).
Of those four contributions, the first two are cancelled by the RDN0 subtraction (and therefore not used for the results of this paper unless stated), leaving the last two terms as potentially important.

\subsubsection{Terms not cancelled by the use of RDN0}\label{sec:notcrosscancelled}

We identified two main contributions to the cross-covariance that remain after RDN0 subtraction, coming from the connected 4 and 6-pt functions.
\\

\begin{figure}[t]
\begin{center}
\hspace*{-0.8cm}
\includegraphics[width=1.0\textwidth]{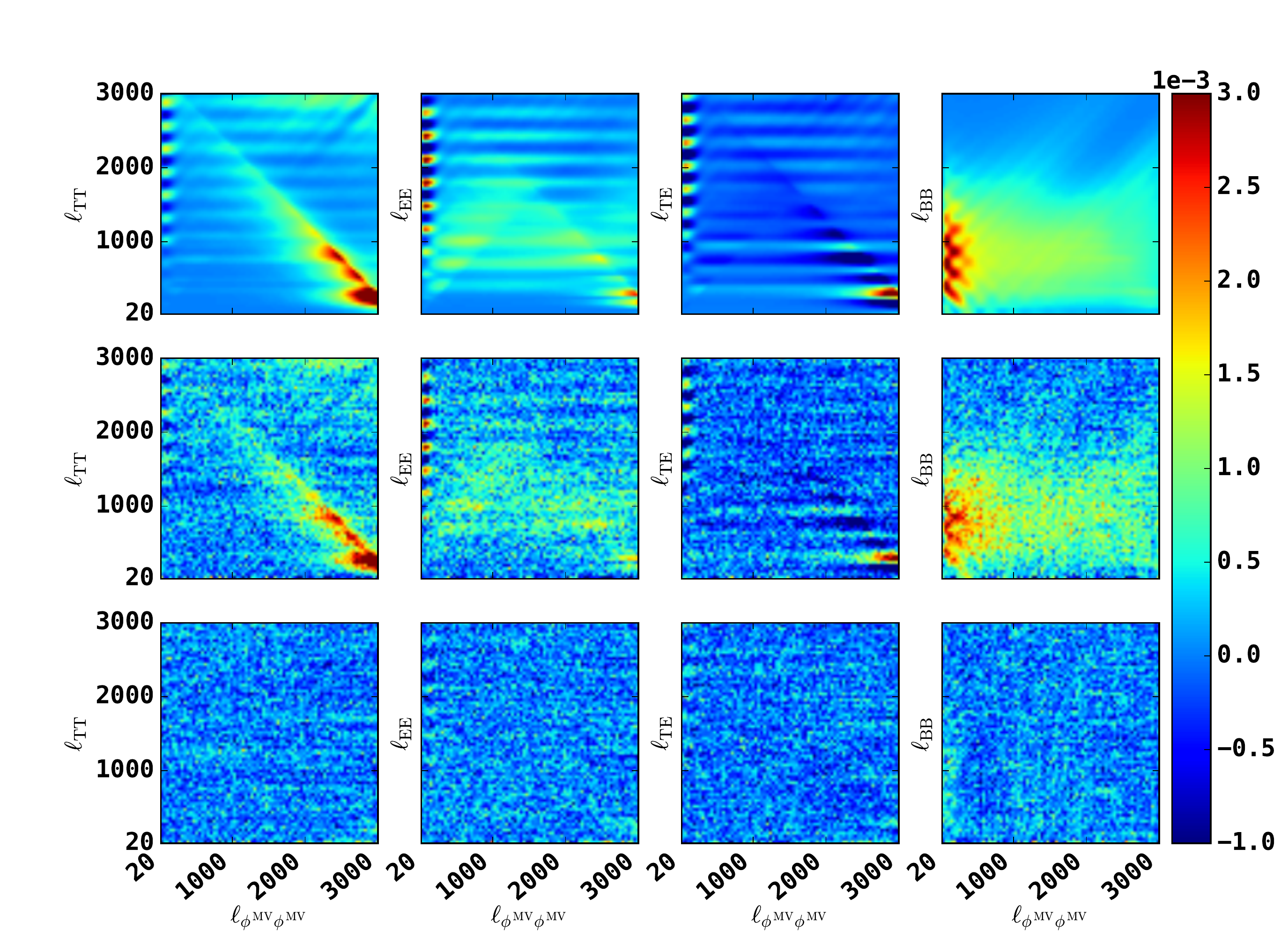}
\caption{Cross-correlation matrices (as defined in Eq.~\ref{eq:corrmat_def}) between the lensed CMB power spectra and the reconstructed minimum variance lensing potential power spectrum without RDN0 subtraction in the case of CMB-S4 like experiment. The top row shows the results from the analytical model. Each panel corresponds to the correlation of the lensing potential power spectrum with a lensed CMB power spectrum: $C_{\ell}^{\tT \tT}$, $C_{\ell}^{\tE \tE}$, $C_{\ell}^{\tT \tE}$, and $C_{\ell}^{\tB \tB}$. The covariance matrices contain all the terms described in the Sec. \ref{sec:covariances} and Appendices~\ref{app:crosscancelled} $\&$ \ref{app:four-point-covariances_rdn0ed}. The middle row shows the results obtained from the set of 5,000 MC simulations. The bottom row is the difference between the analytical model and the simulations. The agreement between the model and the simulation is rather good, except for the B modes where the model tends to mis-estimate the contribution with respect to simulations (see text). This difference has little impact on the results discussed in this paper.
}
\label{fig:corrmat_phixCMB_CMBS4_noise}
\end{center}
\end{figure}

\begin{figure}[t]
\begin{center}
\hspace*{-0.8cm}
\includegraphics[width=1.0\textwidth]{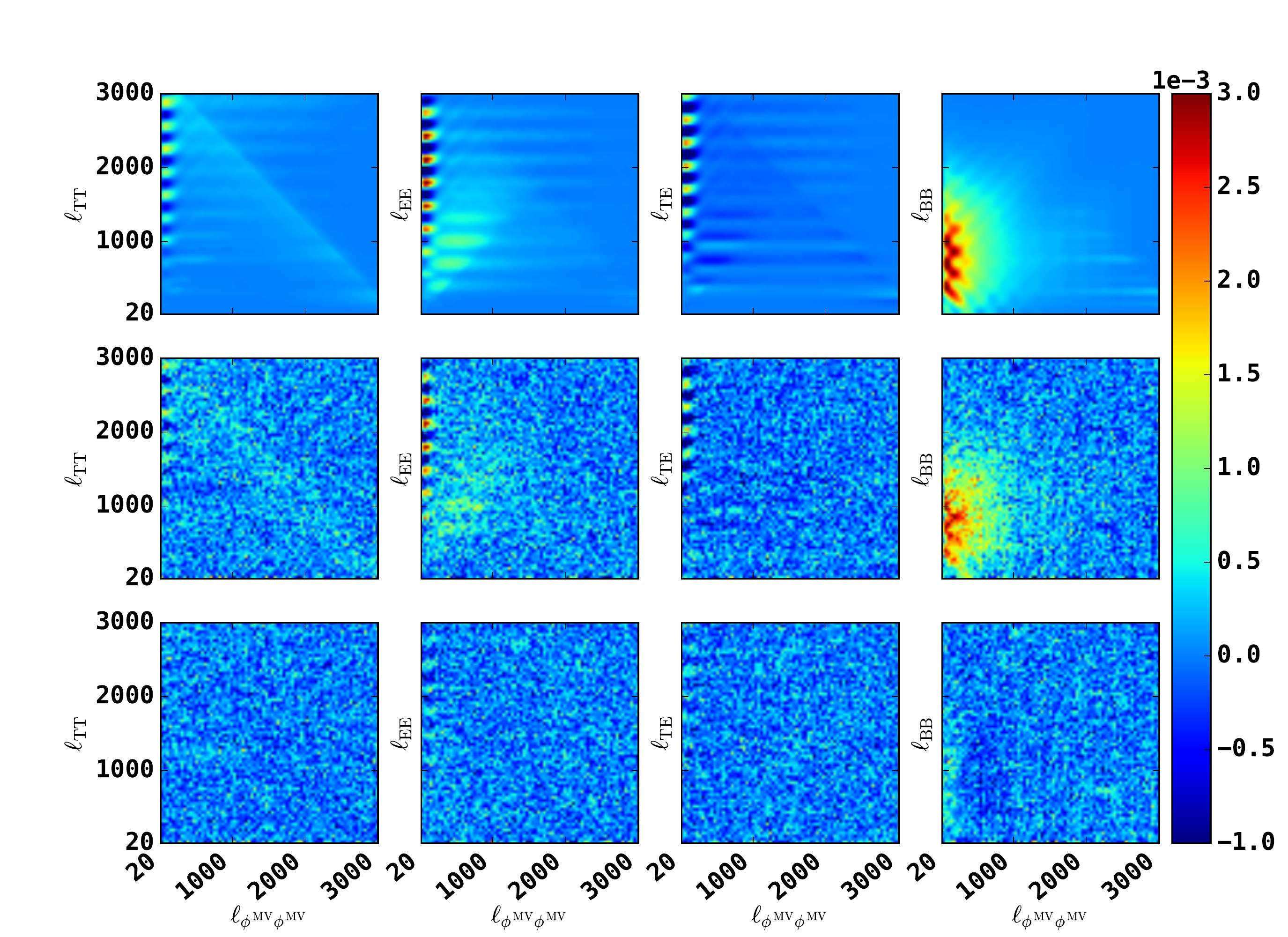}
\caption{Same as Fig. \ref{fig:corrmat_phixCMB_CMBS4_noise}, but we use the realization-dependent noise bias subtraction in the computation of the matrices. The overall correlations at small lensing scales are reduced, but the correlations at large lensing scales remain almost identical.
}
\label{fig:corrmat_phixCMB_CMBS4_noise_RDN0}
\end{center}
\end{figure}

\paragraph{Connected 6-point function: signal term.\label{signal term}}
The connected signal part of the 6-point function comes from the covariance of the realization of the lensing potential power $\hat{C}^{\phi \phi}_{\ell_1}$ and the lensed CMB power, $\hat{C}_{\ell_2}^{\tilde{U}\tilde{V}}$.
Specifically, we define the signal term as being the contraction that appears in the connection by two $\phi$ modes
\begin{equation}
\covPhiCMB{\ell_1}{\ell_2}{XY}{ZW}{U}{V}_{\text{signal}} \equiv
\sum_{\ell_3} \frac{\partial \Cell{\ell_1}{\phi^{XY}}{\phi^{ZW}}}{\partial
\Cell{\ell_3}{\phi}{\phi}} \text{cov}\left( \hat{C}^{\phi\phi}_{\ell_3}, \hat{C}_{\ell_2}^{\tilde{U}\tilde{V}}\right),
\end{equation}
where $\Cell{\ell_1}{\phi^{XY}}{\phi^{ZW}}$ denotes the signal expectation in the fiducial model, including contributions from both the lensing potential power itself and the $N^{(1)}$ bias
\footnote{If we assume no contribution from $N^{(1)}$, then Eq.~\ref{eq:xcorr-signal} reduces to Eq.~E8 in \cite{Schmittfull:2013uea} in the case of temperature. }
\footnote{Note that $\Cell{\ell_1}{\phi^{XY}}{\phi^{ZW}}$ is not directly the estimator mean, which also would have indirect dependence of the lensing power via the response functions.}.
The covariance of the realization powers can be evaluated exactly analytically using the assumed Gaussianity of $\phi$:
\begin{eqnarray}
  \text{cov}\left( \hat{C}^{\phi\phi}_{\ell_1}, \hat{C}_{\ell_2}^{\tilde{U}\tilde{V}}\right)
  &=& 
  \int D\phi\, \hat{C}^{\phi\phi}_{\ell_1} \la \hat{C}_{\ell_2}^{\tilde{U}\tilde{V}}\ra_{UV} \prod_\ell\frac{e^{-(2\ell+1)\hat{C}^{\phi\phi}_{\ell}/(2C^{\phi\phi}_{\ell})}}{(2\pi C^{\phi\phi}_\ell)^{(2\ell+1)/2}}
 - C^{\phi\phi}_{\ell_1} C_{\ell_2}^{\tilde{U}\tilde{V}}
 \nonumber\\
 &=& \frac{2 (C^{\phi\phi}_{\ell_1})^2}{2\ell_1+1} \frac{\partial}{\partial C^{\phi\phi}_{\ell_1}}
 \int D\phi P(\phi)\la \hat{C}_{\ell_2}^{\tilde{U}\tilde{V}}\ra_{UV}
 =\frac{2 (C^{\phi\phi}_{\ell_1})^2}{2\ell_1+1} \frac{\partial C_{\ell_2}^{\tilde{U}\tilde{V}}}{\partial C^{\phi\phi}_{\ell_1}}.
\end{eqnarray}
Using the lensing potential input to our simulations we checked that
the simulation and numerical derivative calculations are consistent with this exact result to high accuracy in the range of multipoles of interest\footnote{And therefore we could conclude that the small mismatch between simulations and model at low multipoles (where the signal term dominates over the others) seen in Figs.~\ref{fig:corrmat_phixCMB_CMBS4_noise}$\&$\ref{fig:corrmat_phixCMB_CMBS4_noise_RDN0} is not due to approximations in the signal term.}.
We then have
\begin{equation}\label{eq:xcorr-signal}
\covPhiCMB{\ell_1}{\ell_2}{XY}{ZW}{U}{V}_{\text{signal}} = \sum_{\ell_3} \dfrac{\partial \Cell{\ell_1}{\phi^{XY}}{\phi^{ZW}}}{\partial \Cell{\ell_3}{\phi}{\phi}} \dfrac{2}{2\ell_3+1}(\Cell{\ell_3}{\phi}{\phi})^2 \dfrac{\partial \Cell{\ell_2}{\tU}{\tV}}{\partial \Cell{\ell_3}{\phi}{\phi}} .
\end{equation}

In the case of temperature and E-polarization, the covariance arises because the same lenses are responsible for the smoothing of the acoustic peaks of the CMB spectrum and for the signal part of the lensing reconstruction power. Both respond to fluctuations in the lensing power, which comes from the cosmic variance of the lenses:
the greater the lensing power in any realization, the greater the smoothing of the CMB power spectrum, and the larger the lensing potential estimator becomes.
This correlation is mostly between large-scale lens modes ($\ell_{\phi \phi} < 500$) and intermediate and small-scale CMB modes that are most affected by the lensing smoothing ($\ell_{UV} > 500$).
For BB the signal covariance produces a broad-band correlation, since the B-mode power has contributions from couplings between a wide range of scales.
For a \textsc{CMB-S4} like experiment, the signal correlation term seems to drive most of the correlations between the reconstructed lensing potential power spectra and the observed lensed CMB power spectra (and entirely dominates after realization-dependent noise subtraction; see Fig.~\ref{fig:corrmat_phixCMB_CMBS4_noise_RDN0}).
We show later in Sec.~\ref{sec:lensing_amplitude} that the signal covariance is almost entirely due to a single mode of the lensing, which can be projected out from the covariance to reduce the correlations.

Note that we have neglected a contribution to the signal term arising from the fact that the normalization response is $\mathcal{O}(C^\phi)$ (from the fluctuation in the response $\tilde{f}$ in the actual realization).
This term is believed to be subdominant with respect to terms already considered in Eq.~\ref{eq:xcorr-signal}.
\\

\paragraph{Connected 4-point function: Type B trispectrum.}

The lensed CMB trispectrum also contributes to the cross-covariance of Eq.~\ref{eq:full-phiphiCMB}.
To second order in the lensing potential power spectrum there are two main contributions: Type A considered in Eq.~\ref{eq:xcorr-trispectrumA} above, and Type B considered here. They are detailed for some specific cases in Appendix~\ref{app:xcorr-trispectrum}.
Following Ref.~\cite{Schmittfull:2013uea}, the Type B term can be split further into two parts, primary and non-primary contributions\footnote{The primary contribution refers to the contribution for which sums over $m$'s in Eq.~\ref{eq:full-phiphiCMB} simplify due to orthogonality relations of Wigner symbols, while the non-primary contribution includes the remaining terms.}.
None of these two Type B contributions are cancelled by the use of the realization-dependent $\hat{N}^{(0)}$ correction, but it has been argued \cite{Hanson:2010rp, Schmittfull:2013uea} that the non-primary contribution is subdominant compared to the primary one.
Therefore, we only focus on the primary contribution for the rest of this paper, given by

\begin{align}
\covPhiCMB{\ell_1}{\ell_2}{XY}{ZW}{U}{V}_{\text{trispectrum}}^{\text{Type B - primary}} &= \dfrac{\Cell{\ell_1}{\phi}{\phi}}{2\ell_2+1} \Big \{ \Big[ \dfrac{\Amplitude{\ell_1}{XY} \CellNoHatLensednoisy{\ell_2}{X}{U}}{2\ell_1+1} \sum_{\ell_3} \gweight{\ell_2}{\ell_3}{\ell_1}{XY}\fweightLensed{\ell_2}{\ell_1}{\ell_3}{VY} + (X \leftrightarrow Y) \Big] \nonumber \\
&+ (X \leftrightarrow Z, Y \leftrightarrow W) \Big \} + (U \leftrightarrow V). \label{eq:xcorr-trispectrumB}
\end{align}
for CMB pairs $ab \in \{ TT, TE, EE, BB \}$.
This term is almost an order of magnitude weaker than the signal contribution described above for the scales of interest.
For some combinations we can express the Type B trispectrum as the noise term Eq.~\ref{eq:xcorr-noise} multiplied by the signal-to-noise ($C^{\phi \phi} / {\cal{A}}$), as shown in Appendix~\ref{app:xcorr-trispectrum} in a handful of cases.
For EE, BB and TE, the signal is mostly at low and intermediate lensing and CMB multipoles ($\ell_{\phi \phi} < 1000$ and $\ell_{UV} < 2000$).
For temperature, the signal is also seen at smaller lensing scales because the signal-to-noise of the reconstruction is rather constant across the multipole range (see Fig.~\ref{fig:n0_n1_bias}).

\subsubsection{Terms cancelled by the use of RDN0}
\label{sec:crosscancelled}
From Eq.~\eqref{eq:clphiphi-RDN0}, the cross-covariance of lensed CMB power spectra with RDN0-corrected lensing power spectra can be expressed in terms of the covariance without any RDN0 correction as follows:
\begin{equation} \label{eq:xcorr-RDNzero}
\text{cov}(\CellHatRDN{\ell_1}{\phi^{XY}}{\phi^{ZW}},\hat{C}_{\ell_2,{\rm expt}}^{\tU\tV}) =
\covPhiCMB{\ell_1}{\ell_2}{XY}{ZW}{U}{V} -
\sum_{(ab), \ell_3} \frac{\partial \Nzero{\ell_1}{XY}{ZW}}{\partial \CellNoHatLensednoisy{\ell_3}{a}{b}}\text{cov}(
\CellLensednoisy{\ell_3}{a}{b},\CellLensednoisy{\ell_2}{U}{V}
).
\end{equation}
The RDN0 correction of the measured lensing power spectrum cancels two terms in Eq.~\ref{eq:contrib-xcorr} that would otherwise appear in the cross-covariance. These two terms are detailed and discussed in Appendix~\ref{app:crosscancelled}.

\subsection{Reconstructed lensing potential power-spectrum auto-correlations}\label{sec:four-point-covariances}

Using the quadratic estimator for the lensing potential defined in Eq.~\ref{eq:estimator-phi} and the estimator for its power spectrum in Eq.~\ref{eq:estimator-phiphi}, the auto-covariance of the reconstructed lensing potential power spectrum is given by the covariance between two CMB 4-point functions.
It can be expressed in a general form as:

\begin{align} \label{eq:full-phiphiphiphi}
\covPhi{\ell_1}{\ell_2}{XY}{ZW}{X^\prime Y^\prime}{Z^\prime W^\prime} &= \dfrac{\Amplitude{\ell_1}{XY} \Amplitude{\ell_1}{ZW} \Amplitude{\ell_2}{X^\prime Y^\prime} \Amplitude{\ell_2}{Z^\prime W^\prime}}{(2\ell_1 +1)(2\ell_2 + 1)} \sum_{\underline{\ell}_3,...,\underline{\ell}_{10}, m_1, m_2} (-1)^{m_1+m_2} \gweight{\ell_3}{\ell_4}{\ell_1}{XY} \gweight{\ell_5}{\ell_6}{\ell_1}{ZW} \gweight{\ell_7}{\ell_8}{\ell_2}{X^\prime Y^\prime} \gweight{\ell_9}{\ell_{10}}{\ell_2}{Z^\prime W^\prime} \nonumber \\
&\times \wigner{\ell_3}{\ell_4}{\ell_1}{m_3}{m_4}{-m_1} \wigner{\ell_5}{\ell_6}{\ell_1}{m_5}{m_6}{m_1} \wigner{\ell_7}{\ell_8}{\ell_2}{m_7}{m_8}{-m_2} \wigner{\ell_9}{\ell_{10}}{\ell_2}{m_9}{m_{10}}{m_2} \nonumber \\
&\times \Big [ \langle \tX_{\underline{\ell}_3} \tY_{\underline{\ell}_4} \tZ_{\underline{\ell}_5} \tW_{\underline{\ell}_6} \tX_{\underline{\ell}_7}^\prime \tY_{\underline{\ell}_8}^\prime \tZ_{\underline{\ell}_9}^\prime \tW_{\underline{\ell}_{10}}^\prime \rangle - \langle \tX_{\underline{\ell}_3} \tY_{\underline{\ell}_4} \tZ_{\underline{\ell}_5} \tW_{\underline{\ell}_6} \rangle \langle \tX_{\underline{\ell}_7}^\prime \tY_{\underline{\ell}_8}^\prime \tZ_{\underline{\ell}_9}^\prime \tW_{\underline{\ell}_{10}}^\prime \rangle \Big],
\end{align}
where $\underline{\ell}_i = (\ell_i, m_i)$.
In the following, we identify the relevant contributions for this analysis.
As in the previous section, we explicitly separate the contributions by whether they are cancelled by the RDN0 subtraction.

\subsubsection{Terms not cancelled by the use of RDN0}\label{sec:four-point-covariances-RDN0ed}

Starting from Eq.~\ref{eq:full-phiphiphiphi}, we identified two terms potentially relevant for our analysis.
\\

\paragraph{Gaussian reconstruction power variance.} \label{sec:phiphi_gaussvar_notrdn0ed}
The first term is the Gaussian reconstruction power variance, which on the full sky is predominantly
\begin{equation} \label{eq:gaussvar-phiphiphiphi}
\text{cov}_G(\CellHat{\ell_1}{\phi^{XY}}{\phi^{ZW}}, \CellHat{\ell_2}{\phi^{X^\prime Y^\prime}}{\phi^{Z^\prime W^\prime}}) = \delta_{\ell_1 \ell_2} \dfrac{1}{2\ell_1 + 1} \Big ( \langle \CellHat{\ell_1}{\phi^{XY}}{\phi^{X^\prime Y^\prime}} \rangle \langle \CellHat{\ell_1}{\phi^{ZW}}{\phi^{Z^\prime W^\prime}} \rangle + \langle \CellHat{\ell_1}{\phi^{XY}}{\phi^{Z^\prime W^\prime}} \rangle \langle \CellHat{\ell_1}{\phi^{ZW}}{\phi^{X^\prime Y^\prime}} \rangle \Big)
\end{equation}
for general lensing reconstruction power. We include $N^{(1)}$ bias in the expectation values here (from Eq.~\ref{eq:phiphi_expectation_value}), but neglect the full off-diagonal signal fluctuation dependence arising from $N^{(1)}$.
\\

\paragraph{Connected 8-point function.} \label{sec:phiphi_8pt_notrdn0ed}
We also have a correlation induced by the connected 8-point function.
The leading order contribution is of order $\OrderStop{\phi}{\phi}{2}$ \cite{Kesden:2002jw}, which makes a full calculation rather involved.
A full analysis is beyond the scope of our paper, but we have checked that some of the simple contractions have a negligible effect on estimates of the lensing amplitude from the lensing power spectrum, and comparison with simulations also shows no evidence for significant missing terms. We therefore drop all connected 8-point function contributions.

\subsubsection{Terms cancelled by the use of RDN0} \label{sec:four-point-covariances_rdn0ed}
Using RDN0 cancels correlations arising because fluctuations in the observed CMB power spectrum induce changes of the Gaussian lensing reconstruction lensing noise.
Explicitly, from Eq.~\ref{eq:clphiphi-RDN0}, the auto-covariance of the measured, RDN0-corrected lensing power spectrum can be expressed in terms of the covariance without RDN0 correction as
\begin{multline} \label{eq:autocorr-RDNzero}
\text{cov}(\CellHatRDN{\ell_1}{\phi^{XY}}{\phi^{ZW}},\CellHatRDN{\ell_2}{\phi^{X'Y'}}{\phi^{Z'W'}}) =
\covPhi{\ell_1}{\ell_2}{XY}{ZW}{X^\prime Y^\prime}{Z^\prime W^\prime}
-
\sum_{(ab),\ell_3}\frac{\partial \Nzero{\ell_1}{XY}{ZW}}{\partial \CellNoHatLensednoisy{\ell_3}{a}{b}}
\covPhiCMBtransp{\ell_2}{\ell_3}{X'Y'}{Z'W'}{a}{b}
\\-
\sum_{(ab),\ell_3}
\covPhiCMB{\ell_1}{\ell_3}{XY}{ZW}{a}{b}
\frac{\partial \Nzero{\ell_2}{X'Y'}{Z'W'}}{\partial \CellNoHatLensednoisy{\ell_3}{a}{b}}
+
\sum_{(ab),(cd),\ell_3, \ell_4}
\frac{\partial \Nzero{\ell_1}{XY}{ZW}}{\partial \CellNoHatLensednoisy{\ell_3}{a}{b}}
\text{cov}(\CellLensednoisy{\ell_3}{a}{b}, \CellLensednoisy{\ell_4}{c}{d} )
\dfrac{\partial \Nzero{\ell_2}{X^\prime Y^\prime}{Z^\prime W^\prime}}{\partial \CellNoHatLensednoisy{\ell_4}{c}{d}}.
\end{multline}
The details of the computation and a discussion about these terms can be found in Appendix~\ref{app:four-point-covariances_rdn0ed}.

\subsection{Detectability of the correlations}

In this section we quantify the detectability of the off-diagonal parts of the covariance.
Let's define our joint data vector $\hat{\textbf{C}}$ as
\begin{equation}\label{eq:joint-data-vector}
\hat{\textbf{C}}_\ell = \left( \hat{C}^{\tT \tT}_{\ell,\text{expt}}, \hat{C}^{\tE \tE}_{\ell,\text{expt}}, \hat{C}^{\tT \tE}_{\ell,\text{expt}}, \hat{C}^{\tB \tB}_{\ell,\text{expt}}, \hat{C}^{\phi^{\rm MV} \phi^{\rm MV}}_{\ell, \rm{RDN0}}
- \Nouane{\ell}{\rm MV}{ } \right).
\end{equation}
The full covariance of this joint data vector (denoted \textbf{cov}) contains all contributions listed above after applying the realization-dependent noise bias subtraction, namely all covariances listed in Sec.~\ref{sec:two-point-covariances} for the CMB auto-covariances, Sec.~\ref{sec:notcrosscancelled} for the cross-covariances, and Sec.~\ref{sec:four-point-covariances-RDN0ed} for the lensing auto-covariances.
We split the total covariance as a Gaussian part $\textbf{cov}^{\rm G}$, and a non-Gaussian part $\textbf{cov}^{\rm NG}$ with unknown amplitude $\alpha$ as
\begin{equation}
\textbf{cov} = \textbf{cov}^{\rm G}+\alpha \textbf{cov}^{\rm NG}.
\end{equation}
The likelihood for the data $\hat{\textbf{C}}$ (in which the $\hat{\textbf{C}}$ are approximated as Gaussian in the fiducial model) reads
\begin{equation}
-2\ln {\cal{L}}(\hat{\textbf{C}}) = \hat{\textbf{C}}^T (\textbf{cov}^{\rm G}+\alpha \textbf{cov}^{\rm NG})^{-1}\hat{\textbf{C}}+ \ln(|\textbf{cov}^{\rm G}+\alpha \textbf{cov}^{\rm NG}|),
\end{equation}
where $|.|$ denotes the determinant of a matrix.
Assuming no prior, the Fisher matrix for the amplitude $\alpha$ of the non-Gaussian part of the covariance is defined as the expectation value
\begin{equation}
F_{\alpha \alpha} =\left. \Big \langle \dfrac{\partial^2 [-\ln {\cal{L}}(\hat{\textbf{C}})]}{\partial \alpha \partial \alpha} \Big \rangle \right|_{\alpha=1} = \dfrac{1}{2} \text{Tr}  \Big[ \textbf{cov}^{\rm NG} \textbf{cov}^{-1} \textbf{cov}^{\rm NG} \textbf{cov}^{-1}\Big],
\end{equation}
where $\text{Tr}$ denotes the trace of a matrix.
The significance (or detectability) of the off-diagonal parts of the covariance is then given by $\sqrt{F_{\alpha\alpha}}$.

For CMB-S4, the off-diagonal parts of the covariance should be detectable with a significance around 6$\sigma$.
This shows that in general the non-Gaussian contributions are not negligible, and must be included to get reliable $\chi^2$ goodness of fit numbers. The impact on cosmological parameters is expected to be much less significant, and we assess this in more detail below after checking agreement with simulations.
Among the lensed CMB spectra, the B modes generate most of the impact. Neglecting B modes in the analysis (auto- and cross-covariance) leads to a lower-significance detection of the off-diagonal parts of the covariance ($\sim 3.5\sigma$).

\section{Simulations} \label{sec:simulations}

\newcommand{\vell}{ {\boldsymbol{\ell}} }

\subsection{Simulation pipeline}

To test and validate our analytical results, we developed a simulation and lensing reconstruction pipeline.
We generate 5,000 periodic square patches of jointly Gaussian unlensed T,Q,U skies of $2500$ $\rm{deg}^2$ from spectra computed using CAMB for a $\Lambda$\textsc{CDM} cosmology based on the latest Planck constraints \cite{Aghanim:2016yuo} with $h = 0.6688$, $\Omega_b h^2 = 0.02214$, $\Omega_c h^2 = 0.1207$, $n_s = 0.9624$, $\sigma_8 = 0.817$, $\tau=0.0581$, and one massive and two massless neutrino eigenstates (sum of the masses $M_\nu \equiv \sum m_\nu=60$ meV).

For convenience, the pipeline uses the flat-sky approximation.  The 2D wavevectors $\boldsymbol {\ell} = (\vell_x,\vell_y)$ of the patch are assigned curved-sky power $C_{|\boldsymbol{\ell}| - 1/2}$.
These maps are then lensed according to the realization of the lensing potential.
The $\phi-T$ and $\phi-E$ correlations are neglected.
The lensing operations are performed numerically using a standard bicubic spline interpolation of the unlensed maps given on a regular grid with resolution $0.7 $ arcmin, which is sufficient given our noise levels and high-$\ell$ cuts.
The lensed spectra agree with the (curved sky) predictions from CAMB (sub-percent accuracy) across the scales we are using for the reconstruction. A Gaussian beam with FWHM 3 arcmin, identical in temperature and polarization, is applied to each lensed sky, together with homogeneous isotropic noise of 1.5 $\mu$K.arcmin (T) and 1.5$\sqrt{2}$ $\mu$K.arcmin (Q,U). We do not include directly real-life complications such as foregrounds, sky-cuts, anisotropic beams and uneven hit-counts etc, that would complicate the lensing reconstruction without being relevant for our purposes. However, we keep only multipoles $20 \leq \ell \leq 3000$ of the simulated maps, which roughly accounts for the loss of modes on large scales due to sky coverage and on very small-scales due to foregrounds.

The lensing reconstruction uses the separability of the weight functions of the quadratic estimator in T,Q,U space. We use a FFT-based real-space implementation for the un-normalized Cartesian components $\alpha_{x,y}$ of the displacement field, which can be written in convenient matrix notation as follows
\begin{equation}
\hat \alpha_{x,y}(\mathbf z) = \sum_{\beta \in \{ T,Q,U \} }\left [ \sum_{\vell}B_\vell \Cov_{\vell}^{-1}\: D_{\vell} \: e^{i\vell \cdot \mathbf z} \right ]^\beta  \left [\sum_{\vell} i\vell_{x,y} C^{\rm{len}}_{\vell} B_\vell \Cov_{\vell}^{-1} D_{\vell}  \: e^{i\vell \cdot \mathbf z} \right ]_\beta.
\end{equation}
In this equation, $D_\vell = (T_\vell,Q_\vell,U_\vell)$ is the data vector input to the MV estimator,  $B_\vell$ is the $3\times 3$ (diagonal) beam matrix with constant diagonal entries $\exp\left [ \ell (\ell + 1) \sigma_{\rm{FWHM}}^2 / (16 \ln 2)\right]$, $\Cov_{\vell} = B_\vell C^{\rm{len}}_{\boldsymbol\ell} B_\vell + N_{\vell}$ the $3 \times 3$ covariance matrix of the harmonic mode $\vell$ of the data (including beam and noise) and $C^{\rm{len}}_{\vell}$ is the $3 \times 3$ spectral matrix of the fields, implemented using the noiseless lensed spectra:
\begin{equation} \label{eq:cl_lens_qe_sims}
C^{\rm{len}}_{\vell} =
\begin{pmatrix}
\CellNoHatLensed{\ell}{T}{T} & \CellNoHatLensed{\ell}{T}{E} \cos 2 \psi_{\vell} & \CellNoHatLensed{\ell}{T}{E} \sin 2 \psi_{\vell} \\
\CellNoHatLensed{\ell}{T}{E} \cos 2 \psi_{\vell} &\CellNoHatLensed{\ell}{E}{E} \cos^{2} 2 \psi_{\vell} +  \CellNoHatLensed{\ell}{B}{B} \sin^{2} 2 \psi_{\vell}  &\left(\CellNoHatLensed{\ell}{E}{E} -\CellNoHatLensed{\ell}{B}{B} \right) \cos 2 \psi_{\vell}  \sin 2 \psi_{\vell}  \\
\CellNoHatLensed{\ell}{T}{E} \sin 2 \psi_{\vell} &\left(\CellNoHatLensed{\ell}{E}{E} -\CellNoHatLensed{\ell}{B}{B} \right)  \cos 2 \psi_{\vell}  \sin 2 \psi_{\vell}  & \CellNoHatLensed{\ell}{E}{E} \sin^{2} 2 \psi_{\vell}+  \CellNoHatLensed{\ell}{B}{B} \cos^{2} 2 \psi_{\vell},
\end{pmatrix}
\end{equation}
with $\psi_\vell$ the phase of the harmonic mode $\vell$.
The Cartesian components are then rotated to curl and potential modes in harmonic space, and normalized by the response to the potential mode, which is identical to the $N^{(0)}$ bias.

This implementation based on a $(T,Q,U)$ description of the data differs (very slightly) from more traditional implementations based on combining the set of estimators built from pairs from $(T,E,B)$, such as the state-of-the-art implementation from the Planck team \cite{Ade:2015zua}. The exact MV weights for the TE estimator are non-separable, making a exact implementation difficult to achieve with good scaling properties,
so the weights are usually approximated. Our implementation, which never calculates the separate estimators, has the advantage of avoiding this small approximation, and is identical to the exact minimum Gaussian variance $\phi$ estimate described in Sec.~\ref{sec:introduction}.
Its numerical cost is approximately proportional to the optimal $f_{\rm{sky}}\times \ell_{\rm{max}}^2$.

Finally, binning is performed over slowly varying quantities:
\begin{equation}
C_{\barb_i}^{XY} = \dfrac{1}{\Delta b_i} \dfrac{1}{[\barb_i (\barb_i + 1)]^{w_{XY}}} \sum_{\ell=b_i}^{b_{i+1} -1} [\ell(\ell+1)]^{w_{XY}} C_{\ell}^{XY},
\end{equation}
where $b_i$ and $\barb_i$ correspond to lower bin boundary and bin centre respectively of the bin number $i$.
The weight powers are chosen so that $w_{UV} =1$ and $w_{\phi \phi}=2$.

As mentioned in Sec.~\ref{sec:lensing_reconstruction}, we found that using lensed temperature power spectrum in the weights of the quadratic estimator was not accurate enough to reconstruct correctly the largest scales (biases of $\sim 10\%$ from the temperature estimator for $\ell_{\rm max}=3000$). To avoid this bias, we instead use the non-perturbative gradient power spectrum $C_{\ell}^{\tT \nabla \tT}$ in Eq.~\ref{eq:cl_lens_qe_sims} for the weights of the quadratic estimator (but we keep the lensed spectra in the covariance matrix for the inverse-filtering operations).
We did not find it necessary to extend this to polarization.

\subsection{Comparison between the model and simulations}

Once we have the set of lensed CMB power spectra and reconstructed lensing potential power spectra, we compute the auto-covariances and cross-covariances.
We show in Fig.~\ref{fig:corrmat_phixCMB_CMBS4_noise} the comparison between the full cross-covariance model developed in the previous section (including terms cancelled and not cancelled by RDN0 subtraction), and the results obtained on simulations.
We show the cross-correlation matrices for a better visualization.
The agreement between both is rather good in temperature, where the model manages to reproduce most of the features seen on simulations.
For the EE and TE power spectra, the model is not as good as for temperature, but nonetheless the agreement is good enough for our purpose ($\sim$10$\%$ difference at large scales, and less than 5$\%$ elsewhere).
The main difference between analytic and simulation results in this case can be seen at large lens scales, where the model tends to overestimate the correlation.
The case of B modes is different.
The model fails to capture correctly the effect at large and intermediate scales (up to 50$\%$ difference at $\ell_{\phi^{\rm MV}\phi^{\rm MV}}=500$).

Fig.~\ref{fig:corrmat_phixCMB_CMBS4_noise_RDN0} shows the same comparison between model and simulations, but we keep only the terms not cancelled by RDN0 subtraction (and therefore used later to derive cosmological parameter uncertainties).
The main reduction of correlation is seen at small lensing scales, and the correlations remain almost untouched at large lensing scales.
This change is expected because fluctuations in the CMB power, which induce larger lensing correlations at small noise-dominated scales than large scales, are suppressed by the RDN0 subtraction.
The agreement between model and simulations is on overall improved, although the difference at large scales is still visible.

These differences between the model and simulations are however not detectable at the $1 \sigma$ level, though the slight overestimation of the correlation in the analytic model reduces the total detectability of the non-Gaussian covariance terms to about $5 \sigma$. However, we note an improvement in the agreement between the model and the simulations if the B modes are discarded from the analysis (both then agree within $0.2 \sigma$). We also found that the final results on cosmological parameters are not greatly affected, with both analytical and simulation results giving similar results (up to few percent differences on the cosmological parameter uncertainties, see Sec.~\ref{sec:cosmo-param}).

\section{Impact of correlations on parameter estimation} \label{sec:impact}

Often likelihoods are approximated as Gaussian in the power spectra, neglecting correlations between the lensing and CMB power spectra. In this section we test this simple approximation against a likelihood using our full covariance, including lensing-induced off diagonal terms and cross-covariances between the spectra.
Our full covariance model includes all contributions listed in Sec.~\ref{sec:covariances} after applying the realization-dependent noise bias subtraction, namely all covariances listed in Sec.~\ref{sec:two-point-covariances} for the CMB auto-covariances, Sec.~\ref{sec:notcrosscancelled} for the cross-covariances, and Sec.~\ref{sec:four-point-covariances-RDN0ed} for the lensing auto-covariances.

\subsection{Lensing amplitude estimates}\label{sec:lensing_amplitude}

We first focus on an overall amplitude parameter $A$ of a fiducial lensing potential power spectrum such that

\begin{equation}
C_\ell^{\phi\phi} = A C_\ell^{\phi\phi} |_{\rm fid},
\end{equation}
keeping all other cosmological parameters fixed.
The lensing amplitude $A$ can be estimated from the reconstructed lensing potential power spectrum using (e.g.~\cite{Schmittfull:2013uea})

\begin{equation} \label{eq:lensing_amplitude_phi}
\hat{A}_{\phi^{XY} \phi^{ZW}} = \dfrac{\sum_{\ell \ell'} C_{\ell}^{\phi \phi}  \Big( \text{cov}^{-1}_{\hat{\phi}^{XY} \hat{\phi}^{ZW}} \Big)_{\ell \ell'} \Big(
\hat{C}_{\ell',\rm{RDN0}}^{\phi^{XY} \phi^{ZW}}
- N_{\ell'}^{(1),XYZW} \Big) }{\sum_{\ell} S_{\ell}^{XYZW}},
\end{equation}
where $S_{\ell}^{XYZW}=\sum_{\ell'} C_{\ell}^{\phi \phi}  \big( \text{cov}^{-1}_{\hat{\phi}^{XY}\hat{\phi}^{ZW}} \big)_{\ell \ell'} C_{\ell'}^{\phi \phi}$ ensures $\langle\hat A\rangle=1$ if data come from the fiducial model, and $( \text{cov}^{-1}_{\hat{\phi}^{XY} \hat{\phi}^{ZW}})$ indicates the matrix inverse of the full lensing auto-covariance matrix including off-diagonal components as defined in Sec.~\ref{sec:four-point-covariances-RDN0ed} (with the realization-dependent bias subtraction).
Note that we have neglected the lensing potential power-spectrum dependence of the $N^{(1)}$ bias, which is subdominant for our purpose.
Similarly, the lensing amplitude can also instead be estimated from the lensed CMB power spectrum
\begin{equation} \label{eq:lensing_amplitude_CMB}
\hat{A}_{\tU \tV} = \dfrac{\sum_{\ell \ell'} \Big( \hat{C}_{\ell,\text{expt}}^{\tU \tV} - {C}_{\ell,\text{expt}}^{UV} \Big) \Big( \text{cov}^{-1}_{\tU \tV,\text{expt}} \Big)_{\ell \ell'} \Big( C_{\ell'}^{\tU \tV} - C_{\ell'}^{UV} \Big)}{\sum_{\ell} S_{\ell}^{\tU \tV}},
\end{equation}
where $U, V \in \{ T, E, B \}$ and $S_{\ell}^{\tU \tV} = \sum_{\ell'} \Big( C_{\ell}^{\tU \tV} - C_{\ell}^{UV} \Big) \Big( \text{cov}^{-1}_{\tU \tV,\text{expt}} \Big)_{\ell \ell'} \Big( C_{\ell'}^{\tU \tV} - C_{\ell'}^{UV} \Big)$, and $\Big( \text{cov}^{-1}_{\tU \tV,\text{expt}} \Big)$ indicates the matrix inverse of the full CMB auto-covariance matrix including off-diagonal components as defined in Sec.~\ref{sec:two-point-covariances}.
The corresponding standard deviations $\sigma_{A_{\phi \phi}}$ and $\sigma_{A_{\tU \tV}}$ of the estimators are the inverse square root of the denominator of each estimator, and they are reported in Table~\ref{tab:correlation-lensing-amplitudes}.
The best constraints come from the BB spectrum and the reconstructed lensing potential power spectrum (an order of magnitude tighter than current measurements).
The TT, EE, and TE spectra perform equally well, with an uncertainty on the lensing amplitude almost three times larger.
Note that these values for the uncertainties agree very well with the values obtained on simulations.

\begin{figure}[!htbp]
\begin{center}
\includegraphics[width=1.0\textwidth]{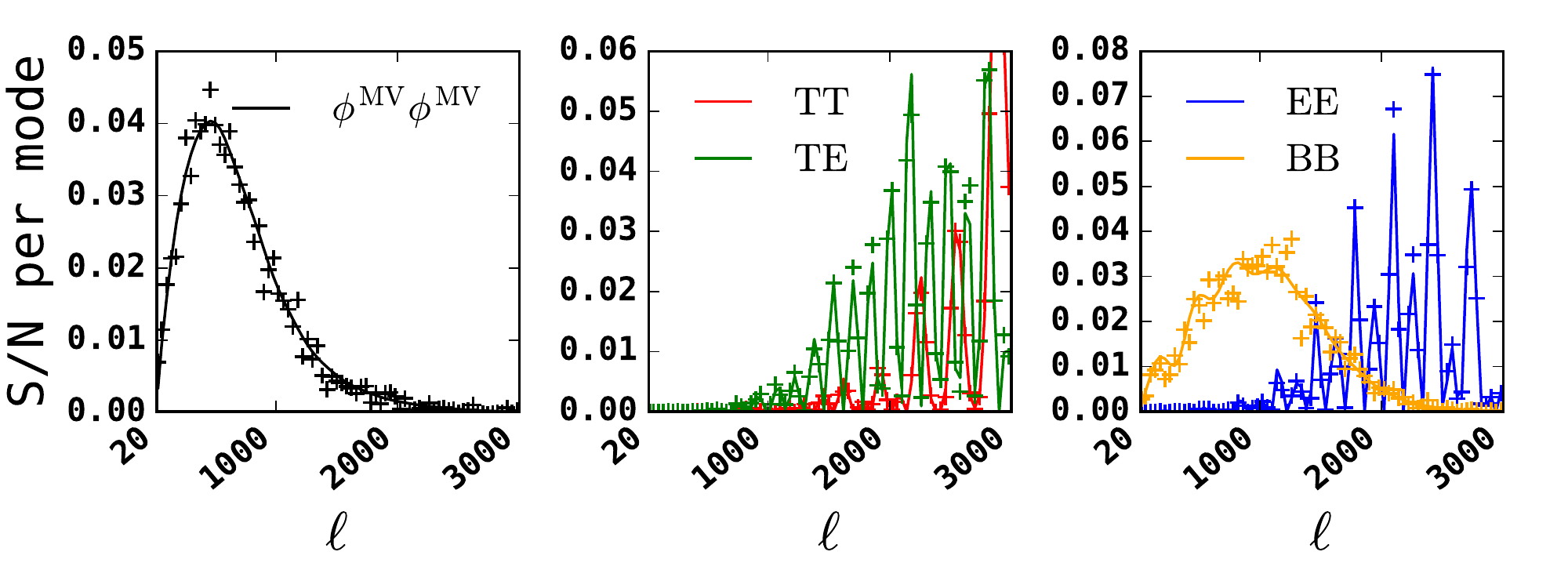}
\caption{Per-$\ell$ contribution $S_{\ell}^{\rm MVMV}$ (left) and $S_{\ell}^{\tU \tV}$ (middle and right) to the total lensing
signal-to-noise squared as defined in Eqs.~\ref{eq:lensing_amplitude_phi} $\&$ \ref{eq:lensing_amplitude_CMB} (solid lines).
For comparison, we overplot the results obtained on simulation (plus signs). The realization-dependent noise bias subtraction has been used to compute $S_{\ell}^{\rm MVMV}$. We show the contribution from minimum variance lensing reconstruction (left panel), lensed temperature (red), lensed E modes (blue), lensed TE power spectrum (green), and lensed B modes (yellow). Since we are interested in the distribution of the signal-to-noise over different scales (and not in the total SNR), each estimator has been normalised so that their integral over $\ell$ is unity.
}
\label{fig:SNR}
\end{center}
\end{figure}

Fig.~\ref{fig:SNR} shows the contributions $S_{\ell}^{\rm MVMV}$ and $S_{\ell}^{\tU \tV}$ to the total signal-to-noise squared ($S/N = \sum_\ell S_\ell$) for the lensing amplitude estimates.
The lensing $S/N$ coming from the 4-point lensing reconstruction comes from a broad range of multipoles, with a maximum value at $\ell \approx 500$.
Lensing $S/N$ from EE and TE power spectra is mainly coming from the acoustic peaks, where the smoothing effect is largest, and the smoothing itself probes mainly lensing modes with $\ell \alt 150$.
However, unlike the case for \planck\ studied in Ref.~\cite{Schmittfull:2013uea} for example, the lensing $S/N$ from the temperature power spectrum for future experiments comes from both the acoustic peaks and the very small scales (where the spectrum starts to become lensing dominated).
Finally, the $S/N$ from B modes comes from the entire range of the lensing B-mode spectrum.
If we extend the multipole range to $\ell_{\text{max}}=4000$, this description is still true for polarization spectra which are noise dominated at high multipoles, but the signal-to-noise for temperature becomes completely dominated by the very small scales ($\ell > 3000$).
However, in practice it may be difficult to clean such high-$\ell$ temperature modes from contaminating foregrounds.

The covariance between the two lensing amplitude estimators is given by

\begin{equation}\label{eq:analytic_cov}
\text{cov}(\hat{A}_{\phi \phi},\hat{A}_{\tU \tV}) = \sigma_{A_{\phi\phi}}^2 \sigma_{A_{\tU \tV}}^2 \sum_{\ell_1, \ell_2 = \ell_{\text{min}}^{\phi \phi}}^{\ell_{\text{max}}^{\phi \phi}} \sum_{\ell_3, \ell_4 = \ell_{\text{min}}^{\tU \tV}}^{\ell_{\text{max}}^{\tU \tV}} C_{\ell_1}^{\phi\phi}  \Big( \text{cov}^{-1}_{\hat{\phi}\hat{\phi}} \Big)_{\ell_1 \ell_2} \text{cov}(\hat{C}_{\ell_2}^{\phi\phi},\hat{C}_{\ell_3,\text{expt}}^{\tU \tV}) \Big( \text{cov}^{-1}_{\tU \tV,\text{expt}} \Big)_{\ell_3 \ell_4} \Big( C_{\ell_4}^{\tU \tV} - C_{\ell_4}^{UV} \Big),
\end{equation}
where we explicitly dropped the indices for the lensing for clarity.
The corresponding correlation is computed via

\begin{figure}[tbp]
\begin{center}
\includegraphics[width=0.5\textwidth]{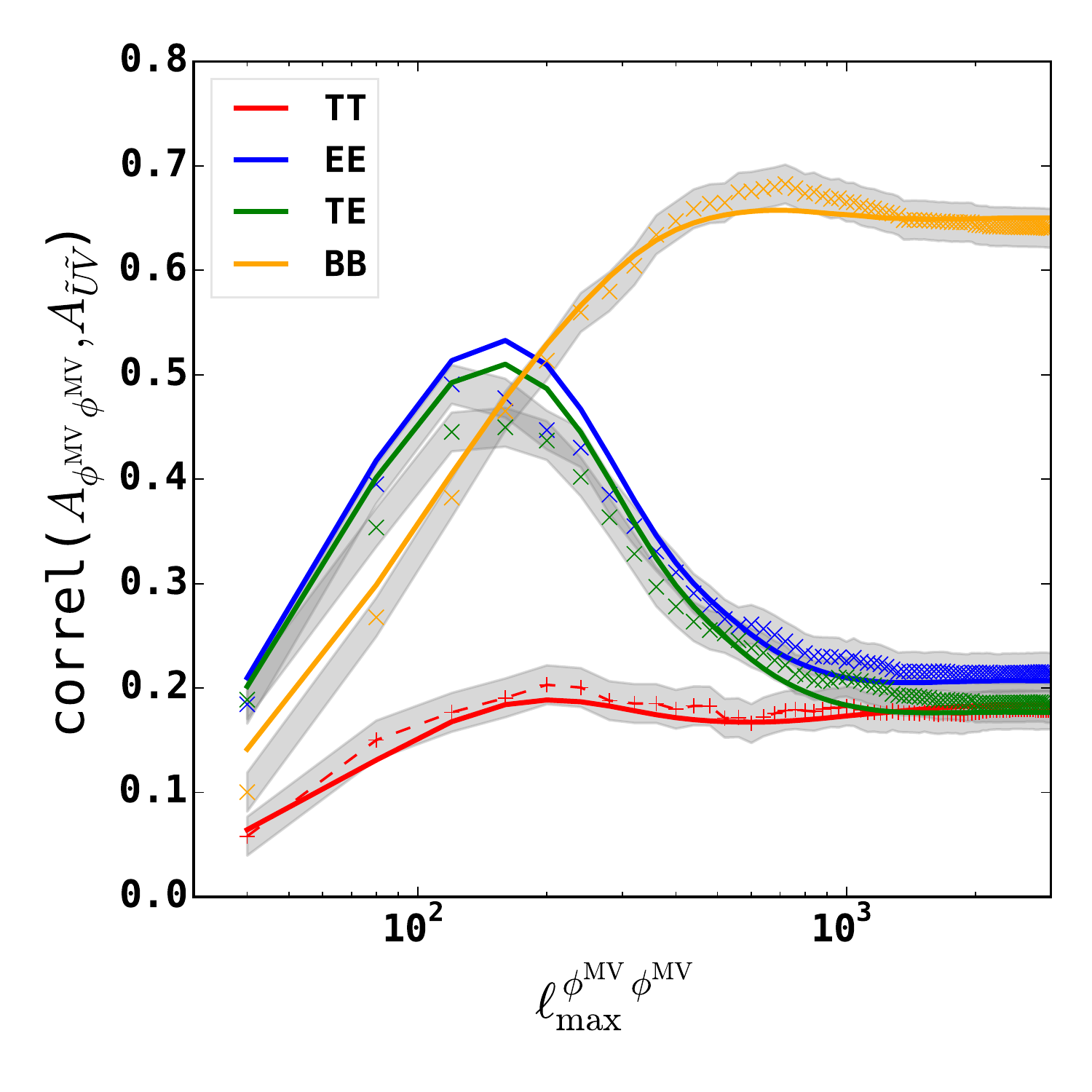}\includegraphics[width=0.5\textwidth]{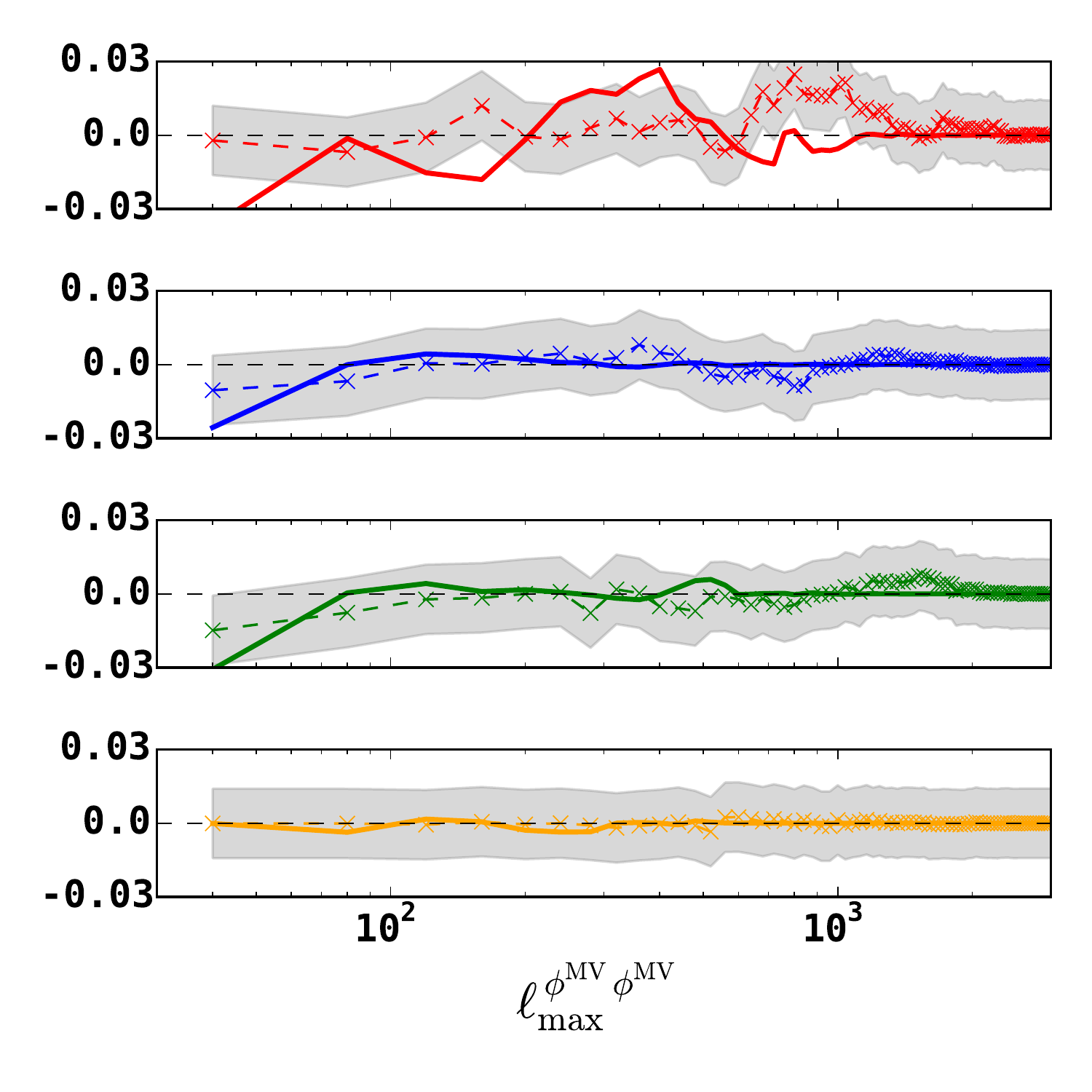}
\includegraphics[width=1.0\textwidth]{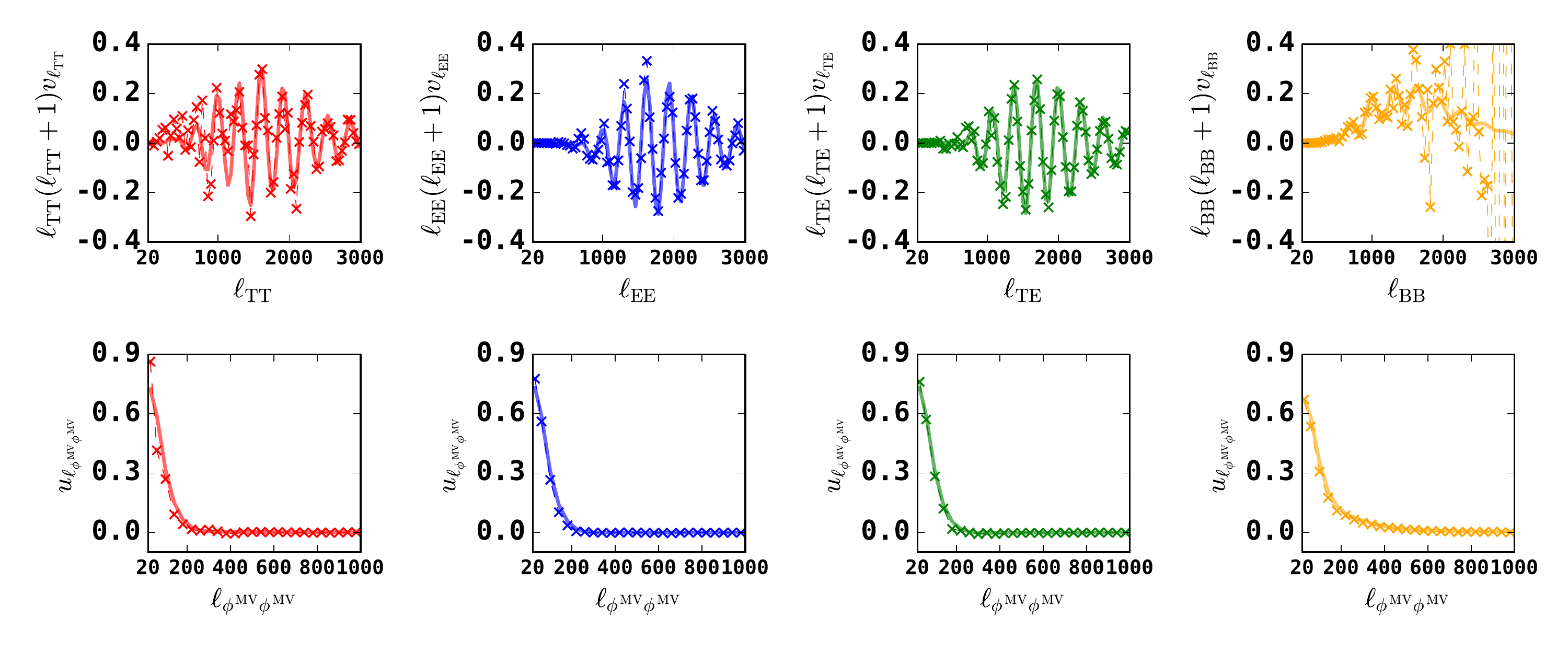}
\caption{\textit{Upper left panel:} Correlation between the 4-point lensing amplitude $\hat A_{\phi \phi}$ (estimated from the measured lensing power spectrum) and 2-point lensing amplitude $\hat A_{\tU \tV}$ (estimated from measured, lensed CMB power spectra), as a function of the maximum multipole $\ell_{\text{max}}^{\phi \phi}$ used for the reconstruction of the lensing potential (keeping $\ell_{\rm max}^{UV}=3000$ fixed).
 \textit{Upper right panel:} Same as upper left panel, but we project out the first singular vector from the cross-covariance matrix. \textit{Lower panel:} First singular vectors from the cross-covariance matrix (top: CMB side, bottom: lensing side). In all plots, the analytical model (solid lines) is compared against simulations (crosses), with TT, EE, TE, and BB shown in red, blue, green, and yellow, respectively.
All results in this figure have been obtained by using realization-dependent noise bias subtraction. The shaded grey regions in the upper left and right panels are the uncertainty coming from the simulations. See text for more discussion.
}
\label{fig:corr-lphimax}
\end{center}
\end{figure}

\begin{equation}
\text{corr}(\hat{A}_{\phi\phi},\hat{A}_{\tU \tV}) = \dfrac{\text{cov}(\hat{A}_{\phi\phi},\hat{A}_{\tU \tV})}{\sigma_{A_{\phi\phi}} \sigma_{A_{\tU \tV}}}.
\end{equation}

We show in Fig.~\ref{fig:corr-lphimax} the evolution of this correlation between the lensing amplitude estimates as a function of the maximum multipole $\ell_{\text{max}}^{\phi \phi}$ for the lensing reconstruction (upper left panel). Here we apply the realization-dependent bias subtraction, and keep $\ell_{\rm max}^{\tilde{U}\tilde{V}}=3000$.

We first notice that unlike the \planck\ case studied in Ref.~\cite{Schmittfull:2013uea}, for CMB-S4 the correlations are no longer negligible.
In the temperature case, Ref.~\cite{Schmittfull:2013uea} found an amplitude correlation of around 3$\%$, while for \textsc{CMB-S4} (red solid line) the correlation between the lensing amplitude estimated from the TT power spectrum and that estimated from the minimum-variance lensing power is as big as 20$\%$ when using the full range of multipoles for lensing reconstruction. Estimating the lensing amplitude from the polarization power spectra EE (blue solid line) or TE (green solid line) instead of TT gives similar correlations for $\ell_{\text{max}}^{\phi \phi}=3000$.
Restricting the amplitude estimates to larger scales leads to higher correlations, reaching up to $50\%$ correlations for $\ell_{\text{max}}^{\phi \phi} \sim 150$.

The lensing amplitude estimator using the BB power spectrum is most affected by correlations (yellow): it is more than $60\%$ correlated with the lensing amplitude estimated from the minimum variance 4-point lensing measurement if $\ell_{\text{max}}^{\phi \phi}\gtrsim 500$.
This is not surprising since the B modes are generated entirely by lensing.
The case of BB is also the case where the analytic model and simulations tend to show some noticeable differences (results obtained on simulations are shown using the cross mark, with the same colour code as the analytic results; shaded grey regions are the uncertainty coming from the simulations).
This difference is also seen in Figs.~\ref{fig:corrmat_phixCMB_CMBS4_noise} $\&$ \ref{fig:corrmat_phixCMB_CMBS4_noise_RDN0}, where we can see residual correlations in the difference between model and simulations.
We investigated the reason of the failure at low multipoles for the lensing reconstruction, but we were not able to find a better agreement within the subset of terms that we calculate.
We note that the model and simulations do reach reasonable agreement if we consider the whole range of multipoles for the lensing reconstruction ($i.e.$ at $\ell^{\phi \phi}_{\text{max}} = 3000$ the model and simulations give similar results).

In Table~\ref{tab:correlation-lensing-amplitudes} we show the impact of correlations on lensing amplitude estimates between the minimum variance reconstruction estimate and the estimate from lensed CMB spectra using the full range of multipoles ($20 \leq \ell \leq 3000$).
The use of the realization-dependent noise bias subtraction has little impact on the correlation (20$\%$ decrease at most).
This small impact tells us that for CMB-S4, the noise contribution is not the dominant one.
After realization-dependent noise bias subtraction, there are only two contributions left in our approximation to the cross-covariance matrix: the signal and the Type B trispectrum contributions (Eqs.~\ref{eq:xcorr-signal} $\&$ \ref{eq:xcorr-trispectrumB}).
We found that the Type B trispectrum contribution is negligible with respect to the signal contribution (almost an order of magnitude lower for relevant scales), i.e.~the signal contribution dominates the lensing amplitude correlation.

The signal contribution to the cross-correlation matrix has a low-rank structure (as shown by Refs.~\cite{Smith:2006nk,Schmittfull:2013uea}).
We performed a Singular Value Decomposition (SVD) of the cross-correlation matrix (after RDN0), and found that most of the information was contained in the first couple of modes\footnote{There is a factor $\sim$50 between the first and the second singular values, and a factor $\sim$10 between the second and the third singular values.}.
In Fig.~\ref{fig:corr-lphimax} (upper right panel), we show the correlation between lensing estimates after projecting out the first mode obtained from the SVD of the cross-covariance matrix.
For polarization spectra, the first mode captures all the correlation, and after projection the residual correlation in all cases becomes extremely small (consistent with zero given the uncertainty of the Monte Carlo simulations).
For the temperature case, one can see some residual correlations at low reconstruction multipoles reaching a few percent, which disappear completely if we also project out the second singular vector from the SVD.
Both analytic predictions (coloured solid lines) and simulations (coloured cross, with shaded region denoting the uncertainty from the MC simulations) are in agreement, despite for example the initial difference seen for the B-mode case if $\ell^{\phi_{\rm MV}\phi_{\rm MV}}_{\rm max}< 2000$.

\begin{table}[btp]
\begin{center}
\parbox{.47\linewidth}{
\caption{\label{tab:correlation-lensing-amplitudes}Impact of correlations on lensing amplitude estimates between the minimum variance estimate and the estimate from lensed CMB spectra, using the whole range of multipoles ($i.e.$ $\ell^{\phi \phi}_{\text{max}} = \ell^{UV}_{\text{max}} = 3000$). Note that for S4 the reconstruction is signal dominated on large scales, hence RDN0 subtraction does not greatly decrease the correlation.
For comparison, the uncertainty in the estimate of the lensing amplitude from minimum variance lensing only is $\sigma_{A_{\phi \phi}} = 0.0037$ ($\sigma_{A_{\phi \phi}} = 0.0057$ if RDN0 is not used).}
\begin{tabular}{lcccc}
\hline
\hline
 & $\sigma_{A_{\tU\tV}}$  & corr($\hat A_{\phi^{\rm MV}\phi^{\rm MV}}, \hat A_{\tU\tV}$) & corr($\hat A_{\phi^{\rm MV}\phi^{\rm MV}}, \hat A_{\tU\tV}$) \\
$UV$ & & RDN0 not used & RDN0 used\\
\hline
TT & 0.011 &  15$\%$ & 17$\%$ \\
EE & 0.012  &  25$\%$ & 21$\%$ \\
TE & 0.013  & 22$\%$ & 18$\%$ \\
BB & 0.0043 &  70$\%$ & 64$\%$ \\
\end{tabular}
}
\hfill
\parbox[0]{.47\linewidth}{
\caption{\label{tab:correlation-lensing-amplitudes_redundancy_complementarity}Constraints on lensing amplitude estimates using the CMB power spectra $( \hat{C}^{\tT \tT}_{\ell,\text{expt}}, \hat{C}^{\tE \tE}_{\ell,\text{expt}}, \hat{C}^{\tT \tE}_{\ell,\text{expt}}, \hat{C}^{\tB \tB}_{\ell,\text{expt}})$ and lensing reconstruction $\phi\phi=( \hat{C}^{\phi^{\rm MV} \phi^{\rm MV}}_{\ell,\rm RDN0} )$, using the whole range of multipoles ($i.e.$ $\ell^{\phi \phi}_{\text{max}} = \ell^{\rm CMB}_{\text{max}} = 3000$) and keeping other cosmological parameters fixed. We show results using Gaussian covariance ($\sigma_{A_{\alpha}}^{\rm G}$) and non-Gaussian covariance ($\sigma_{A_{\alpha}}$), for the joint constraint, from the lensing reconstruction alone, and from the lensed CMB spectra alone.
}
\begin{tabular}{lccc}
\hline
\hline
 & $\sigma_{A_{\alpha}}^{\rm G}$  & $\sigma_{A_{\alpha}}$  \\
\hline
$\alpha$ = CMB+$\phi\phi$ & 0.0020 &  0.0035 \\
$\alpha$ = $\phi\phi$ & 0.0037 &  0.0037 \\
$\alpha$ = CMB & 0.0024 &  0.0039 \\
\end{tabular}
 \vspace{1.3cm}
}
\end{center}
\end{table}

The lower panel of Fig.~\ref{fig:corr-lphimax} shows the first singular vectors coming from the SVD of the cross-covariance matrix.
The upper lower subpanel shows the right singular vectors $v_{\ell_{UV}}$ (normalised by $\ell_{UV}(\ell_{UV}+1)$).
In the case of TT, EE, and TE, they mainly correspond to the difference between lensed and unlensed CMB power spectra (but not completely due to the presence of the Type B trispectrum contribution).
In the case of BB, the first right singular vector has the shape of the E-mode power
spectrum\footnote{Although $v_{\ell_{BB}}$ disagrees between model and simulations at high $\ell_{BB}$, we believe this is not important in practice because the $BB$ power spectrum is mostly noise-dominated on these scales.}.
The lower subpanel shows the left singular vectors $u_{\ell_{\phi^{\rm MV} \phi^{\rm MV}}}$.
In all the cases, the large-scale lenses are the dominant cause of the covariance with the lensed spectra,
and the corresponding signal term is easy to model.

We also performed this analysis by extending the range of multipoles for both CMB spectra and lensing reconstruction up to $\ell_{\rm max}=4000$.
We found that results on polarization do not change (at S4 noise and beam levels, CMB polarization spectra are dominated by the noise for multipoles beyond 3000), so only temperature results change.
At these small scales the lensed temperature power spectrum is driven by the lensing power, and the uncertainty on the lensing amplitude from the lensed temperature spectrum is reduced by a factor of 3 compared to the case where $\ell_{\rm max}=3000$, becoming comparable to constraints from B modes (see Table~\ref{tab:correlation-lensing-amplitudes}).
The correlation between the lensing amplitude estimates is also slightly enhanced, reaching 30$\%$ in the case of temperature (for polarization spectra the correlation already reached a plateau at lower lensing reconstruction multipoles).
However, given that polarization does not provide additional constraints at scales beyond $\ell\sim3000$, and it might well be difficult to access those smaller scales in practice for temperature, we do not further consider scales beyond $\ell_{\rm max} = 3000$.

Given that the lensing amplitudes estimated from the BB power spectrum and the minimum variance lensing power spectrum have similar error bars and are rather correlated, one might worry that lensing reconstruction does not add much independent lensing information to the lensed BB power spectrum (or vice-versa).
Table~\ref{tab:correlation-lensing-amplitudes_redundancy_complementarity} shows how the error on the estimate of the lensing amplitude changes using different data sets: joint estimation from lensed CMB spectra and lensing reconstruction, lensing reconstruction only, and lensed CMB spectra.
Using the non-Gaussian covariance avoids double counting the same lensing information, and therefore
increases the uncertainties on the lensing amplitude compared to treating the constraints as independent.
However, for S4 the lensed CMB does still add additional information on the lensing amplitude, so the measurements are still somewhat complementary\footnote{In the limit of perfect noiseless measurements the lensing field would be reconstructed perfectly, and the lensed CMB power spectra then cannot contain additional (direct) information on the lensing amplitude as there are no additional lensing modes to constrain.
However, in general the perturbative maximum likelihood lensing power spectrum estimator does include both the four-point reconstruction and the response from the lensed power spectra~\cite{Schmittfull:2013uea}.}.
In addition, systematics may affect these measurements in different ways in practice, making it useful to consider both.
Furthermore, iterative/maximum-likelihood lensing reconstruction methods \cite{2003PhRvD..68h3002H} that are more optimal than the quadratic estimator used in this paper are expected to improve the accuracy of the lensing reconstruction significantly.
We therefore always expect the lensing information from lensing reconstruction to be extremely useful.

\subsection{Cosmological parameter estimation} \label{sec:cosmo-param}

In this section, we quantify the effect of the covariances for the estimation of cosmological parameters from the joint data vector described in Eq.~\ref{eq:joint-data-vector}.
As in the previous section, we base our full covariance model on all the contributions that are left after RDN0 subtraction (listed in Sections \ref{sec:two-point-covariances}, \ref{sec:notcrosscancelled} and \ref{sec:four-point-covariances-RDN0ed}).
We first discuss Fisher forecasts, which are fast to evaluate for many combinations of experimental configurations. We then discuss small differences obtained with Markov-Chain Monte Carlo (MCMC) forecasts that better account for the non-Gaussian posterior shape. In both cases we approximate the binned $\hat{\textbf{C}}_\ell$ distribution as having a Gaussian distribution in a fiducial $\Lambda$CDM model, and investigate the change in results when the covariance matrix is approximated using the form expected for Gaussian fields, compared to the approximate more accurate model developed in this paper (accounting for non-Gaussianity of the lensed CMB fields and the lensing reconstruction estimator).

Our base $\Lambda$CDM set of parameters is based on the latest \planck\ constraints \cite{Aghanim:2016yuo}, with $\Omega_{\rm b} h^2 = 0.02214$, $\Omega_{\rm c} h^2 = 0.1207$, $10^9 A_{\rm s} = 2.11788$, $n_{\rm s} = 0.9624$, $\tau = 0.0581$, $100 \theta_{\rm MC} = 1.0411$.
We do not include large-scale \planck\ (or other) CMB data, but do include a $\tau$ prior motivated by the recent \planck\ measurement: $\tau = 0.0581\pm 0.01$ (which we discuss further below).

\subsubsection{Fisher matrix forecasts}

The effect of the covariance on the estimation of a set of cosmological parameters $p_\alpha$ can be estimated using the Fisher matrix
\begin{equation}
F_{\alpha \beta} = \sum_{\ell_1 \ell_2} \dfrac{\partial \textbf{C}_{\ell_1}}{\partial p_\alpha} \left( \textbf{cov}^{-1} \right)_{\ell_1 \ell_2} \dfrac{\partial \textbf{C}_{\ell_2}}{\partial p_\beta},
\end{equation}
where $\textbf{cov}^{-1}$ is the inverse covariance matrix of the joint data vector in Eq.~\ref{eq:joint-data-vector}.
The error $\sigma_{p_\alpha}$ on the parameter $p_\alpha$ is then given by
\begin{equation}
\sigma_{p_\alpha} = \sqrt{\left( F^{-1} \right)_{\alpha \alpha}}.
\end{equation}

We show in Table~\ref{tab:cosmo_param} the impact of non-Gaussian covariances for a base set of flat $\Lambda$CDM cosmological parameters.
The first column shows the 1$\sigma$ bound for the base parameters assuming the covariance matrix of the joint data vector is fully non-Gaussian (containing all blocks, each of which contains off-diagonal elements).
The second column shows the fractional change in this 1$\sigma$ bound when switching between non-Gaussian and Gaussian covariances:
\begin{equation}\label{eq:degradation}
\rm{Degradation} = \dfrac{\sigma_{p_\alpha} - \sigma_{p_\alpha}^{\rm G}}{\sigma_{p_\alpha}^{\rm G}},
\end{equation}
where $\sigma_{p_\alpha}$ and $\sigma_{p_\alpha}^{\rm G}$ are the errors on parameter $p_\alpha$ in the case of non-Gaussian and Gaussian covariances respectively.
The impact of non-Gaussian covariances on the errors is modest, at most 9$\%$ for this set of base parameters.
The third column shows consistency between errors from the the analytic model developed in Sec.~\ref{sec:covariances} and the results obtained from simulations:
\begin{equation}
\rm{Agreement} = \dfrac{\sigma_{p_\alpha} - \sigma_{p_\alpha}^{\rm sims}}{\sigma_{p_\alpha}^{\rm sims}},
\end{equation}
where $\sigma_{p_\alpha}$ and $\sigma_{p_\alpha}^{\rm sims}$ are the errors on parameter $p_\alpha$ obtained from the analytical model and the simulations respectively (both using non-Gaussian covariances).
The agreement between both is good, with a difference on the error at most a few percent, which is sufficient to validate the model developed in this paper. The larger fractional differences in the off-diagonal covariances seen in Figs.~\ref{fig:corrmat_phixCMB_CMBS4_noise} $\&$ \ref{fig:corrmat_phixCMB_CMBS4_noise_RDN0} are not that important because the magnitude of the non-Gaussian terms is small compared to Gaussian terms, so it is not necessary to model them with very high accuracy.

\begin{table}[htpb]
\begin{center}
\parbox{.47\linewidth}{
\caption{ \label{tab:cosmo_param} Impact of non-Gaussian covariances for the base set of parameters from a Fisher matrix analysis using our fiducial CMB-S4 configuration: 1$\sigma$ bound for parameters using full non-Gaussian covariance, degradation with respect to Gaussian covariance, and relative difference between analytical predictions and simulations (error on the error). Parameters used: $\Omega_{\rm b} h^2 = 0.02214$, $\Omega_{\rm c} h^2 = 0.1207$, $10^9 A_{\rm s} = 2.11788$, $n_{\rm s} = 0.9624$, $\tau = 0.0581$, $100 \theta_{\rm MC} = 1.0411$. We assume a Gaussian prior on $\tau$ ($\pm$ 0.01).
}
\begin{tabular}{cccc}
\hline
\hline
 $p_\alpha$& $\sigma_{p_\alpha}$ & Degradation  & Agreement with \\
 & & ($\%$)& simulations ($\%$) \\
\hline
$\Omega_{\rm{b}} h^2$ & 0.00004 & 0.7 & -0.4 \\
$\Omega_{\rm{c}} h^2$ & 0.00066 & 2.4 & -3.2 \\
$10^9 A_{\rm{s}}$ & 0.022 & 9.3 & -1.1 \\
$n_{\rm{s}}$ & 0.0022 & 0.5 & -0.1 \\
$\tau$ & 0.0062 & 6.4 & -1.6 \\
$100\theta_{\rm MC}$ & 0.00010 & -0.2 & -3.9 \\
\end{tabular}
}
\hfill
\parbox[0]{.47\linewidth}{
\caption{
\label{tab:cosmo_param_ext} Constraints on one-parameter extensions to the base $\Lambda$CDM model from a Fisher matrix analysis using our fiducial CMB-S4 configuration: 1$\sigma$ bound for parameters using full non-Gaussian covariance, degradation with respect to Gaussian covariance, and relative difference between analytical predictions and simulations (error on the error). Fiducial parameters used: $M_{\nu}= 100$ meV, $N_{\rm eff} = 3.046$, and $Y_{\rm p} = 0.245$. We assume a Gaussian prior on $\tau$ ($\pm$ 0.01). Each row represents the constraint on the parameter after marginalization over the base set.
}
\begin{tabular}{cccc}
\hline
\hline
$p_\alpha$ & $\sigma_{p_\alpha}$ & Degradation  & Agreement with \\
 &  & ($\%$)& simulations ($\%$) \\
\hline
$M_{\nu}/\rm{meV}$ & 72 & 11 & 0.1 \\
$N_{\rm{eff}}$ & 0.052 & 3.1 & -1.1 \\
$Y_{\rm{p}}$ & 0.0030 & 2.8 & -0.8 \\
\end{tabular}
 \vspace{1.3cm}
}
\end{center}
\end{table}

We also probe one-parameter extensions to the base set of parameters, by considering three other parameters: the sum of neutrino masses, the effective number of relativistic degrees of freedom, and the fraction of baryonic mass in helium. In all three cases we use fiducial values of $M_{\nu}= 100$ meV, $N_{\rm eff} = 3.046$, and $Y_{\rm p} = 0.245$.
The neutrino sector is modelled as three massive neutrinos following a normal hierarchy,
using the measured mass splitting from oscillation experiments\footnote{Physically, we also could use an inverted hierarchy since the lowest possible total mass allowed is around 100 meV.
We approximate the full normal hierarchy by two distinct mass eigenstates, where the lower mass state has degeneracy two. }.
The impact of non-Gaussian covariances from the Fisher matrix is shown in Table~\ref{tab:cosmo_param_ext}.
For these one-parameter extensions, and this experimental configuration, only the sum of neutrino masses is affected by the non-Gaussian covariance: we obtain a 1 $\sigma$ error of 72 meV on the sum of neutrino masses with a mild $\sim 10\%$ degradation from the non-Gaussian covariance.
Notice that if we discard information from the measurement of the lensing potential (using only TT, EE, TE, and BB spectra), the constraint on the sum of neutrino masses is even less sensitive to the choice of covariance ($i.e.$ including or not the off-diagonal elements in the auto-covariance of the lensed CMB spectra makes a change of -0.9$\%$ in the result\footnote{A negative non-Gaussian degradation factor means that the error on the parameter is better when the non-Gaussian covariance is used. This is the case when the parameters are anti-correlated (partially or fully) rather than correlated.}), and the 1 $\sigma$ error becomes $94$ meV.

\begin{figure}[!htbp]
\begin{center}
\includegraphics[width=0.5\textwidth]{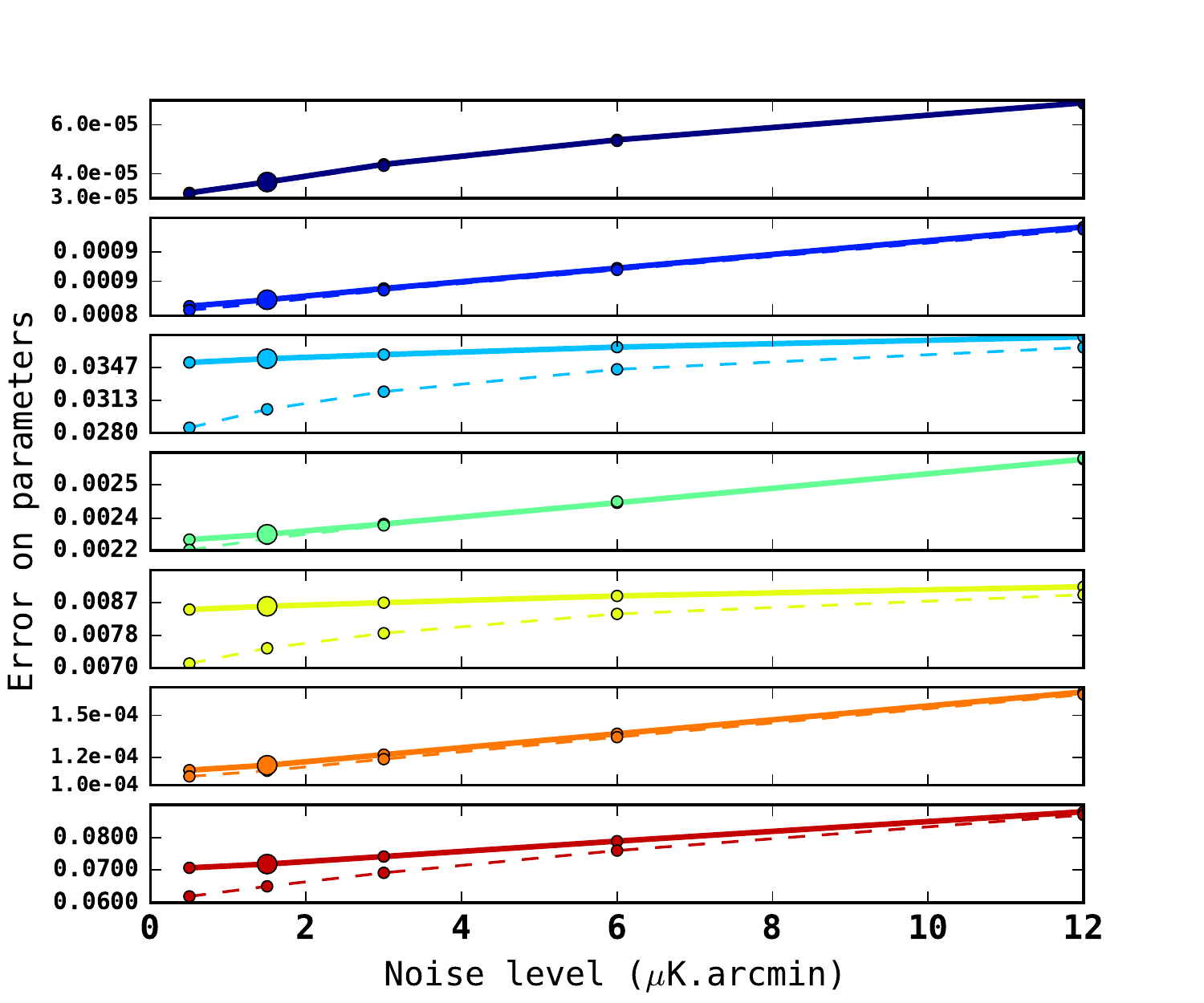}\includegraphics[width=0.5\textwidth]{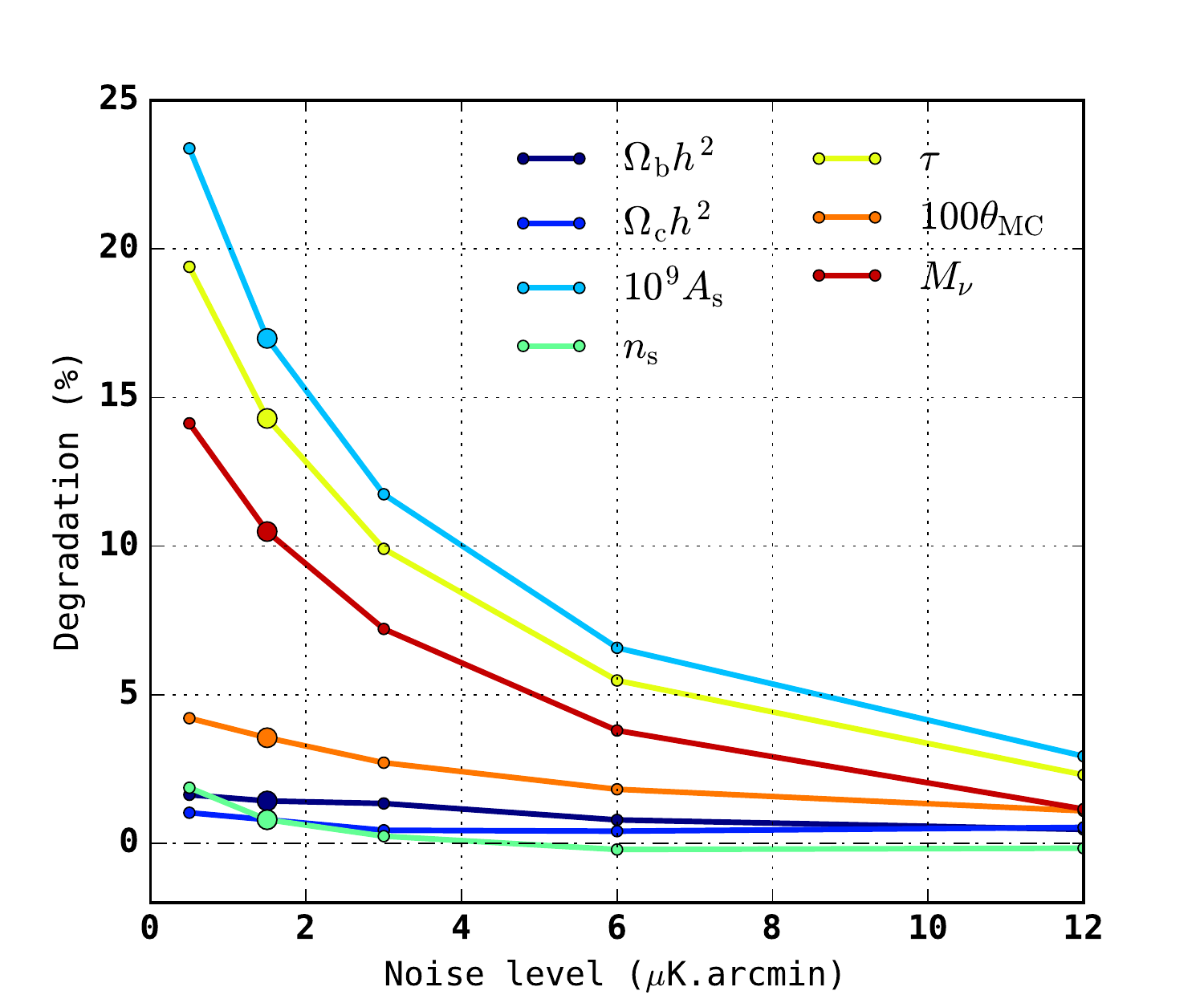}
\caption{Impact of the covariances on the cosmological parameter (Fisher) errors as a function of experimental noise, in the context of a $\Lambda$CDM+$M_\nu$ model. The figures on the left show the errors on the parameters in the case of non-Gaussian covariance (solid curves) and Gaussian covariance (dashed curves). The errors have no units but the sum of neutrino masses in units of eV. The figures on the right show the degradation factor on the error for parameters when using non-Gaussian covariance instead of Gaussian covariance (see Eq.~\ref{eq:degradation}). The CMB-S4 case described in this paper is highlighted with bigger circle markers (temperature noise level of 1.5 $\mu$K.arcmin). We keep the beam width and fraction of observed sky fixed, with values 3.0 arcmin and 40$\%$ respectively, and we apply a Gaussian prior $\tau=0.058 \pm 0.01$. The degradation factor is smaller for bigger values of the experimental noise. See text for more explanations.}
\label{fig:cosmo_param_noise_variation}
\end{center}
\end{figure}

We study the impact of the experimental set-up by looking at the impact of the non-Gaussian covariances for various noise levels.
The results are summarized in Fig.~\ref{fig:cosmo_param_noise_variation}.
We show the errors on parameters and the degradation factors as a function of the instrumental noise level, for noise values from 0.75 $\mu$K.arcmin to 12 $\mu$K.arcmin, and for a $\Lambda$CDM+$M_\nu$ model.
The results for our assumed S4 configuration, the main focus of this paper, are highlighted with bigger circle marks. The degradation increases for smaller noise levels, reaching around 15-20$\%$ for the parameters directly influenced by the lensing (i.e. the optical depth, the scalar amplitude, and the sum of neutrino masses)\footnote{Without a prior on $\tau$, the degradation factor reaches more than 50$\%$ for $\tau$, $A_{\rm s}$, and almost 40$\%$ for $M_\nu$. For a very optimistic experimental setup (and no prior on $\tau$), with noise of 0.5$\mu$K.arcmin, beam size of 1 arcmin and extending the multipole range up to 5000 in polarization, the impact of covariance reaches 50$\%$ on M$_\nu$ and 70$\%$ for $\tau$ and $A_{\rm s}$.}.
And not only does the degradation factor increase with decreasing noise level, but the error on the parameters plateaus.
From Fig.~\ref{fig:cosmo_param_noise_variation}, we observe for example that the error on $\tau$ is rather similar for a noise level of 1.5 $\mu$K.arcmin or 6 $\mu$K.arcmin if the full non-Gaussian covariance is considered.
However as soon as we increase further the level of noise, the degradation factor quickly decreases, with degradation factors for all parameters less than 5$\%$ for a final (temperature) noise of $\agt$10 $\mu$K.arcmin.

CMB-S4 will observe from the ground and therefore the largest scales will be difficult to measure.
We have conservatively set to zero scales with $\ell < 20$, which means that we do not constrain $\tau$ directly from the CMB polarization. Lensing does provide some amplitude information, but relating that to $\tau$ via the observed CMB amplitude is partly degenerate with the effect of other parameters like the neutrino mass. External priors will therefore be crucial to break parameter degeneracies and get tight constraints~\cite{Allison:2015qca}.
For the sum of the neutrino masses, there is a great improvement on the 1 $\sigma$ error if we use the \planck\ $\tau$ prior $\pm 0.01$ (see Table~\ref{table:planck_prior_cosmo_param_error_and_deg}) compared to no prior at all: the error is reduced by almost a factor three.
Moving from the current \planck\ prior on $\tau$ to the lowest achievable bound $\pm 0.002$ by a CMB experiment (assuming an instantaneous reionization process, for a full-sky cosmic variance limited experiment up to $\ell=2500$ as described in Ref.~\cite{Zaldarriaga:2008ap}), as shown in Table~\ref{table:CV_prior_cosmo_param_error_and_deg}, the error is again reduced by almost a factor 1.5.
Introducing a prior on $\tau$ also helps to decrease the non-Gaussian degradation factor in all the cases, and the impact of non-Gaussian covariance is less than 5$\%$ for $\tau$ prior equal to 0.002 (see Table~\ref{table:CV_prior_cosmo_param_error_and_deg}).
In the future, 21 cm experiments could provide even better constraints on $\tau$ as shown in Ref.~\cite{liu:2016eliminating} and therefore further lower the error on cosmological parameters depending on it\footnote{However we found that the improvement on the sum of neutrino masses for values of the prior lower than 0.002 is minor (for the values of noise, beam and bandwidth considered here).}.

\begin{table}[htpb]
\begin{center}
\caption{ \label{table:planck_prior_cosmo_param_error_and_deg} Comparison between the Fisher matrix analysis and the MCMC. For both methods, we show the impact of non-Gaussian covariances for a $\Lambda$CDM+$M_\nu$ model: 1$\sigma$ bound for parameters using full non-Gaussian covariance and degradation with respect to Gaussian covariance. The two columns on the left use only our fiducial CMB-S4 configuration, and the two columns on the right use our fiducial CMB-S4 configuration and forecasted BAO measurements from DESI and Euclid. For all results in this table, we assume a Gaussian prior on $\tau$ of $\pm$ 0.01.
}
\begin{tabular}{c||cc|cc||cc|cc}
\hline
\hline
& \multicolumn{4}{c||}{CMB-S4 alone} & \multicolumn{4}{c}{CMB-S4 + DESI/Euclid BAO} \\
\hline
& $\sigma_{p_\alpha}$ & Degradation & $\sigma_{p_\alpha}$ & Degradation & $\sigma_{p_\alpha}$ & Degradation & $\sigma_{p_\alpha}$ & Degradation  \\
 \raisebox{2ex}{ $p_\alpha$}& Fisher  & Fisher&MCMC  & MCMC & Fisher  & Fisher&MCMC  & MCMC \\
\hline
$\Omega_{\rm{b}} h^2$ & 0.000037 & 1$\%$ & 0.000035 &-4$\%$ & 0.000035  & 1$\%$   & 0.000035  & 3$\%$ \\
$\Omega_{\rm{c}} h^2$ & 0.00083 & 1$\%$ & 0.00074 & 1$\%$  & 0.00054  & 3$\%$   & 0.00075  & 8$\%$  \\
$10^9 A_{\rm{s}}$          & 0.036     & 17$\%$  & 0.033     & 11$\%$   & 0.028      & 11$\%$    & 0.031      & 13$\%$\\
$n_{\rm{s}}$                   & 0.0023   & 1$\%$ & 0.0023 & -2$\%$ & 0.0020   & 1$\%$   & 0.0019    & 2$\%$ \\
$\tau$                             &  0.0086  & 14$\%$  & 0.0081 & 10$\%$   & 0.0072   & 9$\%$   & 0.0080    & 11$\%$ \\
$100\theta_{\rm MC}$    & 0.00011 & 4$\%$ & 0.00011 & 0$\%$   & 0.000090 & 0$\%$ & 0.000088  & 2$\%$ \\
$M_{\nu}/\rm{meV}$                       & 72          & 11$\%$  & 69          & 5$\%$   & 24          & 4$\%$  & 34           & 5$\%$ \\
\end{tabular}
\end{center}

\begin{center}
\caption{\label{table:CV_prior_cosmo_param_error_and_deg} Same as Table~\ref{table:planck_prior_cosmo_param_error_and_deg}, but we assume a Gaussian prior on $\tau$ of $\pm$ 0.002.\hfill
}
\begin{tabular}{c||cc|cc||cc|cc}
\hline
\hline
& \multicolumn{4}{c||}{CMB-S4 alone} & \multicolumn{4}{c}{CMB-S4 + DESI/Euclid BAO} \\
\hline
& $\sigma_{p_\alpha}$ & Degradation & $\sigma_{p_\alpha}$ & Degradation & $\sigma_{p_\alpha}$ & Degradation & $\sigma_{p_\alpha}$ & Degradation  \\
 \raisebox{2ex}{ $p_\alpha$}& Fisher  & Fisher&MCMC  & MCMC & Fisher  & Fisher&MCMC  & MCMC \\
\hline
$\Omega_{\rm{b}} h^2$     & 0.000036 & 1$\%$ & 0.000035 & -2$\%$ & 0.000035 & 0$\%$ & 0.000035 & 0$\%$ \\
$\Omega_{\rm{c}} h^2$     & 0.00083 & 2$\%$ & 0.00068 & -2$\%$ & 0.00030 & -1$\%$ & 0.00023 & 2$\%$ \\
$10^9 A_{\rm{s}}$              & 0.0098 	 & 4$\%$ & 0.0092   & -3$\%$ & 0.0082   & 4$\%$ & 0.0084 	& 4$\%$ \\
$n_{\rm{s}}$                       & 0.0023 	& 3$\%$ &  0.0022   & -2$\%$ & 0.0018   & 6$\%$ & 0.0017 	& 6$\%$ \\
$\tau$                                & 0.0020 	& 1$\%$ & 0.0020    & 1$\%$  & 0.0020    & 1$\%$ & 0.0020 	& 3$\%$ \\
$100\theta_{\rm MC}$ 	& 0.00011 & 3$\%$ & 0.00011  & 4$\%$  & 0.000087 & 1$\%$ & 0.000086 	& -1$\%$ \\
$M_{\nu}/\rm{meV}$ 			& 53 		& 3$\%$ & 55 	   & -1$\%$ & 17 	      & 1$\%$ & 19 		& 1$\%$ \\
\end{tabular}

\end{center}
\end{table}

We also include a prior from future DESI and Euclid BAO measurements (see Tables V and VI in Ref.~\cite{font:2014desi} for the numbers used).
These combined constraints are shown in Tables~\ref{table:planck_prior_cosmo_param_error_and_deg} $\&$ \ref{table:CV_prior_cosmo_param_error_and_deg}, for two different priors on the optical depth $\tau$.
The inclusion of BAO priors lowers the errors, and the biggest effect is seen for the sum of neutrino masses, $\theta_{\rm MC}$ (or $H_0$), and $A_{\rm s}$, for which a measurement of BAO helps to strongly break the geometric degeneracy in the CMB data~\cite{Efstathiou:1998xx}.
The non-Gaussian degradation factor becomes smaller in the case without the BAO as the lensing becomes less important as the results are degeneracy limited.

\subsubsection{Monte Carlo Markov Chain posterior likelihood estimation}

To check the results obtained with the Fisher analysis, we make a direct likelihood exploration using a Monte Carlo Markov Chain (MCMC) approach by using the mean log likelihood evaluated at the fiducial model.
We focus on the $\Lambda$CDM+$M_{\nu}$ model, and run MCMC chains using \cosmomc\footnote{\url{http://cosmologist.info/cosmomc/}} for the following cases: CMB-S4 like experiment, Gaussian or non-Gaussian covariances, with or without inclusion of external BAO measurements, and two different priors on $\tau$ (0.01 and 0.002).

\begin{figure}[t]
\begin{center}
\includegraphics[width=0.5\textwidth]{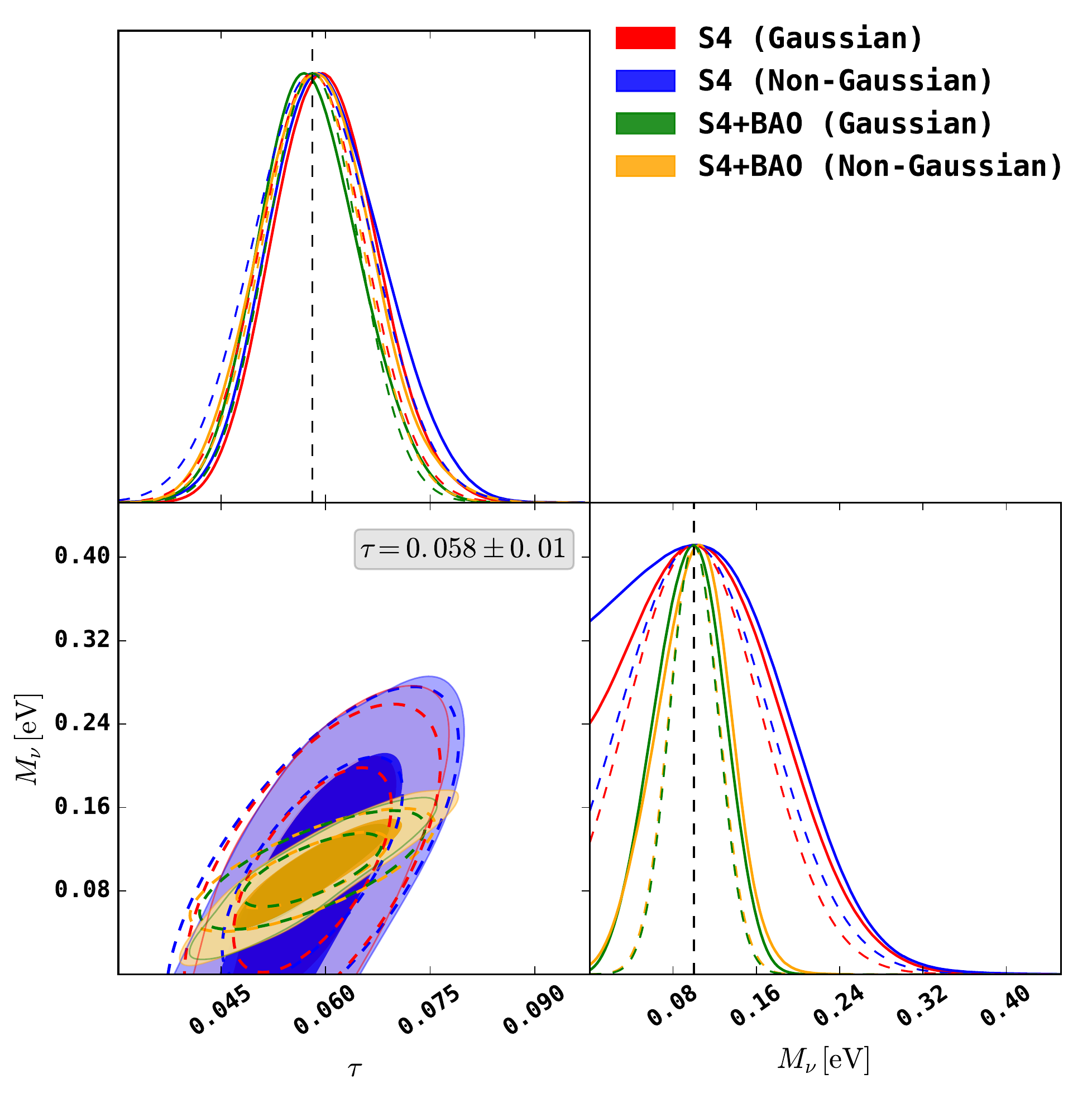}\includegraphics[width=0.5\textwidth]{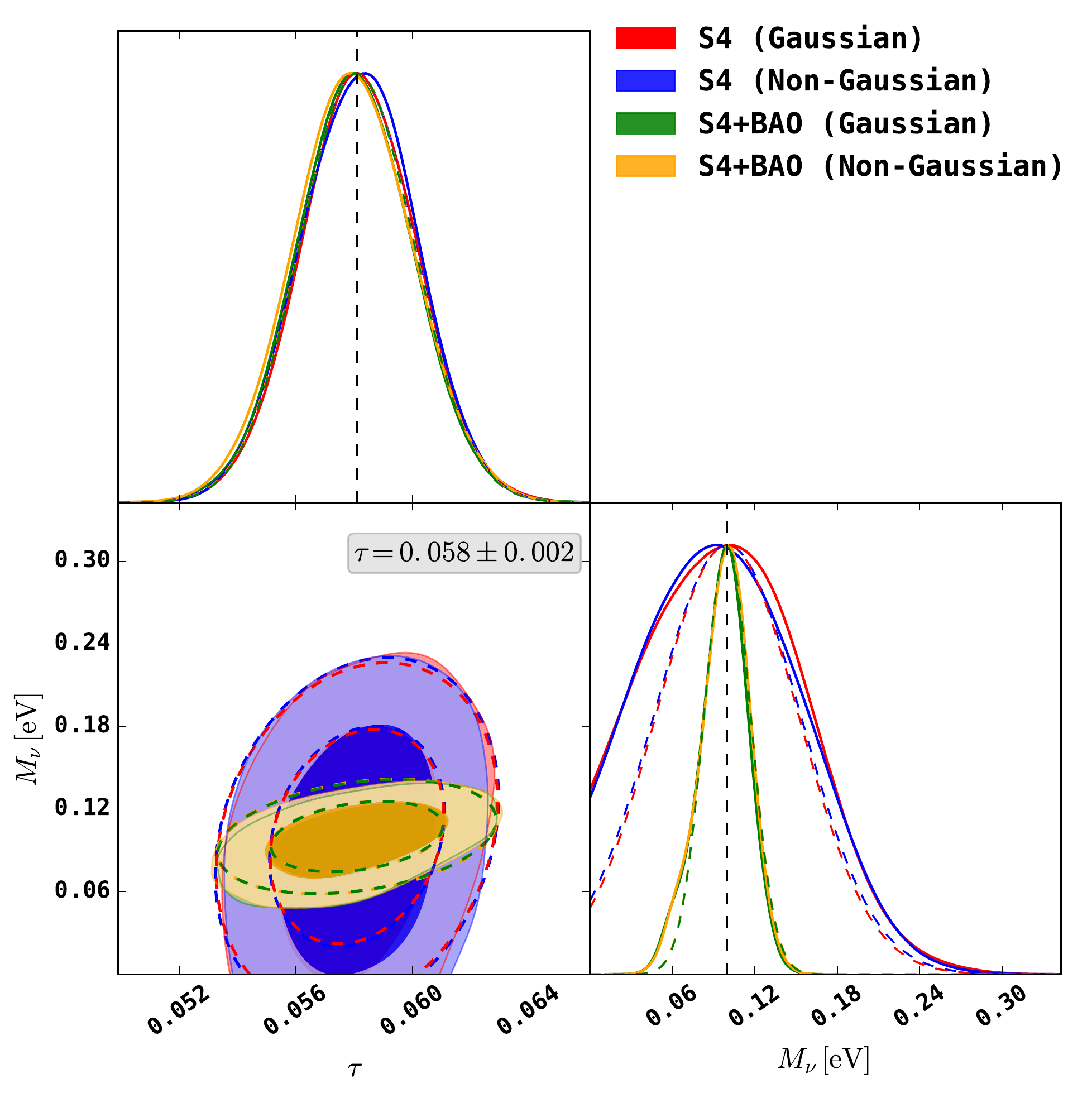}
\caption{One- and two-dimensional joint marginalized posterior distributions at 68$\%$ and 95$\%$ confidence level in the $\tau - M_\nu$ plane from a $\Lambda$CDM+$M_{\nu}$ set of parameters. The left panel assumes a prior on $\tau$ from the current \planck\ measurement ($\pm$ 0.01), and the right panel assumes the lowest bound from CMB measurement ($\pm$ 0.002). 
In each panel, we plot the contours for CMB-S4 alone (red filled ellipses for the Gaussian covariance; blue filled ellipse for the non-Gaussian covariance) and CMB-S4+BAO measurements (green filled ellipses for the Gaussian covariance; orange filled ellipse for the non-Gaussian covariance). For comparison, we also show the results from the Fisher matrix (dashed ellipses). The disagreement between MCMC and Fisher results is smaller when BAO measurements are added. For the sum of neutrino masses, as we do not account for the excluded parameter region $M_\nu < 0$ in the Fisher approach, making a direct comparison with MCMC of one-dimensional joint marginalized posterior distributions difficult (see Tables~\ref{table:planck_prior_cosmo_param_error_and_deg} $\&$ \ref{table:CV_prior_cosmo_param_error_and_deg} for marginalized 1 $\sigma$ errors on parameters).
}
\label{fig:cosmo_param_MCMC}
\end{center}
\end{figure}

Fig.~\ref{fig:cosmo_param_MCMC} shows one- and two-dimensional joint marginalized posterior distributions in the $\{\tau, M_{\nu} \}$ parameter space (using the 0.01 $\tau$ prior on the left and the 0.002 $\tau$ prior on the right).
The corresponding 1 $\sigma$ errors and non-Gaussian degradation factors are shown in Table~\ref{table:planck_prior_cosmo_param_error_and_deg} (using the 0.01 $\tau$ prior) and Table~\ref{table:CV_prior_cosmo_param_error_and_deg} (using the 0.002 $\tau$ prior).
These tables also list results for the other cosmological models of the $\Lambda$CDM+$M_{\nu}$ model. Note that the numerical sampling error for the MCMC is a few percent, so the degradation factors are reported at this level of precision and small percent-level changes should not be over-interpreted. To ease the comparison between both methods we also report the results from the Fisher method at this level of precision.

The left panel of Fig.~\ref{fig:cosmo_param_MCMC} shows that if we use the prior on $\tau$ from the recent \planck\ measurement, the MCMC results (filled ellipses and solid lines) are not in good agreement with the Fisher matrix results (dashed).
This disagreement is seen for Gaussian and non-Gaussian covariances, with and without BAO information.
While the directions of degeneracy axes are similar, the parameter errors inferred from the posterior distributions differ significantly between the Fisher matrix and MCMC analyses in some cases (hinting at a non-Gaussian posterior which is better captured by the MCMC analysis as discussed at the end of this section).

From Table~\ref{table:planck_prior_cosmo_param_error_and_deg}, 
we find the MCMC results to be less sensitive to the impact of non-Gaussian covariance than the results from Fisher matrix for CMB-S4 alone, although the orders of magnitude remain the same. One notable difference is on the sum of the neutrino masses, where the Fisher method indicates twice the degradation of MCMC.
If we include the BAO measurement, parameter errors shrink in both cases, although differences between Fisher analysis and MCMC remain in some cases (most notably for $\Omega_c h^2$ and $M_\nu$). Significant non-Gaussian degradations tend to agree better in most cases if BAO information is included.
Generically, better agreement between Fisher and MCMC analyses when including BAO information might be related to the fact that the inclusion of BAO helps to break the degeneracies present in the CMB data, which should make the likelihood more Gaussian.

If we now put a tighter prior on $\tau$ by assuming the lowest possible bound from CMB measurement (right panel of Fig.~\ref{fig:cosmo_param_MCMC}, and Table~\ref{table:CV_prior_cosmo_param_error_and_deg}), the agreement between the MCMC and Fisher results are in better agreement, although still slightly different.
For the sum of neutrino masses, the difference in 1 $\sigma$ error is decreased slightly with $\sigma_{M_\nu} = 53$ meV from Fisher and $\sigma_{M_\nu} = 55$ meV from MCMC results in the CMB-S4 case, and $\sigma_{M_\nu} = 17$ meV from Fisher and $\sigma_{M_\nu} = 19$ meV in the case of CMB-S4+BAO.
The corresponding degradation factors do change, but remain small (less than 5$\%$) for both methods (comparable to the sampling error in the MCMC case).

Several works,  e.g.~Refs.~\cite{Lesgourgues:2005yv,2012JCAP...09..009W,2012MNRAS.425.1170H}, already pointed out that a Fisher analysis may not give very accurate error estimates for some combinations of parameters, noting significant discrepancies between Fisher matrix estimation and MCMC results.
When there are curving degeneracies or excluded regions in the parameter space (like $M_\nu<0$) the shape is very non-Gaussian, and Fisher results are expected to be unreliable.
To get more reliable Fisher estimates, we could find a set of more Gaussian parameters for which the Fisher errors could be calculated.
For example Ref.~\cite{2012JCAP...09..009W} proposed physically motivated Gaussian parameterizations (although still strongly advocating for the use of MCMC), or Ref.~\cite{Joachimi:2011iq} proposed performing Box-Cox transformations on the parameter space.
However, these extensions are beyond the scope of this paper. The shape of actual data posteriors would of course also depend on the true model, and the actual realization of the data obtained.

\section{Conclusions}  \label{sec:conclusions}

We developed a detailed (but approximate) model for the auto and cross-covariances of the CMB and CMB lensing power spectra and tested it against simulations. The main conclusions are that:
\begin{itemize}
\item There are correlations between the CMB and lensing reconstruction power spectra, as well as small off-diagonal non-Gaussian contributions to the covariances of auto power spectra. The correlation and non-Gaussian contributions are small, but are easily detectable in a fit to future S4-like data (at $\sim 5 \sigma$, improving the model fit to the data).
\item Non-Gaussian contributions to the CMB auto-correlation are small, but can be accurately modelled using the approximations of Sec.~\ref{sec:two-point-covariances}~\cite{Smith:2006nk,Li:2006pu, Rocher:2006fh, BenoitLevy:2012va, Schmittfull:2013uea}. B-modes produce the dominant off-diagonal contributions to the auto-covariance. Temperature and E modes perform equally well, with a combined impact as large as the B modes alone.
\item Using estimators with realization-dependent reconstruction noise subtraction, the auto-covariance of the lensing reconstruction power spectrum is well approximated by the simplest Gaussian model.
\item The cross-correlation with the CMB is well modelled by the approximation
    \begin{multline}
\covPhiCMB{\ell_1, \rm RDN0}{\ell_2}{XY}{ZW}{U}{V} \approx \sum_{\ell_3} \dfrac{\partial \Cell{\ell_1}{\phi^{XY}}{\phi^{ZW}}}{\partial \Cell{\ell_3}{\phi}{\phi}} \dfrac{2}{2\ell_3+1}(\Cell{\ell_3}{\phi}{\phi})^2 \dfrac{\partial \Cell{\ell_2}{\tU}{\tV}}{\partial \Cell{\ell_3}{\phi}{\phi}} \\
+ \dfrac{\Cell{\ell_1}{\phi}{\phi}}{2\ell_2+1} \Big \{ \Big[ \dfrac{\Amplitude{\ell_1}{XY} \CellNoHatLensednoisy{\ell_2}{X}{U}}{2\ell_1+1} \sum_{\ell_3} \gweight{\ell_2}{\ell_3}{\ell_1}{XY}\fweightLensed{\ell_2}{\ell_1}{\ell_3}{VY} + (X \leftrightarrow Y) \Big] + (X \leftrightarrow Z, Y \leftrightarrow W) \Big \} + (U \leftrightarrow V),
\end{multline}
where the first term dominates\footnote{In the case of CMB-S4 alone and a prior on $\tau$ of 0.01, neglecting the second term in the covariance leads to a change at most of 1$\%$ of the non-Gaussian degradation factor.}. In practice, one can quickly test the impact of the non-Gaussian covariance by using the simplified expression for the cross-covariance:
\begin{equation} \label{eq:cross-simplified}
\covPhiCMB{\ell_1,\rm RDN0}{\ell_2}{XY}{ZW}{U}{V} \approx \sum_{\ell_3} \dfrac{\partial \Cell{\ell_1}{\phi^{XY}}{\phi^{ZW}}}{\partial \Cell{\ell_3}{\phi}{\phi}} \dfrac{2}{2\ell_3+1}(\Cell{\ell_3}{\phi}{\phi})^2 \dfrac{\partial \Cell{\ell_2}{\tU}{\tV}}{\partial \Cell{\ell_3}{\phi}{\phi}}.
\end{equation}
For the considered possible CMB-S4 experimental setup, the correlation is most important for large lensing scales ($\ell_{\phi \phi} < 500$), but affects a large range of CMB scales ($\ell_{\rm{TT}, \rm{EE}, \rm{TE}} > 1000$ and $20<\ell_{\rm BB} < 2000$).
\item These simple analytic models match S4-like simulations with excellent accuracy for the auto-correlation, and good, but not perfect, accuracy for the cross-correlation.
\item The impact of the non-Gaussian covariance is not negligible for a CMB-S4 experiment: neglecting it would lead to an underestimation of errors on the lensing amplitude because the lensed CMB and lensing reconstruction are double counting the same information (the estimates are correlated by up to 60\% if the BB spectrum is included). However, the correlation is dominated by one or two eigenmodes, which could be projected out to reduce the correlation.
\item Using a Fisher matrix analysis, we show that correlations can affect standard cosmological parameter errors from the CMB by up to several tens of percent for an CMB-S4 experiment. For lower sensitivity observations the correlations should be safely negligible. Using additional external data (priors on $\tau$, BAO measurements) also makes the correlations negligible with CMB-S4. A more accurate analysis using MCMC gives similar conclusions but also highlights some inaccuracies of the Fisher matrix analysis.
\item Although the impact is small for standard cosmological parameters if external data is included, an accurate likelihood model should include the correlations, which is easy using the model presented here. The code is available at \url{https://github.com/JulienPeloton/lenscov}.
\end{itemize}

In practice, the full covariance could be estimated from simulations, in which case there would be no need for an analytic model.
However, accurate covariance estimates require running a large number of CMB realizations (typically corresponding to a total area of hundreds of full skies), which is very computationally expensive.
In particular, the entire suite of simulations needs to be rerun from scratch every time experiment specifications such as noise level, beam size or sky coverage change.
Making these changes is much simpler using our analytic covariance model by changing for example noise level or beam size when evaluating the equations.
Another potential disadvantage of simulated covariances is that they are always somewhat noisy due to the finite number of simulations used, which can lead to sub-optimal error bars.
Based on our covariance model and comparisons with simulations,
one might be able to obtain less noisy covariance estimates by exploiting the fact that the covariances are dominated by a few singular modes, which could be estimated from much fewer realizations than the covariance between all multipole bins. The analytic model could also be used as a prior or regulator for simulation based estimators.

A potential disadvantage of the analytic covariance model is that it relies on the quadratic estimator of the lensing reconstruction, and it is not immediately clear how to extend it to more optimal lensing reconstruction estimators that rely on iterative estimates \cite{2003PhRvD..68h3002H}. However, the leading cross-covariance term (Eq.~\ref{eq:cross-simplified}) is caused by a signal covariance which would likely remain of the same form, and would therefore probably still dominate the non-Gaussian covariance.

We have not modelled the covariance with delensed CMB spectra. While the delensing process should reduce correlations between CMB and lensing, we would still expect the residual covariances to contain some non-Gaussianities (see for example Ref.~\cite{Namikawa:2015tba} for the detailed B-modes case, or Ref.~\cite{Green:2016cjr} which proposes an analytical model of the covariance of the delensed spectra).
Future work could investigate this further with simulations, or extend the analytic model of this paper.

\ifwordcount
\else
\acknowledgements
The authors would like to thank Blake Sherwin for discussion and his contributions at an early stage of the project.
The authors also thank Daniel Green, Joel Meyers, Uros Seljak and Alexander van Engelen for useful discussions.
JP, AL and JC acknowledge support from the European Research Council under
the European Union's Seventh Framework Programme (FP/2007-2013) / ERC Grant Agreement No.~[616170].
JP and AL acknowledge support
from the Science and Technology Facilities Council {[}grant number
ST/L000652/1{]}.
MS acknowledges support from the Bezos Fund through the Institute for Advanced Study.
This research used resources of the National Energy Research Scientific Computing Center, a DOE Office of Science User Facility supported by the Office of Science of the U.S.~Department of Energy under Contract No.~DE-AC02-05CH11231.

\begin{appendix}

\section{Derivations of terms in covariances} \label{app:cross-cov}

\subsection{Auto-covariance of the CMB} \label{app:two-point-covariances}

Here we give a partial derivation for some of the terms in the auto-covariance of the CMB (see Sec.~\ref{sec:two-point-covariances}) to build some intuition for where the various terms come from.

First note that the lensed CMB is linear in the unlensed CMB, so $\tilde{U}_{\ell m}  =\sum_{X, \ell' m'} X_{\ell' m'}\frac{\partial \tilde{U}_{\ell m}}{\partial X_{\ell' m'}}$, where the unlensed fields are $X\in \{T,E,B\}$.
Contractions over unlensed fields can therefore be done using $\sum_{XY}\delta_{\ell \ell'}\delta_{m m'}\frac{1}{2}C^{XY}_{\ell}\frac{\partial^2}{\partial X_{\ell m}\partial Y_{\ell' m'}^*}$.
For a set of isotropic Gaussian fields with covariance matrix $\mC_\ell$, we can also relate power spectrum derivatives to expectations of field derivatives using
\begin{equation}\frac{\partial \la A \ra}{\partial C^{XY}_\ell}  = \frac{(2\ell+1)}{2} \left\la  \left[(\mC^{-1}_\ell\hat{\mC}_\ell\mC^{-1}_\ell)_{XY}-(\mC_\ell^{-1})_{XY}\right]A\right\ra = \sum_{m}\left\la \frac{1}{2}\frac{\partial^2 A}{\partial X_{\ell m}\partial Y_{\ell m}^*} \right\ra,
\end{equation}
where $A$ is any function of the Gaussian fields. Isotropy of expectation values then implies for example that
\begin{equation}
  \left\la \frac{1}{2}\frac{\partial^2 \hat{C}^{\tU\tV}_{\ell_1}}{\partial X_{\ell m}\partial Y_{\ell' m'}^*} \right\ra   = \frac{\delta_{\ell\ell'}\delta_{m m'}}{2\ell+1} \frac{\partial C^{\tU\tV}_{\ell_1}}{\partial C_\ell^{XY}}.
  \label{eq:diffrelation}
\end{equation}

To analyse the auto-covariance we expand it into a Gaussian piece,
\begin{equation}\label{eq:app-variance-CMB}
\text{cov}_G(\CellLensednoisy{\ell_1}{U}{V}, \CellLensednoisy{\ell_2}{U^\prime}{V^\prime}) \equiv \delta_{\ell_1 \ell_2} \dfrac{1}{2\ell_1 + 1} \Big ( \CellNoHatLensednoisy{\ell_1}{U}{U^\prime} \CellNoHatLensednoisy{\ell_1}{V}{V^\prime} + \CellNoHatLensednoisy{\ell_1}{U}{V^\prime} \CellNoHatLensednoisy{\ell_1}{V}{U^\prime} \Big)
\end{equation}
plus fully connected non-Gaussian pieces as follows
\begin{align}
\covlensCMB{U}{V}{U^\prime}{V^\prime} &= \dfrac{1}{(2\ell_1+1)(2\ell_2+1)}  \sum_{m_1 m_2} \langle \tU_{\ell_1 m_1} {\tV_{\ell_1 m_1}}^* \tU_{\ell_2 m_2}^\prime \tV_{\ell_2 m_2}^{\prime *} \rangle - \CellNoHatLensednoisy{\ell_1}{U}{V} \CellNoHatLensednoisy{\ell_2}{U^\prime}{V^\prime}\\
 \nonumber\\&=
\text{cov}_G(\CellLensednoisy{\ell_1}{U}{V}, \CellLensednoisy{\ell_2}{U^\prime}{V^\prime}) +
\text{cov}_{\text{conn}}(\CellLensednoisy{\ell_1}{U}{V}, \CellLensednoisy{\ell_2}{U^\prime}{V^\prime}).
\label{eq:autoexpansion}
\end{align}
The leading-order connected piece is given by
\begin{multline}
\dfrac{1}{(2\ell_1+1)(2\ell_2+1)} \sum_{\ell_3 m_3,m_1,m_2}
C^\phi_{\ell_3}\left\la \frac{\partial (\tU_{\ell_1m_1}\tU_{\ell_2m_2}')}{\partial \phi_{\ell_3m_3}^*}\right\ra
\left\la \frac{\partial (\tV_{\ell_1m_1}^*\tV_{\ell_2m_2}^{'*})}{\partial \phi_{\ell_3m_3}}\right\ra + \text{perm}
\\= \dfrac{1}{(2\ell_1+1)(2\ell_2+1)} \sum_{\ell_3} \Cell{\ell_3}{\phi}{\phi} (\tilde{f}_{\ell_1 \ell_3 \ell_2}^{UU^\prime}\tilde{f}_{\ell_1 \ell_3 \ell_2}^{VV^\prime{}^*} + \tilde{f}_{\ell_1 \ell_3 \ell_2}^{UV^\prime}\tilde{f}_{\ell_1 \ell_3 \ell_2}^{VU^\prime{}^*}),
\end{multline}
where we used the definition of the non-perturbative response functions~\cite{Lewis:2011fk}
\begin{equation}
\left\la \frac{\partial (\tU_{l_1m_1}\tV_{l_2m_2})}{\partial \phi_{l_3m_3}^*}\right\ra \equiv \wigner{l_1}{l_2}{l_3}{m_1}{m_2}{m_3} \tilde{f}_{l_1l_3l_2}^{UV}.
\end{equation}
This leading order term is generally small, but for
$\covlensCMB{X}{Y}{B}{B}$ where $X,Y\in \{T,E\}$ it is more important and can be included in alternative ways.
For example, for the covariance between EE and BB, to lowest order we have
\begin{equation}
  \dfrac{1}{(2\ell_1+1)(2\ell_2+1)} \sum_{\ell_3} \Cell{\ell_3}{\phi}{\phi} (\tilde{f}_{\ell_1 \ell_3 \ell_2}^{EB}\tilde{f}_{\ell_1 \ell_3 \ell_2}^{EB{}^*} + \tilde{f}_{\ell_1 \ell_3 \ell_2}^{EB}\tilde{f}_{\ell_1 \ell_3 \ell_2}^{EB{}^*}) \approx  \frac{2 \left(\Cell{\ell_1}{{E}}{{E}}\right)^2}{2\ell_1+1}\frac{\partial \Cell{\ell_2}{\tilde{B}}{\tilde{B}}}{\partial \Cell{\ell_1}{E}{E}}.
\end{equation}
This term can also be included more generally by instead doing CMB cross-contractions using
\begin{multline}
\frac{1}{4} \sum_{\substack{XY,X'Y' \\ \ell_3m_3,\ell_4m_4}}
C_{\ell_3}^{XY}C_{\ell_4}^{X'Y'}
\left(\left\la  \frac{\partial^2 \hat{C}^{\tU\tV}_{\ell_1} }{\partial X_{\ell_3 m_3} \partial X^{*'}_{\ell_4 m_4}}
 \right\ra
\left\la  \frac{\partial^2 \hat{C}^{\tilde{B}\tilde{B}}_{\ell_2}}{\partial Y^*_{\ell_3 m_3} \partial Y'_{\ell_4 m_4}}
 \right\ra
+
\left\la  \frac{\partial^2 \hat{C}^{\tU\tV}_{\ell_1} }{\partial X_{\ell_3 m_3} \partial Y^{*'}_{\ell_4 m_4}}
 \right\ra
\left\la  \frac{\partial^2 \hat{C}^{\tilde{B}\tilde{B}}_{\ell_2}}{\partial Y^*_{\ell_3 m_3} \partial X'_{\ell_4 m_4}}
 \right\ra\right)\\
 = \sum_{XY,\ell} \dfrac{\partial \Cell{\ell_1}{\tU}{\tV}}{\partial \Cell{\ell}{X}{Y} }
\covCMBNoLens{\ell}{\ell}{X}{Y}{E}{E} \dfrac{\partial \Cell{\ell_2}{\tilde{B}}{\tilde{B}}}{\partial \Cell{\ell}{E}{E} },
\end{multline}
where in the second line we used Eq.~\eqref{eq:diffrelation}. This includes both leading and next order terms in $\Cell{\ell}{\phi}{\phi}$, and is non-perturbative in the derivative response functions.
Unfortunately, it is not straightforward to generalize this
to terms not involving BB, since complications then arise with disconnected terms appearing.

In all cases at next order there is also a simple term connected by two lensing fields given by
\begin{equation}
\frac{1}{2} \sum_{\ell m,\ell'm'}
C_{\ell}^{\phi\phi}C_{\ell'}^{\phi\phi}
\left\la \frac{\partial^2 \hat{C}^{\tU\tV}_{\ell_1}}{\partial \phi_{\ell m}\partial \phi^{*'}_{\ell' m'}} \right\ra
\left\la \frac{\partial^2 \hat{C}^{\tU'\tV'}_{\ell_2}}{\partial \phi^*_{\ell m}\partial \phi'_{\ell' m'}} \right\ra
=\sum_{\ell} \dfrac{\partial \Cell{\ell_1}{\tU}{\tV}}{\partial \Cell{\ell}{\phi}{\phi}}
\frac{2 (\Cell{\ell}{\phi}{\phi})^2}{2\ell+1}
\dfrac{\partial \Cell{\ell_2}{\tU'}{\tV'}}{\partial \Cell{\ell}{\phi}{\phi}},
\end{equation}
where we used Eq.~\eqref{eq:diffrelation} with $X=Y=\phi$.
The term is numerically important, and accounts for the correlated fluctuations in the lensed CMB induced by fluctuations in the lensing power.
We neglect various other contractions at this order that do not simplify into simple power spectrum derivatives.

\subsection{CMB and lensing cross-covariances: terms cancelled by the use of RDN0} \label{app:crosscancelled}

As seen in Sec.~\ref{sec:crosscancelled}, The RDN0 correction of the measured lensing power spectrum cancels two terms.
\\

\paragraph{Fully disconnected 6-point functions: noise term.}
The first identified contribution comes from fully disconnected terms (disconnected 6-point functions of zeroth order in $C^{\phi \phi}$), the leading covariance from the reconstruction noise being dependent on the CMB power.
Taking Gaussian contractions between two power spectra and taking the Gaussian noise part of the expectation, we have
\begin{multline}
\label{eq:xcorr-noise}
\covPhiCMB{\ell_1}{\ell_2}{XY}{ZW}{U}{V}_{\text{noise}} =
\frac{1}{2}\sum_{(ab),\ell m,\ell' m'}
\left\la \dfrac{\partial \hat{C}_{\ell_1}^{\phi^{XY}\phi^{ZW}}}{\partial \tilde{a}^*_{\ell m,\rm{expt}}\tilde{b}_{\ell' m',\rm{expt}}}\right\ra
\left( \CellNoHatLensednoisy{\ell}{a}{U}\CellNoHatLensednoisy{\ell'}{b}{V} + \CellNoHatLensednoisy{\ell'}{b}{U}\CellNoHatLensednoisy{\ell}{a}{V}\right)\delta_{\ell\ell_2}\delta_{\ell'\ell_2}
\\=
\sum_{(ab)} \dfrac{\partial \Nzero{\ell_1}{XY}{ZW}}{\partial \CellNoHatLensednoisy{\ell_2}{a}{b}} \text{cov}_G(\CellLensednoisy{\ell_2}{a}{b}, \CellLensednoisy{\ell_2}{U}{V}),
\end{multline}
for CMB pairs $(ab) \in \{ TT, TE, EE, BB \}$ and Gaussian covariance $\mathrm{cov}_G$.
This term reflects the fact that both the noise bias of the reconstructed lensing potential and the lensed CMB fields share the same CMB fields: if the CMB power fluctuates high, both the estimated lensed CMB power and Gaussian lensing reconstruction noise $N^{(0)}$ fluctuate high. It produces broadband correlations between wide ranges of scales, and dominates the full covariance with small-scale lensing power.

The large-scale ($\ell_{\phi \phi} \alt 1000$) lensing modes are reconstructed with high signal to noise for S4, and hence noise covariance only makes a small contribution to the total correlation there.
On smaller lensing reconstruction scales, $\ell_{\phi \phi} \gtrsim 1000$, fluctuations in the CMB power do induce larger correlations, with
the relative importance of the correlation for the different CMB power spectra depending on the weights of the different CMB modes in the minimum variance lensing estimator.
The MV lensing estimator gives most weight to the EB estimator at low $\ell_{\phi \phi}$, but the temperature estimator become relatively more important for the reconstruction of smaller-scale lenses (see Fig.~\ref{fig:n0_n1_bias}).
Low $\ell_{\phi \phi}$ are therefore only very weakly correlated to the TT spectrum on large scales even ignoring the signal variance, with BB and EE correlations dominating the noise contributions there; the TT and TE correlations dominate the noise contributions at high $\ell_{\phi \phi}$.

Additionally,
the lensing reconstruction depends on CMB modes satisfying the triangle constraint up to an $\ell_{\rm max}^{UV}$ where they become noise dominated (and hence are cut off by the weights), unless an $\ell_{\rm max}^{UV}$ cut-off scale is imposed.  For $\ell_{\phi \phi} \ll \ell_{\rm max}^{UV}$ most of the temperature (and E-polarization) lensing reconstruction is from squeezed shapes involving only high-$\ell_{UV}$ modes which have the largest weights and lowest cosmic variance; the noise correlation at moderate $\ell_{\phi \phi}$ is therefore mainly with high $\ell_{\rm max}^{UV}$ for TT.
For $\ell_{\phi \phi} \sim \ell_{\rm max}^{UV}$ the triangles can no longer be squeezed, and a wide range of larger-scale CMB modes contribute (with small weight on large scales being compensated by the other mode being near $\ell_{\rm max}^{UV}$). The cosmic variance of the large-scale
CMB modes is however much larger than that of the small-scale ones, resulting in a strong correlation structure around $\ell_{UV} \sim 200$ and the first few peaks where there is the smallest number of modes. The noise correlation with EE has a similar structure at high $\ell_{\phi \phi}$, but at lower $\ell_{\phi \phi}$ there is a broader range of correlation scales (as for BB) because the E modes also enter the EB estimator.

As demonstrated for temperature in Ref.~\cite{Schmittfull:2013uea}, and shown in Figs.~\ref{fig:corrmat_phixCMB_CMBS4_noise} $\&$ \ref{fig:corrmat_phixCMB_CMBS4_noise_RDN0} in the general case, this leading-order noise term can be efficiently mitigated by the use of the realization-dependent $\hat{N}^{(0)}$ correction\footnote{See also first line of Eq.~\eqref{eq:clphiphi-RDN0}.}.
\\

\paragraph{Connected 4-point function: Type A trispectrum.}

In addition to the noise term, there is another contribution from the CMB trispectrum to the cross-covariance, denoted Type A trispectrum hereafter.
It is a correction term to the noise contribution in Eq.~\ref{eq:xcorr-noise} containing higher order terms in $C^{\phi \phi}$, and its general form is given by

\begin{equation}\label{eq:xcorr-trispectrumA}
\covPhiCMB{\ell_1}{\ell_2}{XY}{ZW}{U}{V}_{\text{trispectrum}}^{\text{Type A}} = \sum_{(ab), \ell_3} \dfrac{\partial \Nzero{\ell_1}{XY}{ZW}}{\partial \CellNoHatLensednoisy{\ell_3}{a}{b}} \covlensCMBells{\ell_3}{\ell_2}{a}{b}{U}{V}_{\text{conn.4pt}}.
\end{equation}
See Appendix~\ref{app:xcorr-trispectrum} for more details on the derivation of this term.

\subsection{Cross-covariances between 2-point and 4-point functions: connected 4-point functions} \label{app:xcorr-trispectrum}

We are interested in connected 4-point function contribution to the 6-point function in the last line of Eq.~\ref{eq:full-phiphiCMB}.
This is basically the contribution from the lensed CMB trispectrum to the cross-covariance.
We split the 6-point function in three groups $G$ of two fields:
\begin{equation} \label{eq:split-phixCMB}
\underbrace{\tX_{\underline{\ell}_3} \tY_{\underline{\ell}_4}}_{G1} \quad | \quad \underbrace{\tZ_{\underline{\ell}_5} \tW_{\underline{\ell}_6}}_{G2} \quad \Big | \quad \underbrace{\tU_{\ell_2 m_2} \tV_{\ell_2, -m_2}}_{G3}
\end{equation}
Following Ref.~\cite{Schmittfull:2013uea}, we can show that the non-vanishing contributions are of the form\footnote{The notation $\langle G1G2 \rangle$ means that we correlate one field from group G1 with one field from group G2.} $\langle G1G2 \rangle \times \langle G1G2G3G3\rangle_{c}$ (Type A) and $[\langle G1G3\rangle \times \langle G1G2G2G3 \rangle_{c}) + \langle G2G3\rangle \times \langle G1G1G2G3\rangle_{c}]$ (Type B).
\\

\paragraph{Type A:} One can express the full expression in a rather simple form by noticing that the pairing $\langle G1G2 \rangle$ gives rise to the lensed CMB power spectrum, while the pairing $\langle G1G2G3G3\rangle_{c}$ is the connected 4-point function contribution to the lensed CMB auto-correlation (see Sec. \ref{sec:two-point-covariances}).
Then, using sum over $m$'s and the Eq.~\ref{eq:amplitude-estimator}, we obtain:

\begin{equation}\label{eq:app-xcorr-trispectrumA}
\covPhiCMB{\ell_1}{\ell_2}{XY}{ZW}{U}{V}_{\text{trispectrum}}^{\text{Type A}} = \sum_{(ab), \ell_3} \dfrac{\partial \Nzero{\ell_1}{XY}{ZW}}{\partial \CellNoHatLensednoisy{\ell_3}{a}{b}} \covlensCMBells{\ell_3}{\ell_2}{a}{b}{U}{V}_{\text{conn.4pt}},
\end{equation}
for CMB pair $ab \in \{ TT, TE, EE, BB \}$.
We can interpret the covariance Eq.~\ref{eq:app-xcorr-trispectrumA} as the correction to the noise contribution of Eq.~\ref{eq:xcorr-noise} due to the non-diagonal parts of the lensed CMB auto-covariance at higher orders in $C^{\phi \phi}$ (Eq.~\ref{eq:xcorr-noise} is $\OrderStop{\phi}{\phi}{0}$, while Eq.~\ref{eq:app-xcorr-trispectrumA} contains in our development $\OrderStop{\phi}{\phi}{1}$ and $\OrderStop{\phi}{\phi}{2}$ terms).
\\

\paragraph{Type B:} In general, there are 8 non-vanishing terms. To compute this term, we need the expression for the lensed CMB trispectrum.
First let's notice that the connected 4-point function $\langle \tY_{\underline{\ell}_4} \tZ_{\underline{\ell}_5} \tW_{\underline{\ell}_6}  \tV_{\ell_2, -m_2}  \rangle_{c}$ at $\OrderStop{\phi}{\phi}{1}$ contains only terms $\langle \tY_{\underline{\ell}_4} \tZ_{\underline{\ell}_5} \delta W_{\underline{\ell}_6}  \delta V_{\ell_2, -m_2}  \rangle_{c}$ and permutations thereof ($\langle \tY_{\underline{\ell}_4} \tZ_{\underline{\ell}_5} \tW_{\underline{\ell}_6}  \delta^2 V_{\ell_2, -m_2}  \rangle_{c}$ and permutations are cancelled by the disconnected part).
Therefore, using Eq.~\ref{eq:taylor-CMB-multipoles-order1} we have
\begin{equation} \label{eq:lensed-CMB-trispectrum-order2}
\langle \tY_{\underline{\ell}_4} \tZ_{\underline{\ell}_5} \tW_{\underline{\ell}_6}  \tV_{\ell_2, -m_2} \rangle_{c}^{(2)} = \dfrac{1}{8} \sum_{\ell m} (-1)^{m} \wigner{\ell_3}{\ell_4}{\ell}{m_3}{m_4}{-m} \wigner{\ell_5}{\ell_6}{\ell}{m_5}{m_6}{m} \Cell{\ell}{\phi}{\phi} \fweightLensed{\ell_3}{\ell}{\ell_4}{YZ}\fweightLensed{\ell_5}{\ell}{\ell_6}{WV} \text{ + all permutations,}
\end{equation}
where the superscript (2) indicates that we stopped the development at $\OrderStop{\phi}{\phi}{1}$.
The general formula for the lensed CMB trispectrum contribution to the covariance does not have a simple expression (see Eq.~\ref{eq:xcorr-trispectrumB} for the general case), so we focus here on only few combinations that simplify.

Let's consider $X=Y=Z=W=U=V$, then inserting Eq.~\ref{eq:lensed-CMB-trispectrum-order2}, Eq.~\ref{eq:amplitude-estimator} and \ref{eq:gweights} in the cross-covariance equation leads to the final result
\begin{equation}
\covPhiCMB{\ell_1}{\ell_2}{XX}{XX}{X}{X}_{\text{trispectrum}}^{\text{Type B - primary}} = 2 \dfrac{\Cell{\ell_1}{\phi}{\phi}}{\Amplitude{\ell_1}{XX}} \dfrac{\partial \Nzero{\ell_1}{XX}{XX}}{\partial \CellNoHatLensednoisy{\ell_2}{X}{X}} \dfrac{2}{2\ell_2 +1} (\CellNoHatLensednoisy{\ell_2}{X}{X})^2.
\label{eq:xcorr-trispectrumB-diag}
\end{equation}
If we consider $X=Z=E$, $Y=W=B$, and $U=V \in \{ E,B \}$, we have only 4 non-vanishing terms (terms with $\langle EB \rangle$ will vanish), and we obtain:

\begin{equation}
\covPhiCMB{\ell_1}{\ell_2}{EB}{EB}{U}{U}_{\text{trispectrum}}^{\text{Type B - primary}} = 2 \dfrac{\Cell{\ell_1}{\phi}{\phi}}{\Amplitude{\ell_1}{EB}} \dfrac{\partial \Nzero{\ell_1}{EB}{EB}}{\partial \CellNoHatLensednoisy{\ell_2}{U}{U}} \dfrac{2}{2\ell_2 +1} (\CellNoHatLensednoisy{\ell_2}{U}{U})^2.
\end{equation}
These simplified forms are useful to gain some intuition: we can see that the effect of the Type B trispectrum will be the same as the noise term from Eq.~\ref{eq:xcorr-noise}, modulated by the signal-to-noise of the reconstruction $(C^{\phi \phi}/{\cal{A}})$.

\subsection{Lensing auto-covariances: terms cancelled by the use of RDN0} \label{app:four-point-covariances_rdn0ed}

By expanding the second term on the RHS of Eq.~\ref{eq:autocorr-RDNzero} in terms of the different contributions from Sec.~\ref{sec:cross-covariances}, the last term on the RHS is cancelled because of contributions coming from Eq.~\ref{eq:xcorr-noise} and \ref{eq:xcorr-trispectrumA} from Appendix~\ref{app:crosscancelled}, and we are left with
\begin{align}
\text{cov}(\CellHatRDN{\ell_1}{\phi^{XY}}{\phi^{ZW}},\CellHatRDN{\ell_2}{\phi^{X'Y'}}{\phi^{Z'W'}}) =\,&
\covPhi{\ell_1}{\ell_2}{XY}{ZW}{X^\prime Y^\prime}{Z^\prime W^\prime}
-
\sum_{(ab),\ell_3}\frac{\partial \Nzero{\ell_1}{XY}{ZW}}{\partial \CellNoHatLensednoisy{\ell_3}{a}{b}}
\covPhiCMBtransp{\ell_2}{\ell_3}{X'Y'}{Z'W'}{a}{b}_{\text{Sec.\ref{sec:notcrosscancelled}}}
\nonumber\\
&-
\sum_{(ab),\ell_3}
\covPhiCMB{\ell_1}{\ell_3}{XY}{ZW}{a}{b}_{\text{Sec.\ref{sec:notcrosscancelled}}}
\frac{\partial \Nzero{\ell_2}{X'Y'}{Z'W'}}{\partial \CellNoHatLensednoisy{\ell_3}{a}{b}}
\nonumber\\
&-
\sum_{(ab),(cd),\ell_3, \ell_4}
\frac{\partial \Nzero{\ell_1}{XY}{ZW}}{\partial \CellNoHatLensednoisy{\ell_3}{a}{b}}
\text{cov}(\CellLensednoisy{\ell_3}{a}{b}, \CellLensednoisy{\ell_4}{c}{d} )
\dfrac{\partial \Nzero{\ell_2}{X^\prime Y^\prime}{Z^\prime W^\prime}}{\partial \CellNoHatLensednoisy{\ell_4}{c}{d}}.
\label{eq:autocorr-RDNzero_cross}
\end{align}
The second and third terms of this equation contain the signal and Type B trispectrum contributions from the Sec.~\ref{sec:notcrosscancelled} (Eqs.~\ref{eq:xcorr-signal} $\&$ \ref{eq:xcorr-trispectrumB} respectively).
The last term of the RHS of this equation contains the noise and Type A trispectrum contributions from the Appendix~\ref{app:crosscancelled} (Eqs.~\ref{eq:xcorr-noise} $\&$ \ref{eq:xcorr-trispectrumA} respectively) that we explicitly expand to make the auto-covariance of the CMB to appear.
\\

\paragraph{Contribution from reconstruction noise fluctuating with the lensed CMB power.}
\label{sec:phiphi_8pt_rdn0ed}
The last term on the RHS of Eq.~\ref{eq:autocorr-RDNzero_cross} contains the auto-covariance of the lensed CMB.
It cancels contractions that appear in the raw estimator
 power spectrum estimators, specifically those terms arising from connecting two lensed CMB fields in each of the estimators:
\begin{multline}
\frac{1}{4}
\sum_{\substack{(abcd)\\ \{\ell'm'\}}}
\left(\la a_{\ell_1'm_1',\rm{expt}} b^*_{\ell_2'm_2',\rm{expt}} c_{\ell_3'm_3',\rm{expt}} d^*_{\ell_4'm_4',\rm{expt}}\ra
-\la a_{\ell_1'm_1',\rm{expt}} b^*_{\ell_2'm_2',\rm{expt}} \ra\la c_{\ell_3'm_3',\rm{expt}} d^*_{\ell_4'm_4',\rm{expt}}\ra\right)
\\
\times
\left\la\frac{\partial^2 \hat{C}_{\ell_1}^{\phi^{XY}\phi^{ZW}}}{\partial \tilde{a}_{\ell_1' m_1',\rm{expt}}\partial \tilde{b}^*_{\ell_2'm_2',\rm{expt}}}\right\ra
\left\la\frac{\partial^2 \hat{C}_{\ell_3}^{\phi^{X'Y'}\phi^{Z'W'}}}{\partial \tilde{c}_{\ell_3' m_3',\rm{expt}}\partial \tilde{d}^*_{\ell_4'm_4',\rm{expt}}}\right\ra
\\
\approx
\sum_{(ab),(cd),\ell_3, \ell_4}
\frac{\partial \Nzero{\ell_1}{XY}{ZW}}{\partial \CellNoHatLensednoisy{\ell_3}{a}{b}}
\text{cov}(\CellLensednoisy{\ell_3}{a}{b}, \CellLensednoisy{\ell_4}{c}{d} )
\dfrac{\partial \Nzero{\ell_2}{X^\prime Y^\prime}{Z^\prime W^\prime}}{\partial \CellNoHatLensednoisy{\ell_4}{c}{d}},
\label{eq:8ptnoisecontri}
\end{multline}
where the CMB auto-covariance includes both Gaussian and non-Gaussian parts, but we keep only the Gaussian reconstruction noise part of the estimator expectations.
\\

\begin{figure}[tbp]
\begin{center}
\includegraphics[width=1.0\textwidth]{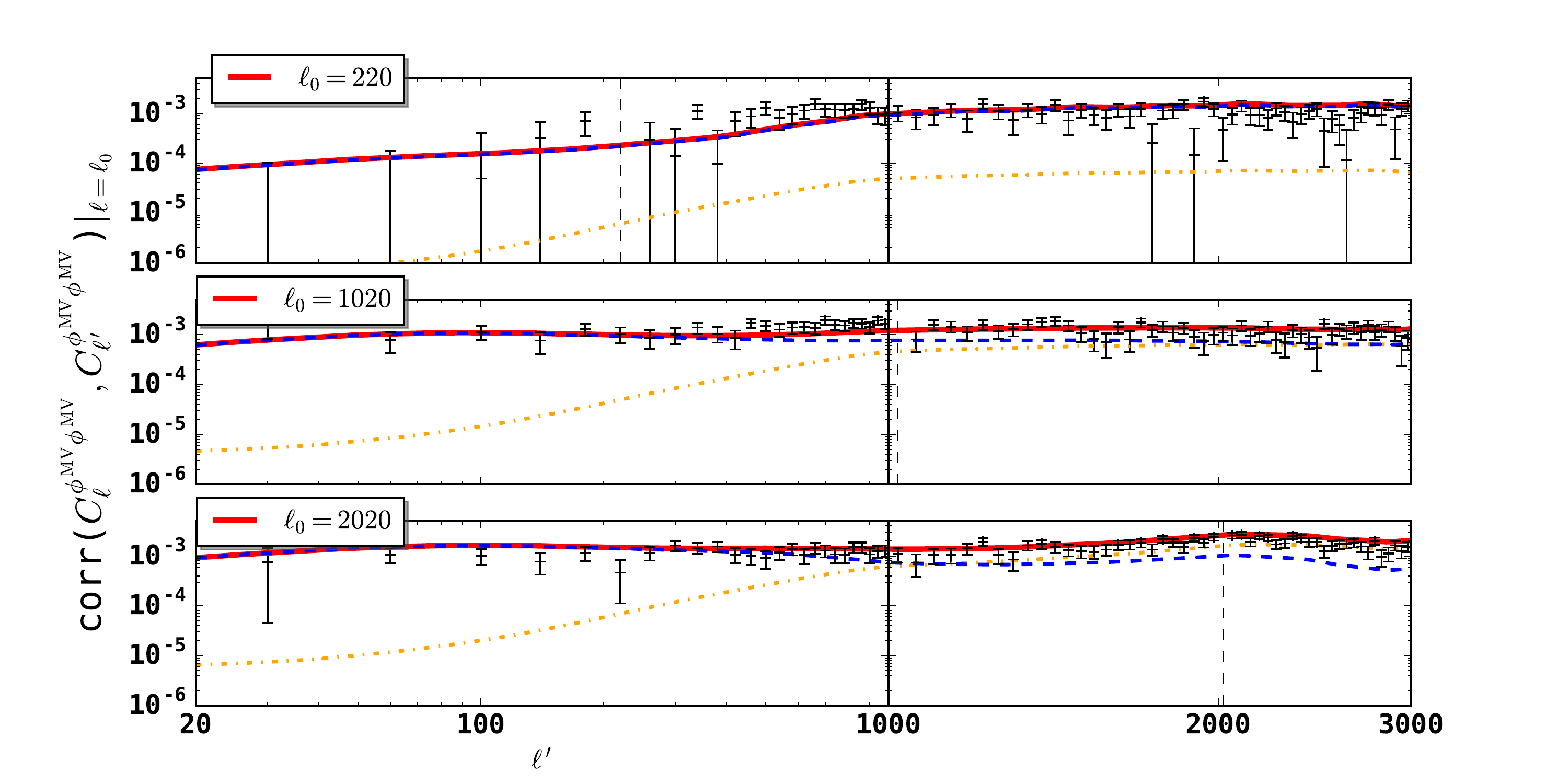}
\caption{
Different slices of the reconstructed lensing potential power spectrum correlation matrix: $\ell_0=220$ (top), $\ell_0=1020$ (middle), $\ell_0=2020$ (bottom). The analytic model (solid red curves) contains the different contributions listed in Appendix~\ref{app:four-point-covariances_rdn0ed} (contributions removed by the use of RDN0), namely the contributions from Appendix~\ref{sec:phiphi_8pt_rdn0ed} (yellow dashed-dotted curves), and from Appendix~\ref{sec:phiphi_crossterm_rdn0ed} (dashed blue curves).
For visualisation purpose, we subtract the reconstruction variance (see Sec.~\ref{sec:phiphi_gaussvar_notrdn0ed}) and we emphasise the scale $\ell^\prime=\ell_0$ by a vertical line. We also show the results obtained on simulation (black crosses). The model and the simulations are in relatively good agreement although some difference are seen. After RDN0 subtraction, only the Gaussian reconstruction variance is considered in our model. See text for discussions.
}
\label{fig:corrmat_phixphi_CMBS4}
\end{center}
\end{figure}

\paragraph{Contribution from signal and trispectrum terms in the lensing reconstruction-CMB cross-covariance.} \label{sec:phiphi_crossterm_rdn0ed}
The second and third terms on the RHS of Eq.~\ref{eq:autocorr-RDNzero} cancel terms appearing in the raw estimator covariance involving terms described in Sec.~\ref{sec:crosscancelled}.
We expect that these terms might be relevant because the connected 6-point function and the primary (Type B) connected 4-point function have a large effect on the cross-covariance between CMB and reconstruction power spectra.
We expect the dominant terms in Eq.~\ref{eq:full-phiphiphiphi} that multiply a CMB 2-point with a CMB 6-point function to be
$\langle \tX_{\underline{\ell}_3}\tZ_{\underline{\ell}_5} \rangle \langle\tY_{\underline{\ell}_4} \tW_{\underline{\ell}_6} \hat C^{\phi\phi}_{\ell_2}\rangle$,
$\langle \tX_{\underline{\ell}_3}\tW_{\underline{\ell}_6}\rangle \langle \tY_{\underline{\ell}_4}\tZ_{\underline{\ell}_5} \hat C^{\phi\phi}_{\ell_2} \rangle$,
$\langle \tY_{\underline{\ell}_4} \tZ_{\underline{\ell}_5} \rangle \langle \tX_{\underline{\ell}_3} \tW_{\underline{\ell}_6} \hat C^{\phi\phi}_{\ell_2} \rangle$ and
$\langle \tY_{\underline{\ell}_4} \tW_{\underline{\ell}_6}\rangle \langle \tX_{\underline{\ell}_3} \tZ_{\underline{\ell}_5} \hat C^{\phi\phi}_{\ell_2} \rangle$,
and four similar terms obtained from $\ell_1 \leftrightarrow \ell_2$, as
\footnote{The last line can be obtained as follows. In the permutation written out, $\langle \tX_{\underline{\ell}_3} \tZ_{\underline{\ell}_5}\rangle$ enforces $l_3=l_5$ and $m_3=-m_5$.
The sum over $m_1$ and $m_3$ then enforces $\ell_4=\ell_6$ and $m_4=-m_6$.
Summing over $m_4$ leads to $\langle\hat C^{\tY\tW}_{\ell_4}\hat C^{\phi^{X^\prime Y^\prime}\phi^{Z^\prime W^\prime}}_{\ell_2}\rangle^{(1)+(2)}$, which can be expressed in terms of contributions to the covariance between measured lensed CMB and lensing reconstruction power spectra from Sec.~\ref{sec:notcrosscancelled}. The sum over $\ell_3$ leads to $\partial N^{(0)}/\partial C^{\tilde{a}\tilde{b}}_\text{expt}$.
}
\begin{multline} \label{eq:phiphixphiphi-cross}
\dfrac{\Amplitude{\ell_1}{XY} \Amplitude{\ell_1}{ZW} }{(2\ell_1 +1)} \sum_{\underline{\ell}_3,...,\underline{\ell}_{6}, m_1} (-1)^{m_1} \gweight{\ell_3}{\ell_4}{\ell_1}{XY} \gweight{\ell_5}{\ell_6}{\ell_1}{ZW}  \wigner{\ell_3}{\ell_4}{\ell_1}{m_3}{m_4}{-m_1} \wigner{\ell_5}{\ell_6}{\ell_1}{m_5}{m_6}{m_1}  \\
\times  \Big [ \langle \tX_{\underline{\ell}_3} \tZ_{\underline{\ell}_5} \rangle \langle\tY_{\underline{\ell}_4}  \tW_{\underline{\ell}_6}
\hat C^{\phi^{X^\prime Y^\prime}\phi^{Z^\prime W^\prime}}_{\ell_2}
 \rangle^{(1)+(2)} + \text{3 similar} \Big ] \;+ \;(\ell_1 \leftrightarrow \ell_2) \\
= \sum_{(ab),\ell_4}\frac{\partial \Nzero{\ell_1}{XY}{ZW}}{\partial \CellNoHatLensednoisy{\ell_4}{a}{b}}
\covPhiCMBtransp{\ell_2}{\ell_4}{X'Y'}{Z'W'}{a}{b}_{\text{Sec.~\ref{sec:notcrosscancelled}}} + \sum_{(ab),\ell_4}
\covPhiCMB{\ell_1}{\ell_4}{XY}{ZW}{a}{b}_{\text{Sec.~\ref{sec:notcrosscancelled}}}
\frac{\partial \Nzero{\ell_2}{X'Y'}{Z'W'}}{\partial \CellNoHatLensednoisy{\ell_4}{a}{b}},
\end{multline}
where (1)+(2) means that only $\OrderStop{\phi}{\phi}{1}$ connected 4-point contributions (Eq.~\ref{eq:xcorr-trispectrumB}), and $\OrderStop{\phi}{\phi}{2}$ matter cosmic variance contribution (Eq.~\ref{eq:xcorr-signal}) from the connected 6-point function are included in the cross-covariance between the reconstructed lensing potential power spectrum and lensed CMB power spectra in the last line (excluding the $\OrderStop{\phi}{\phi}{0}$ noise contribution because it is already included in Eq.~\ref{eq:8ptnoisecontri}).
We show the contribution of Eq.~\ref{eq:phiphixphiphi-cross} to the total auto-correlation in Fig. \ref{fig:corrmat_phixphi_CMBS4} (dashed blue curves).
This term is responsible of large off-diagonal correlations, and it dominates over other contributions at small scales, but also at large-scales because of the signal term from Eq.~\ref{eq:xcorr-signal}.
\\

\paragraph{Comparison against simulations}
In Fig.~\ref{fig:corrmat_phixphi_CMBS4} we plot the different contributions removed by the RDN0 subtraction (from Appendices~\ref{sec:phiphi_8pt_rdn0ed} $\&$ \ref{sec:phiphi_crossterm_rdn0ed}).
For comparison, we overplot the results obtained using our set of simulations.
The agreement is on overall good, although some discrepancies can be seen especially at large scales ($\ell_{\phi \phi} < 1000$).
Those differences appear to be unimportant for our purpose (as seen in Sec.~\ref{sec:cosmo-param}) since after applying the realization-dependent bias subtraction the auto-covariance is predominantly dominated by diagonal elements.
The agreement of diagonal elements (considering only Eq.~\ref{eq:gaussvar-phiphiphiphi}) between the model and the simulations is at the sub-percent level.

\section{Evaluating power spectrum derivatives} \label{app:corr_func}

Following \cite{Lewis:2006fu}, the lensed CMB temperature power spectrum can be written as

\begin{equation}\label{eq:cllensed-corrfunc}
C^{\tT \tT}_{\ell} = 2\pi \int_{-1}^{1} \xi(\beta) \wignerd{0}{0}{\ell}(\beta) d(\cos\beta),
\end{equation}
where $\xi$ is the correlation function and $\wignerd{m}{m^{\prime}}{\ell}$ is the Wigner (small) d-matrix \footnote{The integration in Eq.~\ref{eq:cllensed-corrfunc} (and same for polarization) is done using a Gauss-Legendre quadrature rule:

\begin{equation}
C^{\tT \tT}_{\ell} = 2\pi \int_{-1}^{1} \xi(\beta) \wignerd{0}{0}{\ell}(\beta) d(\cos\beta) \approx 2\pi \sum_{i=1}^{n} \xi(x_i) \wignerd{0}{0}{\ell}(x_i) w_i
\end{equation}
where here $w_i$ are Gauss-Legendre integration weights.}.
The correlation function can be expressed in terms of the unlensed spectrum and the lensing potential power spectrum to good accuracy as

\begin{equation}
\xi(\beta) \approx \sum_{\ell_1} \dfrac{2\ell_1 +1}{4\pi} C^{T T}_{\ell_1} e^{-\ell_1 (\ell_1 +1)\sigma(\beta)^2/2} \Big( \wignerd{0}{0}{\ell_1}(\beta)  + \dfrac{\ell_1(\ell_1 +1)}{2} C_{{\rm gl},2}(\beta) \wignerd{1}{-1}{\ell_1}(\beta) \Big),
\end{equation}
where $\sigma(\beta)^2 = C_{\rm gl}(0) - C_{\rm gl}(\beta)$, and

\begin{align}
C_{\rm gl}(\beta) &= \sum_{\ell} \dfrac{(2\ell +1)(\ell+1)\ell}{4\pi} C_{\ell}^{\phi \phi} \wignerd{1}{1}{\ell}(\beta) \\
C_{{\rm gl},2}(\beta) &= \sum_{\ell} \dfrac{(2\ell +1)(\ell+1)\ell}{4\pi} C_{\ell}^{\phi \phi} \wignerd{1}{-1}{\ell}(\beta).
\end{align}
Then we obtain the derivative of the lensed CMB power spectrum using

\begin{equation}
\dfrac{\partial C^{\tT \tT}_{\ell_1}}{\partial C_{\ell_2}^{X Y}} = 2\pi \int_{-1}^{1} \dfrac{\partial \xi(\beta)}{\partial C_{\ell_2}^{X Y}} \wignerd{0}{0}{\ell_1}(\beta) d(\cos\beta),
\end{equation}
where $X,Y \in \{ T,E,B,\phi \}$.
Extending these results to polarization gives:

\begin{align} \label{eq:derivative-cltt-clpp-corrfunc}
\dfrac{\partial C^{\tE \tE}_{\ell}}{\partial C_{\ell_2}^{X Y}} - \dfrac{\partial C^{\tB \tB}_{\ell}}{\partial C_{\ell_2}^{X Y}} &= 2\pi \int_{-1}^{1} \dfrac{\partial \xi^{-}(\beta)}{\partial C_{\ell_2}^{X Y}}  \wignerd{2}{-2}{\ell}(\beta) d(\cos\beta), \\
\dfrac{\partial C^{\tE \tE}_{\ell}}{\partial C_{\ell_2}^{X Y}}  + \dfrac{\partial C^{\tB \tB}_{\ell}}{\partial C_{\ell_2}^{X Y}}  &= 2\pi \int_{-1}^{1} \dfrac{\partial \xi^{+}(\beta)}{\partial C_{\ell_2}^{X Y}}  \wignerd{2}{2}{\ell}(\beta) d(\cos\beta), \\
\dfrac{\partial C^{\tT \tE}_{\ell_1}}{\partial C_{\ell_2}^{X Y}} &= 2\pi \int_{-1}^{1} \dfrac{\partial \xi^{\times}(\beta)}{\partial C_{\ell_2}^{X Y}} \wignerd{2}{0}{\ell_1}(\beta) d(\cos\beta), \label{eq:derivative-cltt-clpp-corrfunc2}
\end{align}
where $\xi^{+}$, $\xi^{-}$, and $\xi^{\times}$ are defined in Ref.~\cite{Lewis:2006fu}.
This method of calculating the lensed power spectrum derivatives is more accurate than using the leading-order series-expansion method (see for example Ref.~\cite{Hu:2000ee}),
and prevents an artificially high correlation between the lensed CMB power spectra and the reconstructed lensing potential power spectrum.
The fact that the correlations are enhanced in the case of the series expansion can be partly understood by noticing that series-expansion method tends to over-lens the signal with respect to the correlation function method, as shown in Fig. \ref{fig:comp-left-singular-vectors}.
In the acoustic region for example, the features due to lensing are artificially enhanced if we use the series-expansion, and therefore the correlation between the lensing amplitude estimates becomes (artificially) stronger.

\begin{figure}[!htbp]
\begin{center}
\includegraphics[width=1.0\textwidth]{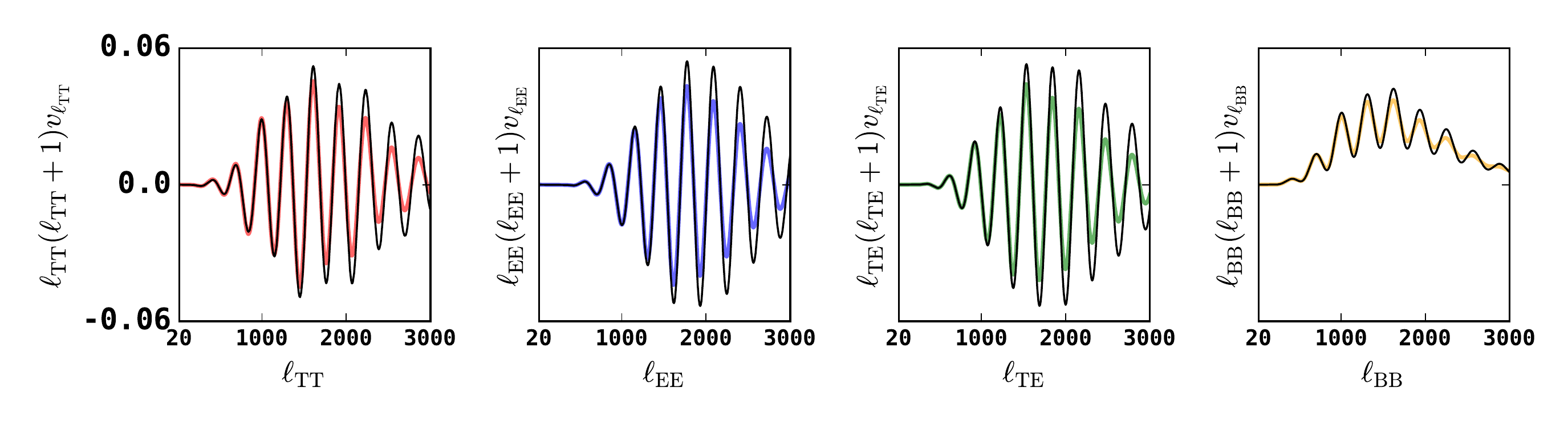}
\caption{First singular vector of the signal covariance matrix (Eq. \ref{eq:xcorr-signal}) when using correlation function (coloured lines, with TT, EE, TE, BB in red, blue, green, and yellow respectively) or series-expansion methods (black lines) to compute the derivative of the lensed CMB spectra with respect to the lensing potential power spectrum. This first singular vector represents mainly the difference between lensed and unlensed CMB power spectra (or E modes in the case of the B modes). The computation of derivatives using series expansion over-lenses the power spectra compared to the correlation function method (Eqs.~\ref{eq:derivative-cltt-clpp-corrfunc}--\ref{eq:derivative-cltt-clpp-corrfunc2}).}
\label{fig:comp-left-singular-vectors}
\end{center}
\end{figure}

\section{N1 deconvolution}

\label{app:n1-deconvolution}
The estimation of the lensing potential power spectrum suffers from several biases.
The influence of the $N^{(0)}$ bias on the covariance is reduced using realization-dependent bias subtraction (see Eq.~\ref{eq:RDN0_gaussian_bias}).
However, there is still an $N^{(1)}$ bias, which is mostly seen at very small scales.
This can be modelled analytically, and subtracted from the lensing potential power spectrum estimate (with perturbative corrections to account for model dependence when used for a likelihood~\cite{Ade:2015zua}).
However, subtracting it does not reduce the off-diagonal elements of the covariance of the estimator ($N^{(1)}_\ell$ depends on the lensing spectrum over a wide range of $\ell$).
Instead we could deconvolve the estimator from the bias using
\begin{equation}
\hat{C}_{\ell,\text{decon}}^{\phi^{\rm MV} \phi^{\rm MV}} = \sum_{\ell^{\prime}}  \Big( \mathbf{I} + \mathbf{N}^{(1),\rm MV} \Big)_{\ell \ell^\prime}^{-1} (\CellHat{\ell^\prime}{\phi^{\rm MV}}{\phi^{\rm MV}} - \NzeroHatCAL{\ell^\prime}{\rm MV}{ }),
\end{equation}
where the elements of the $\mathbf{N}^{(1),\rm MV}$ matrix are given by

\begin{equation}
\mathbf{N}^{(1),\rm MV}_{\ell \ell^\prime} = \sum_{XY,ZW} w_{\ell}^{XY} w_{\ell}^{ZW} \dfrac{\partial N^{(1),XYZW}_\ell}{\partial C_{\ell^\prime}^{\phi \phi}}.
\end{equation}
We show in Fig.~\ref{fig:N1_deconvolution} the effect of the $N^{(1)}$ deconvolution on the auto-correlation of the reconstructed lensing potential power-spectrum, with and without the realization-dependent noise subtraction applied.
One can see that the effect of the $N^{(1)}$ deconvolution on the off-diagonal elements takes place at very small scales (where $N^{(1)}$ is relatively important), but also between large and intermediate lensing scales.
However, given the experimental setup chosen here, and the smallness of the off-diagonal elements prior to the deconvolution, the deconvolution has a negligible impact on cosmological parameter estimation.
This is consistent with our neglect of the off-diagonal $N^{(1)}$ covariance in the covariance model of Sec.~\ref{sec:four-point-covariances}.

\begin{figure}[!htbp]
\begin{center}
\includegraphics[width=0.7\textwidth]{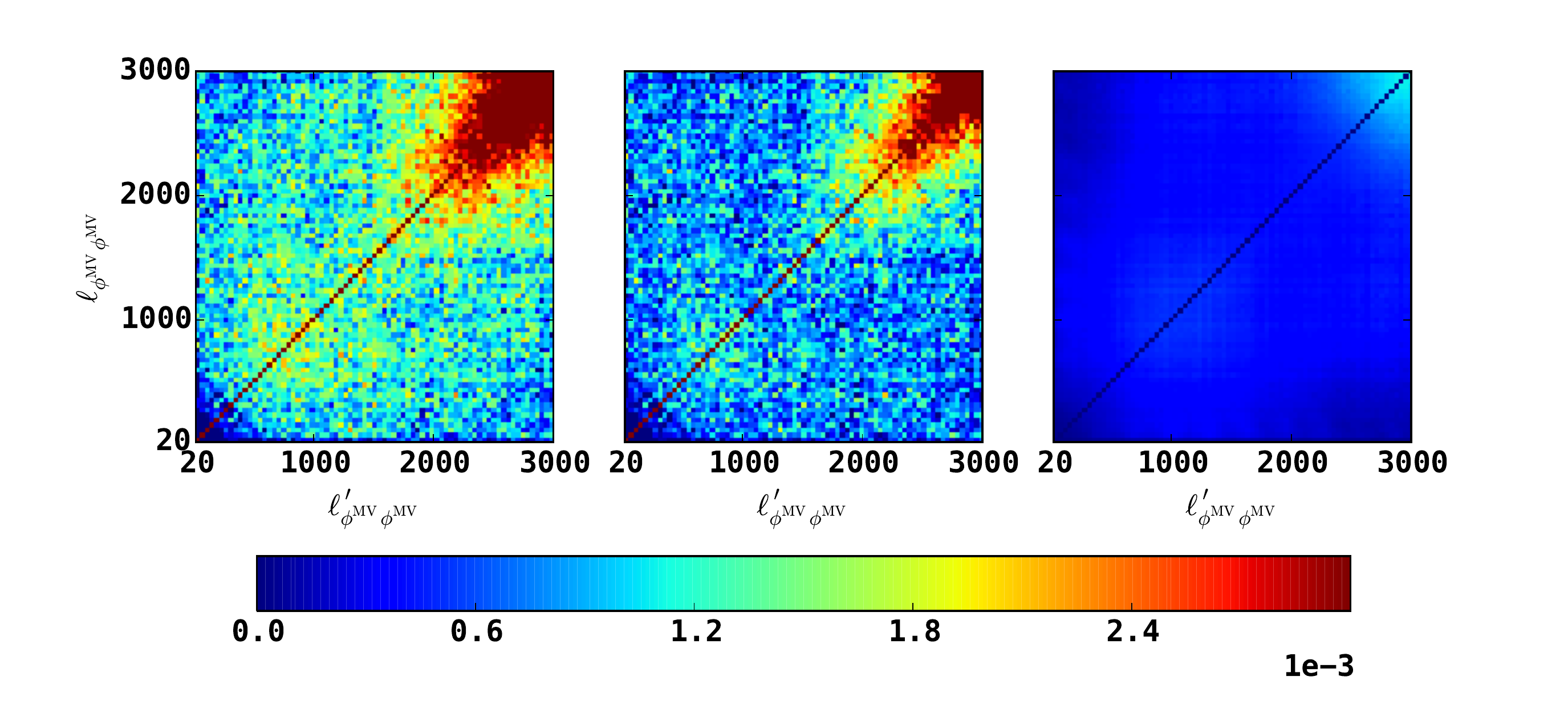}
\includegraphics[width=0.7\textwidth]{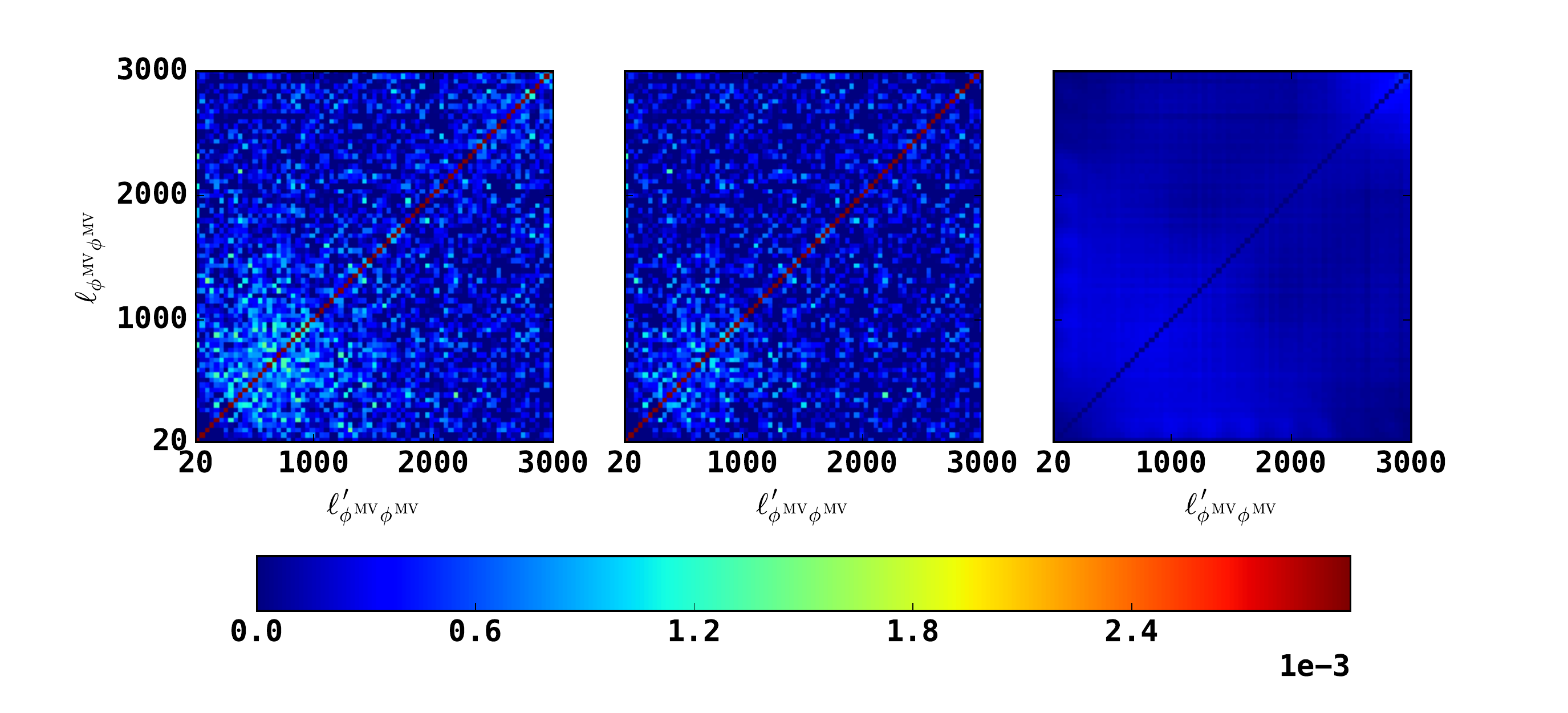}
\caption{Effect of the $N^{(1)}$ deconvolution on the auto-correlation of the reconstructed lensing potential power-spectrum. \textit{Top}: The left sub-panel shows the auto-correlation of the reconstructed lensing potential power-spectrum (no RDN0 subtraction), the middle sub-panel shows the auto-correlation of the reconstructed lensing potential power-spectrum with the $N^{(1)}$ deconvolution applied, and the right sub-panel is the difference between both. The off-diagonal elements are reduced. The reduction takes place at very small scales (where $N^{(1)}$ is relatively important), but also between large and intermediate lensing scales. \textit{Bottom}: Same, but the realization-dependent noise subtraction has been applied prior to the deconvolution.}
\label{fig:N1_deconvolution}
\end{center}
\end{figure}

\end{appendix}

\bibliography{antony,cosmomc}
\fi

\end{document}